\documentclass[pre,aps,twocolumn]{revtex4}
\usepackage{graphicx}
\usepackage{dcolumn}
\usepackage{bm,physics}
\usepackage{amsmath,amssymb,blkarray}
\usepackage{color,soul}
\usepackage[toc,page]{appendix}

\begin{document}

\title{Unconventional thermal and magnetic-field-driven changes of a bipartite entanglement of a  mixed spin-(1/2,$S$) Heisenberg dimer with an uniaxial single-ion anisotropy}
\author{Hana Vargov\'a$^{1,\footnote{Corresponding author: hcencar@saske.sk}}$ and Jozef Stre\v{c}ka$^2$}
\affiliation{
$^1$Institute of Experimental Physics, Slovak Academy of Sciences, Watsonova 47, 040 01 Ko\v {s}ice, Slovakia \\ 
$^2$ Department of Theoretical Physics and Astrophysics, Faculty of Science, P. J. \v{S}af\'{a}rik University, Park 
Angelinum 9, 040 01 Ko\v{s}ice, Slovakia}

\begin{abstract}
The concept of negativity is adapted in order to explore the quantum and thermal entanglement  of  the mixed spin-(1/2,$S$) Heisenberg dimers in  presence of an external magnetic field. The mutual   interplay between the spin size $S$, XXZ exchange  and uniaxial single-ion anisotropy is thoroughly examined  with a goal to tune the degree and thermal stability of the pairwise entanglement.  It turns out that the antiferromagnetic spin-(1/2,$S$) Heisenberg dimers  exhibit  higher degree of entanglement and higher threshold temperature in comparison with their ferromagnetic counterparts when assuming   the same set of model parameters. The increasing spin magnitude $S$ accompanied with an easy-plane uniaxial single-ion anisotropy can enhance not only the thermal stability  but simultaneously the degree of entanglement. It is additionally shown  that the further enhancement of a bipartite entanglement  can be achieved  in the mixed spin-(1/2,$S$) Heisenberg dimers, involving half-odd-integer spins $S$.  Under this condition the thermal negativity  saturates at low-enough temperatures in its maximal value regardless of the magnitude of half-odd-integer spin $S$. The   magnetic field induces consecutive discontinuous  phase transitions in the mixed spin-(1/2,$S$) Heisenberg dimers with $S\!>\!1$, which are manifested in a  surprising oscillating magnetic-field dependence  of the negativity  observed at low enough temperature. 
\end{abstract} 

\maketitle

\section{Introduction}
\label{sec:introduction}

 Extraordinary correlations between subsystems of a quantum-mechanical ensemble, known as entanglement, belongs to the most fascinating phenomena attracting a lot of attention during the last few decades.  A huge concerment in this field of study  closely relates to  perspective applications of this phenomenon in a quantum computing~\cite{Loss,Hayashi,Shulman,Shi}, quantum information~\cite{Jones,Delbecq}, and quantum memory circuits decoding~\cite{Watson,Berger,Li17}. However, a long period before, it was believed that the entanglement could exist exclusively on the atomic scale and completely vanishes at macroscopic scales as a consequence of the decoherence  arising from the  interactions between  a large number of matter constituents and their environment. It was presumed, in addition, that arbitrarily small thermal fluctuations rapidly smear out the quantum correlations  and thus, the entanglement cannot exist at   non-zero temperatures. Some theoretical  predictions~\cite{Gunlycke,Arnesen,Wang,Zhou,Wang2,Wang3} have presented  relevant  arguments that the entanglement could surprisingly  exist  even at  finite temperatures, but rapidly falls down as temperature increases.  
 
From a theoretical perspective   low-dimensional  Heisenberg spin models seem to be a reasonable theoretical ground, which allows an exact study of the  quantum and thermal entanglement depending on  external stimuli such as the magnetic field and/or temperature~\cite{Asoudeh,Canosa,Torrico,Luitz,Verissimo,Souza19,
Ananikian,Rojas,Arian,Karlova19,Wang2006,Cenci2,Varga21}.   Beside  the most intensively analysed spin-1/2 case,  a few theoretical works were focused on the entanglement of the mixed-spin Heisenberg systems~\cite{Wang2006,Cenci2,Varga21,Sun06,Hao,Guo,Wang08, Huang2008,Yang,Zhu, Sun09,  Wang09,Guo10, Guo11,  Solano1,  Solano2, Li, Xu, Guo2014, Zhou15,  Zhou16,  Han, Adamyan}.  The particular interest in this field of study has been motivated by the pioneering work by Wang et al.~\cite{Wang2006}, which demonstrated   that a higher difference between dissimilar spin constituents can slightly enhance the thermal entanglement at higher temperatures due to the respective shift of its threshold temperature. Unfortunately, the enlargement of a threshold temperature is simultaneously  accompanied by the reduction of the  strength of  mutual quantum correlations. In order to minimize the reduction of the degree of entanglement, many of subsequent studies were concentrated on  an extended mixed-spin Heisenberg chain involving the  Dzyaloshinskii-Moriya interaction (DMI)~\cite{Li,Xu,Zhou15,Zhou16}, the effect of nonuniform magnetic field~\cite{Yang,Wang09,Guo10,Guo11,Guo2014,Zhou15,Adamyan}, long-range interaction~\cite{Han} and uniaxial single-ion anisotropy~\cite{Solano1,Solano2}, respectively.  It was verified for   the mixed spin-(1/2,1), spin-(1/2,3/2) and spin-(1/2,5/2) Heisenberg chains that the inhomogeneity of the external magnetic field can be suitable tuning parameter for enhancing the thermal entanglement in a high-temperature region.  On the other hand, the DMI can enhance the low-temperature entanglement of the antiferromagnetic mixed spin-(1/2,3/2)  Heisenberg dimer~\cite{Zhou16}, nevertheless  the enhancement of entanglement in the mixed spin-(1/2,1) Heisenberg dimer is possible only for the ferromagnetic exchange coupling~\cite{Li}.  Based on the results obtained for the mixed  spin-(1/2,1) Heisenberg dimer~\cite{Solano1,Solano2}, the uniaxial single-ion anisotropy seems to be   another relevant driving force for an enhancement of entanglement in presence of thermal fluctuations.  However, the comprehensive analysis of other mixed-spin Heisenberg dimers with  higher spins $S\!>\!1$ is still absent. This fact motivated us to study the mixed spin-(1/2,$S$) Heisenberg dimers involving the   uniaxial single-ion anisotropy and magnetic field simultaneously with the special goal to verify whether the interplay between the uniaxial single-ion anisotropy, magnetic field  and the  spin magnitude $S$  can enhance   thermal entanglement at high enough temperatures.  It should be emphasized  that  the uniaxial single-ion anisotropy may be relevant for several heterodinuclear complexes as for instance in magnetic compounds~\cite{Holleitner} like the [MnCu(pbaOH)(H$_2$O)$_3$]$\cdot n$H$_2$O (pbaOH = 2 - hydroxy - 1,3 - propylenebis (oxamato))~\cite{Kahn}, [Ni(dpt)(H$_2$O)$_3$Cu(pba)]$\cdot$H$_2$O (pba=1,3 - propylenebis(oxamato) and dpt=bis - (3 - aminopropyl)amine)~\cite{Hagiwara99} or  [NiCu(pba)(D$_2$O)$_3$]$\cdot$2D$_2$O~\cite{Hagiwara98}.

The paper is organized as follows.  The investigated mixed spin-(1/2,$S$) Heisenberg dimer will be defined in Sec.~\ref{sec:model} together with a few details of the calculation procedure used in our rigorous study. The most interesting results concerned with   the quantum and thermal entanglement under the influence of increasing spin  magnitude $S$ will be discussed in Sec.~\ref{sec:result}. Besides,   the effect of an applied external magnetic field on the thermal entanglement  and the threshold temperature of the mixed spin-(1/2,$S$) Heisenberg dimers will be also discussed for an arbitrary spin-$S$. Finally, some concluding remarks are given in Sec.~\ref{sec:conclusion} and a few  details of  analytical derivations are presented in Appendices~\ref{App A}-\ref{App C}. 

\section{Model and Method}
\label{sec:model}

Let us consider the mixed spin-(1/2,$S$) Heisenberg dimers with an uniaxial single-ion anisotropy under the influence of the external   magnetic field defined through the following   Hamiltonian 
\begin{eqnarray}
\hat{\cal H}\!\!\!&=&\!\!\!J\left[\Delta(\hat{\mu}^x\hat{S}^x\!+\!\hat{\mu}^y\hat{S}^y)\!+\!\hat{\mu}^z\hat{S}^z\right]\!+\!D(\hat{S}^z)^2
\nonumber\\
\!\!\!&-&\!\!\!Bg\mu_B\left(\hat{\mu}^z\!+\!\hat{S}^z\right).
\label{eq1}
\end{eqnarray} 
In above, the  symbols $\hat{\mu}^{\alpha}$ and $\hat{S}^{\alpha}$ ($\alpha\!=\!x,y,z$) correspond to spatial components of  spin-1/2 and spin-$S$ ($S\!\geq\!1$)   operators, $J$ is  the XXZ exchange interaction with an exchange anisotropy $\Delta$,  $D$ is  the uniaxial single-ion anisotropy acting on the spin-$S$ magnetic ion only.
Finally, the model under the investigation accounts for the effect of  external magnetic  field $B$  applied along the $z$-direction,  $g$ denotes the  gyromagnetic Land\'e $g$-factor and $\mu_B$ is the Bohr magneton.

In order to study the quantum and thermal entanglement of the  mixed spin-(1/2,$S$) Heisenberg dimers we will employ the concept of Peres-Horodecki~\cite{Peres, Horodecki}, according to  which   negative eigenvalue of a partially transposed density matrix is a necessary condition for the  onset of entanglement. To quantify the strength of quantum and thermal entanglement one may therefore utilize the quantity known as negativity~\cite{Vidal}. The negativity of the mixed state given by the density matrix $\rho$ is by definition  the sum of all negative eigenvalues $\lambda_i$ of partially transposed density matrix ${\hat{\rho}}^{T_{1/2}}$ 
\begin{eqnarray}
{\cal N}(\rho)\!=\!\sum_{\lambda_i<0}|\lambda_i|.
\label{eq2}
\end{eqnarray} 
It is worthwhile to remark that the  negativity of the maximally entangled state is equal to one-half (${\cal N}\!=\!1/2$) for the mixed spin-(1/2,$S$) Heisenberg dimers, whereas  the negativity completely vanishes  (${\cal N}\!=\!0$) in the separable states without the bipartite entanglement..  

Before  calculating the negativity  it is necessary to derive the eigenvalues and eigenvectors of the  Hamiltonian~\eqref{eq1}, which can  be  easily calculated in the standard  orthonormal  basis  $\lvert \mu^z,S^z\rangle$ constructed from all available   eigenvectors of $z$-components of two constituent spins with eigenvalues
 $ \mu^z\!=\!\pm1/2$ and $ S^z\!=\!-S,-S\!+\!1,\dots,S\!-\!1,S$. For this purpose, let us introduce first  the notation for raising and lowering   ladder operators $
\hat{S}^{\pm}\!=\!\hat{S}^x\!\pm\!i\hat{S}^y$ and $ \hat{\mu}^{\pm}\!=\!\hat{\mu}^x\!\pm\!i\hat{\mu}^y$,
which allows us to rewrite the  Hamiltonian~\eqref{eq1} into the more convenient form
\begin{align}
\hat{\cal H}&\!=\!J\left[ \frac{\Delta}{2}\left(\hat{S}^+\hat{\mu}^-\!+\!\hat{S}^-\hat{\mu}^+ \right)\!+\!\hat{S}^z\hat{\mu}^z\right]\!+\!D(\hat{S}^z)^2
\nonumber\\
&\!-\!Bg\mu_B(\hat{\mu}^z\!+\!\hat{S}^z).
\label{eq4}
\end{align} 
As a result, one  immediately realizes that the $z$-component of the spin operators $\hat{S}^z$ ($\hat{\mu}^z$) trivially act on   the original basis states $\hat{S}^{z}\vert S^z\rangle\!=\!S^z\vert S^z\rangle$ and $\hat{\mu}^{z}\vert \mu^z\rangle\!=\!\mu^z\vert \mu^z\rangle$, whereas the  raising and lowering    ladder operators $\hat{S}^+$ ($\hat{\mu}^+$) and $\hat{S}^-$ ($\hat{\mu}^-$) shift the relevant quantum spin number by unity
\begin{align}
&\hat{S}^{\mp}\vert S^z\rangle\!=\!\sqrt{S(S\!+\!1)\!-\!S^z(S^z\!\mp\!1)}\vert S^z\!\mp\!1\rangle,\nonumber\\
&\hat{\mu}^{\mp}\vert \mu^z\rangle\!=\!\sqrt{\frac{3}{4}\!-\!\mu^z(\mu^z\!\mp\!1)}\vert \mu^z\!\mp\!1\rangle.
\label{eq4a}
\end{align} 
Subsequently, the application of the Hamiltonian~\eqref{eq4} on the basis state $\vert \pm 1/2,S^z\rangle$ leads to the identity
\begin{align}
\hat{\cal H}\lvert\pm\tfrac{1}{2},S^z\rangle&\!=\!\frac{J\Delta}{2}\sqrt{S(S\!+\!1)\!-\!S^z(S^z\!\pm\!1)}\,\lvert\mp\tfrac{1}{2},S^z\!\pm\!1\rangle
\nonumber\\
&\hspace*{-0.4cm}\!+\!\frac{1}{2}\left[ S^z(\pm J\!+\!2DS^z)\!-\!h(2S^z\!\pm\!1)\right]\lvert\pm\tfrac{1}{2},S^z\rangle,
\label{eq4b}
\end{align}
where $S^z\!=\!-S,-S\!+\!1,\dots,S\!-\!1,S$ and $h\!=\!Bg\mu_B$. The non-zero matrix elements define the block diagonal structure of the Hamiltonian,  which consists of  two one-by-one blocks and 2$S$  two-by-two blocks characterized by a specific value of the $z$-component of the total spin $S_t^z\!=\!S^z\!+\!\mu^z$ running from $-S\!-\!1/2$ to $S\!+\!1/2$.  Consequently, one can easily derive the respective eigenvalues and eigenvectors. The extremal  values of $S_t^z\!=\!\pm(S\!+\!1/2)$ define  two  one-by-one blocks, whose element is identical with its eigenvalue and the respective standard basis state designates the corresponding eigenvector
\begin{align}
\langle \pm\tfrac{1}{2},\pm S\rvert \hat{\cal H}\lvert \pm\tfrac{1}{2},\pm S\rangle&\!=\!\frac{1}{2} \left[S(J\!+\!2DS)\!\mp\!h(2S\!+\!1)\right],\nonumber\\
\varepsilon_{S,\pm(S+\tfrac{1}{2})}&\!=\!\frac{1}{2} \left[S(J\!+\!2DS)\!\mp\!h(2S\!+\!1)\right],\nonumber\\
\lvert \pm(S\!+\!\tfrac{1}{2})\rangle&\!=\!\lvert \pm\tfrac{1}{2},\pm S\rangle.
\label{eq6}
\end{align}
The remaining sectors with the total spin momentum $S^z_t\!=\!-S\!+\!1/2,-S\!+\!3/2,\dots,S\!-\!1/2$ ($S^z\!=\!-S\!+\!1,\dots,S\!-\!1$) form two-by-two blocks 
\begin{align}
&\left(
\begin{array}{cc}
\langle \frac{1}{2},S^z\vert \hat{\cal H}\vert \frac{1}{2},S^z\rangle  & \langle \frac{1}{2},S^z\vert \hat{\cal H}\vert \!-\!\frac{1}{2},S^z\!+\!1\rangle  \\\\
\langle \!-\!\frac{1}{2},S^z\!+\!1\vert \hat{\cal H}\vert \frac{1}{2},S^z\rangle  & \langle \!-\!\frac{1}{2},S^z\!+\!1\vert \hat{\cal H}\vert \!-\!\frac{1}{2},S^z\!+\!1\rangle  
\end{array}
\right)
\label{eq7}
\end{align}
with the matrix elements  explicitly defined as
\begin{align}
&\langle \tfrac{1}{2},S^z\rvert \hat{\cal H}\lvert \tfrac{1}{2},S^z
\rangle\!=\!
\frac{1}{2}\left[S^z(J\!+\!2DS^z)\!-\!h(2S^z\!+\!1)\right],
\nonumber\\
&\langle -\tfrac{1}{2},S^z\!+\!1\rvert \hat{\cal H}\lvert -\tfrac{1}{2},S^z\!+\!1
\rangle
\nonumber\\
&\hskip 1cm\!=\!
\frac{1}{2}\left[-(S^z\!+\!1)(J\!-\!2DS^z\!-\!2D)\!-\!h(2S^z\!+\!1)\right],
\nonumber\\
&\langle \tfrac{1}{2},S^z\rvert \hat{\cal H}\lvert -\tfrac{1}{2},S^z\!+\!1
\rangle\!=\!\langle -\tfrac{1}{2},S^z\!+\!1\rvert \hat{\cal H}\lvert \tfrac{1}{2},S^z
\rangle\nonumber\\
&\hskip 1cm\!=\!\frac{J\Delta}{2}\sqrt{S(S\!+\!1)\!-\!S^z(S^z\!+\!1)}.
\label{eq7a}
\end{align}
The respective couples of eigenvalues and eigenvectors within those two-by-two orthogonal subspaces read
\begin{align}
\varepsilon^{\mp}_{S,S^z_t}&=-\frac{P_{S^z_t}}{4}\!\mp\!\frac{1}{4}\sqrt{R_{S^z_t}^2\!+\!Q_{S,S^z_t}},
\label{eq10}\\
\vert\left({S^z_t}\right)_{\mp}\rangle&=c^{\mp}_{S,S^z_t}\lvert\tfrac{1}{2},S^z\rangle\!\mp\! c^{\pm}_{S,S^z_t}\lvert \!-\tfrac{1}{2},S^z\!+\!1\rangle.\nonumber
\end{align}
For brevity, we have introduced  in   Eq.~(\ref{eq10}) the new functions $P_{S_t^z}$, $R_{S_t^z}$, $Q_{S,S_t^z}$, $c^{\mp}_{S,S_t^z}$    denoting the following expressions
\allowdisplaybreaks
\begin{align}
&P_{S^z_t}=(J\!-\!2D)\!-\!D(2S^z_t\!-\!1)(2S^z_t\!+\!1)
\!+\!4hS^z_t,
\label{eq11}\\
&R_{S^z_t}=2(J\!-\!2D)S^z_t,
\label{eq12}\\
&Q_{S,S^z_t}=(J\Delta)^2 [4S(S\!+\!1)\!-\!(2S^z_t\!-\!1)(2S^z_t\!+\!1)],
\label{eq12a}\\
& c_{S,S^z_t}^{\mp}\!=\!\frac{1}{\sqrt{2}}\sqrt{1\!\mp\!\frac{R_{S^z_t}}{\sqrt{R_{S^z_t}^2\!+\!Q_{S,S^z_t}
}}\hspace*{0.3cm}}.
\label{eq13}
\end{align}
Based on the knowledge of  a complete energy spectrum of eigenvalues $\varepsilon_{S,\pm(S+1/2)}, \varepsilon^{\mp}_{S,S^z_t}$  and corresponding eigenvectors $\vert \pm(S\!+\!1/2)\rangle, \vert (S^z_t)_{\mp}\rangle$ ($S^z_t\!=\!-S\!+\!1/2,\dots,S\!-\!1/2$), one is  able to construct the relevant density operator ${\hat{\rho}}$ according to the formula 
\begin{align}
{\hat{\rho}}&=\frac{1}{\cal Z}\left\{
{\rm e}^{-\beta\varepsilon_{S,S+\frac{1}{2}}}
\lvert S\!+\!\tfrac{1}{2}\rangle\langle S\!+\!\tfrac{1}{2}\rvert
\right.\label{eq14}\\
&\hskip 1cm\!+\!{\rm e}^{-\beta\varepsilon_{S,-(S+\frac{1}{2})}}
\lvert -\left(S\!+\!\tfrac{1}{2}\right)\rangle\langle -\left(S\!+\!\tfrac{1}{2}\right)\rvert
\nonumber\\
&\hskip 1cm\!+\!\sum_{S^z_t=-S+1/2}^{S-1/2} \left[{\rm e}^{-\beta \varepsilon_{S,S^z_t}^- }\lvert(S^z_t)_-\rangle \langle (S^z_t)_-\rvert \right.
\nonumber\\
&\hskip 3cm\!+\!\left.\left.
{\rm e}^{-\beta \varepsilon_{S,S^z_t}^+ }\lvert(S^z_t)_+\rangle \langle (S^z_t)_+\rvert
\right]
\right\},
\nonumber
\end{align} 
where $\beta\!=\!1/(k_BT)$, $T$ is an absolute temperature, $k_B$ is a Boltzmann's constant and ${\cal Z}$ is the partition function
\begin{align}
{\cal Z}&\!=\!2\left\{{\rm e}^{-\frac{\beta S}{2}(J+2DS)}\cosh\left[\frac{\beta h}{2}(2S\!+\!1)\right]\right.
\nonumber\\
&\hspace*{-0.2cm}+\!\!\!\left.\sum_{S^z_t=-S+1/2}^{S-1/2}\!\!{\rm e}^{\frac{\beta}{4} P_{S^z_t}}\cosh\left( \frac{\beta}{4} \sqrt{R_{S^z_t}^2\!+\!Q_{S,S^z_t}
}\right)\right\}.
\label{eq15}
\end{align} 
The density matrix ${\hat{\rho}}$ representing a matrix representation of the density operator~\eqref{eq14}  again has in the standard basis the same   block diagonal form (classified according to the $S^z_t$  value)  involving two one-by-one blocks with the extremal values of the total spin $S^z_t\!=\!\pm(S\!+\!1/2)$ and 2$S$ two-by-two blocks with $S^z_t\!=\!-S\!+\!1/2,\dots,S\!-\!1/2$. All non-zero elements of the density matrix can be commonly  expressed through the following general formulas
\begin{align}
&\langle \tfrac{1}{2},S^z\rvert\hat{\rho}\lvert \tfrac{1}{2},S^z\rangle\!=\!
\nonumber\\
&=\!\left\{
\begin{array}{ll}
\frac{1}{\cal Z}{\rm e}^{-\beta \varepsilon_{S,S+\frac{1}{2}}}, & \hskip -4cm{\mbox{if\;\;}S^z\!=\!S; S^z_t\!=\!S\!+\!\frac{1}{2}; }\\\\
\frac{1}{\cal Z}\left[(c_{S,S^z_t}^-)^2{\rm e}^{-\beta \varepsilon_{S,S^z_t}^-}\!+\!(c_{S,S^z_t}^+)^2{\rm e}^{-\beta \varepsilon_{S,S^z_t}^+}\right], & \\
& \hskip -4.cm {\mbox{if\;\;} S^z\!=\!-S,-S\!+\!1,\dots,S\!-\!1},\\
& \hskip -4.cm {\mbox{\;\;\;\;} S^z_t\!=\!-S\!+\!\frac{1}{2},\dots,S\!-\!\frac{1}{2}},
\end{array}
\right.
\label{eq16}\\
&\langle -\tfrac{1}{2},S^z\rvert\hat{\rho}\lvert -\tfrac{1}{2},S^z\rangle\!=\!
\nonumber\\
&=\!\left\{
\begin{array}{ll}
\frac{1}{\cal Z}{\rm e}^{-\beta \varepsilon_{S,-(S+\frac{1}{2})}}, &  \hskip -4.cm{\mbox{if\;\;}S^z\!=\!-S;S^z_t\!=\!-S\!-\!\frac{1}{2}; }\\\\
\frac{1}{\cal Z}\left[(c_{S,S^z_t}^+)^2{\rm e}^{-\beta \varepsilon_{S,S^z_t}^-}\!+\!(c_{S,S^z_t}^-)^2{\rm e}^{-\beta \varepsilon_{S,S^z_t}^+}\right], &\\
 & \hskip -4.9cm {\mbox{if\;\;} S^z\!=\!-S\!+\!1,-S\!+\!2,\dots,S; S_t^z\!=\!S^z\!-\!\frac{1}{2}},\\
 & \hskip -4.9cm {\mbox{\;\;\;\;} S^z_t\!=\!-S\!+\!\frac{1}{2},\dots,S\!-\!\frac{1}{2}},
\end{array}
\right.
\label{eq17}\\
&\langle \tfrac{1}{2},S^z\rvert\hat{\rho}\lvert -\tfrac{1}{2},S^z\!+\!1\rangle
\!=\!\langle -\tfrac{1}{2},S^z\!+\!1\rvert\hat{\rho}\lvert \tfrac{1}{2},S^z\rangle
\nonumber\\
&=\!
\frac{c_{S,S^z_t}^-c_{S,S^z_t}^+}{\cal Z}\left[{\rm e}^{-\beta \varepsilon_{S,S^z_t}^+}\!-\!{\rm e}^{-\beta \varepsilon_{S,S^z_t}^-}\right],
\nonumber\\
& \hskip 3cm {\mbox{if\;\;} S^z\!=\!-S,-S\!+\!1,\dots,S\!-\!1},\nonumber\\
& \hskip 3cm {\mbox{\;\;\;\;} S^z_t\!=\!-S\!+\!\frac{1}{2},\dots,S\!-\!\frac{1}{2}}.
\label{eq18}
\end{align}
For a completeness, the readers can find the explicit form of the density-matrix elements for a few selected mixed spin-(1/2,$S$) Heisenberg dimers ($S\!=\!1,3/2,2,5/2$) in the Appendix~\ref{App A}. 

In order to calculate the  density matrix $\hat{\rho}^{T_{1/2}}$ partially transposed with respect the spin-1/2 subsystem, it is sufficient to replace the   bra and ket state vectors referred to the spin-1/2 subsystem. It is clear, that the diagonal elements derived from  Eqs.~\eqref{eq16}-\eqref{eq17} remain unchanged, whereas off-diagonal ones are moved to other positions. The partial transposition does not conserve the total spin momentum $S^z_t$, but it conserves the staggered spin momentum $S^z_{tm}\!=\!S^z\!-\!\mu^z$ running from $-S\!-\!1/2$ to $S\!+\!1/2$. Subsequently, the non-zero elements of the partially transposed density matrix $\hat{\rho}^{T_{1/2}}$ expressed in term of $S^z_{tm}$ have the form
\begin{align}
&\langle \tfrac{1}{2},S^z\rvert\hat{\rho}^{T_{1/2}}\lvert \tfrac{1}{2},S^z\rangle\!=\!
\nonumber\\
&=\resizebox{0.98\columnwidth}{!}{$\displaystyle
\!\left\{
\begin{array}{ll}
\frac{1}{\cal Z}{\rm e}^{-\beta \varepsilon_{S,S+\frac{1}{2}}}, & \hskip -5.cm{\mbox{if\;\;}S^z\!=\!S; S^z_{tm}\!=\!S\!-\!\frac{1}{2}}\\\\
\frac{1}{\cal Z}\left[(c_{S,S^z_{tm}+1}^-)^2{\rm e}^{-\beta \varepsilon_{S,S^z_{tm}+1}^-}\!+\!(c_{S,S^z_{tm}+1}^+)^2{\rm e}^{-\beta \varepsilon_{S,S^z_{tm}+1}^+}\right], & \\
& \hskip -5.cm {\mbox{if\;\;} S^z\!=\!-S,-S\!+\!1,\dots,S\!-\!1},\\
& \hskip -5.cm {\mbox{\;\;\;\;} S^z_{tm}\!=\!-S\!-\!\frac{1}{2},\dots,S\!-\!\frac{3}{2}},
\end{array}
\right.
$}
\label{eq19}\\
&\langle -\tfrac{1}{2},S^z\rvert\hat{\rho}^{T_{1/2}}\lvert -\tfrac{1}{2},S^z\rangle\!=\!
\nonumber\\
&=\resizebox{0.98\columnwidth}{!}{$\displaystyle
\!\left\{
\begin{array}{ll}
\frac{1}{\cal Z}{\rm e}^{-\beta \varepsilon_{S,-(S+\frac{1}{2})}}, &  \hskip -5cm{\mbox{if\;\;}S^z\!=\!-S; S^z_{tm}\!=\!-S\!+\!\frac{1}{2}}\\\\
\frac{1}{\cal Z}\left[(c_{S,S^z_{tm}-1}^+)^2{\rm e}^{-\beta \varepsilon_{S,S^z_{tm}-1}^-}\!+\!(c_{S,S^z_{tm}-1}^-)^2{\rm e}^{-\beta \varepsilon_{S,S^z_{tm}-1}^+}\right], &\\
 & \hskip -5cm {\mbox{if\;\;} S^z\!=\!-S\!+\!1,-S\!+\!2,\dots,S},\\
 & \hskip -5cm {\mbox{\;\;\;\;} S^z_{tm}\!=\!-S\!+\!\frac{3}{2},\dots,S\!+\!\frac{1}{2}},
\end{array}
\right.
$}
\label{eq20}\\
&\langle -\tfrac{1}{2},S^z\rvert\hat{\rho}^{T_{1/2}}\lvert \tfrac{1}{2},S^z\!+\!1\rangle
\!=\!\langle \tfrac{1}{2},S^z\!+\!1\rvert\hat{\rho}^{T_{1/2}}\lvert -\tfrac{1}{2},S^z\rangle
\nonumber\\
&=\!
\frac{c_{S,S^z_{tm}}^-c_{SS,^z_{tm}}^+}{\cal Z}\left[{\rm e}^{-\beta \varepsilon_{S,S^z_{tm}}^+}\!-\!{\rm e}^{-\beta \varepsilon_{SS,^z_{tm}}^-}\right],
\nonumber\\
& \hskip 3.2cm {\mbox{if\;\;} S^z\!=\!-S,-S\!+\!1,\dots,S\!-\!1},\nonumber\\
& \hskip 3.2cm {\mbox{\;\;\;\;} S^z_{tm}\!=\!-S\!+\!\frac{1}{2},\dots,S\!-\!\frac{1}{2}}.
\label{eq21}
\end{align}
Note that the density matrix $\hat{\rho}^{T_{1/2}}$  is a block diagonal  with a maximal block's  size of 2$\times$2 achieving a specific $S^z_{tm}$ value. Two one-by-one blocks with extremal $S^z_{tm}\!=\!\pm(S\!+\!1/2)$
involve a single element
\begin{align}
&\langle \pm\tfrac{1}{2},\mp S\rvert\hat{\rho}^{T_{1/2}}\lvert \pm\tfrac{1}{2},\mp S\rangle
%\nonumber\\&
\!=\!\frac{1}{\cal Z}\left[\left(c_{S,\mp(S-\frac{1}{2})}^{\mp}\right)^2{\rm e}^{-\beta \varepsilon_{S,\mp(S-\frac{1}{2})}^-}\right.
\nonumber\\
&\hskip 3.cm\left.\!+\!\left(c_{S,\mp(S-\frac{1}{2})}^{\pm}\right)^2{\rm e}^{-\beta \varepsilon_{S,\mp(S-\frac{1}{2})}^+}\right], \!
\label{eq21a}
\end{align}
which directly determines  two positive eigenvalues $\lambda_{\mp(S\!+\!1/2)}$  of a partially transposed density matrix
\begin{align}
\lambda_{\mp (S+\frac{1}{2})}\!=\!\frac{1}{\cal Z}&\left[\left(c_{S,\mp(S-\frac{1}{2})}^{\mp}\right)^2{\rm e}^{-\beta \varepsilon_{S,\mp(S-\frac{1}{2})}^-}\right.
\nonumber\\
\!&+\!\left.\left(c_{S,\mp(S-\frac{1}{2})}^{\pm}\right)^2{\rm e}^{-\beta \varepsilon_{S,\mp(S-\frac{1}{2})}^+}\right].
\label{eq22}
\end{align}
Other 2$S$ two-by-two blocks determined by basis state vectors  with $S^z_{tm}\!=\!-S\!+\!1/2,\dots,S\!-\!1/2$
\begin{align}
%\allowdisplaybreaks
&\resizebox{1\columnwidth}{!}{$
M(S^z_{tm})\!=\!\left(
\begin{array}{cc }
 \langle \frac{1}{2}, S^z\rvert\hat{\rho}^{T_{1/2}}\lvert \frac{1}{2},S^z\rangle& \langle \frac{1}{2}, S^z\rvert\hat{\rho}^{T_{1/2}}\lvert -\frac{1}{2},S^z\!-\!1\rangle\\\\
\langle -\frac{1}{2}, S^z\!-\!1\rvert\hat{\rho}^{T_{1/2}}\lvert \frac{1}{2},S^z\rangle &\langle -\frac{1}{2}, S^z\!-\!1\rvert\hat{\rho}^{T_{1/2}}\lvert -\frac{1}{2},S^z\!-\!1\rangle \\
%\end{block}
\end{array}\right)
$}\nonumber\\\nonumber\\
&\hskip1cm\!=\!\left(
\begin{array}{cc }
m_{11} & m_{12}\\
m_{21} & m_{22}
\end{array}\right)
\label{eq21b}
\end{align}
immediately result to the remaining couple of eigenvalues 
\begin{align}
\hspace*{-0.2cm}\lambda_{S^z_{tm}}^{\mp}\!=\!\frac{1}{2}\left[
\left(m_{11}\!+\!m_{22}\right)\!\mp\!\sqrt{(m_{11}\!-\!m_{22})^2\!+\!4m_{12}m_{21}}\right]\!.\!\!\!
\label{eq24}
\end{align}
Due to lengthy of explicit form of  the~\eqref{eq24}, the readers can find them in Appendix~\ref{App B}. At the same time, the complete list of partially transposed density matrices ${\hat{\rho}}^{T_{1/2}}$ for the mixed spin-(1/2,$S$) Heisenberg dimers with specific spin values $S\!=\!1,3/2,2,5/2$ is given in Appendix~\ref{App C}. 
Analysing  Eq.~\eqref{eq24} in detail one identifies that  only eigenvalues $\lambda^-_{S^z_{tm}}$ can be negative,  and hence, the respective bipartite entanglement is  in accordance to the definition \eqref{eq2} determined  by the formula
\begin{align}
{\cal N}(\rho)&=-\sum_{S^z_{tm}=-S+\frac{1}{2}}^{S-\frac{1}{2}}\min(0,\lambda_{S^z_{tm}}^-).
\label{eq23}
\end{align} 

\section{Results and discussion}
\label{sec:result}
In order to minimize the huge parametric space, all further discussions will be limited to    the physically most interesting  case with an isotropic exchange interaction defined through the parameter  $\Delta\!=\!1$. For simplicity, the gyromagnetic factor $g$  of both types of magnetic ions  is set equal to two ($g\!=\!2$). It is worthwhile to note that other choice of the unequal $g$ factor has only the quantitative, but not qualitative, impact on  all obtained observations. 
\subsection{Quantum negativity}

\begin{figure}[t!]
{\includegraphics[width=.475\columnwidth,trim=0.8cm 9.2cm 4cm 8.cm, clip]{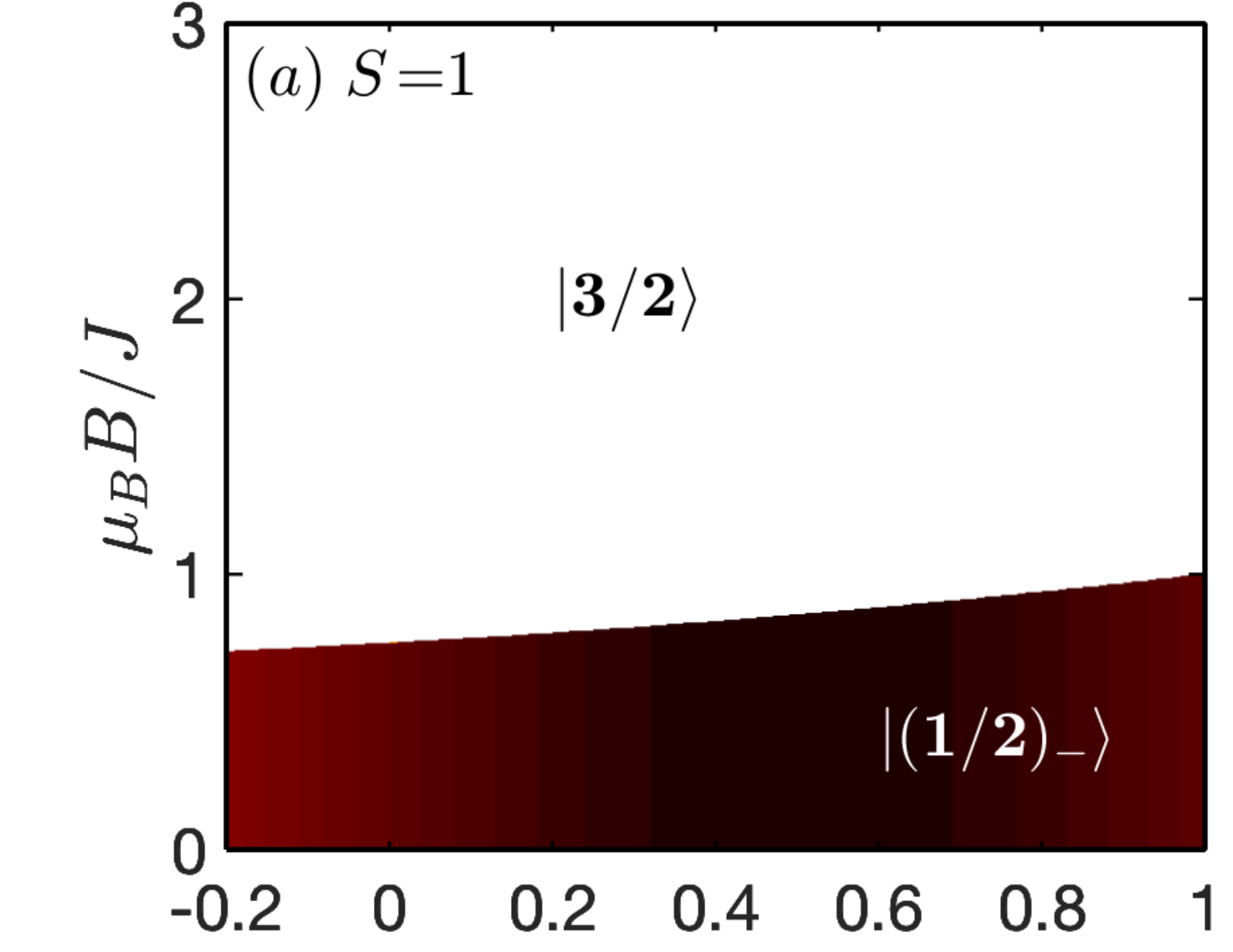}}
{\includegraphics[width=.505\columnwidth,trim=2.9cm 9.2cm 1cm 8.cm, clip]{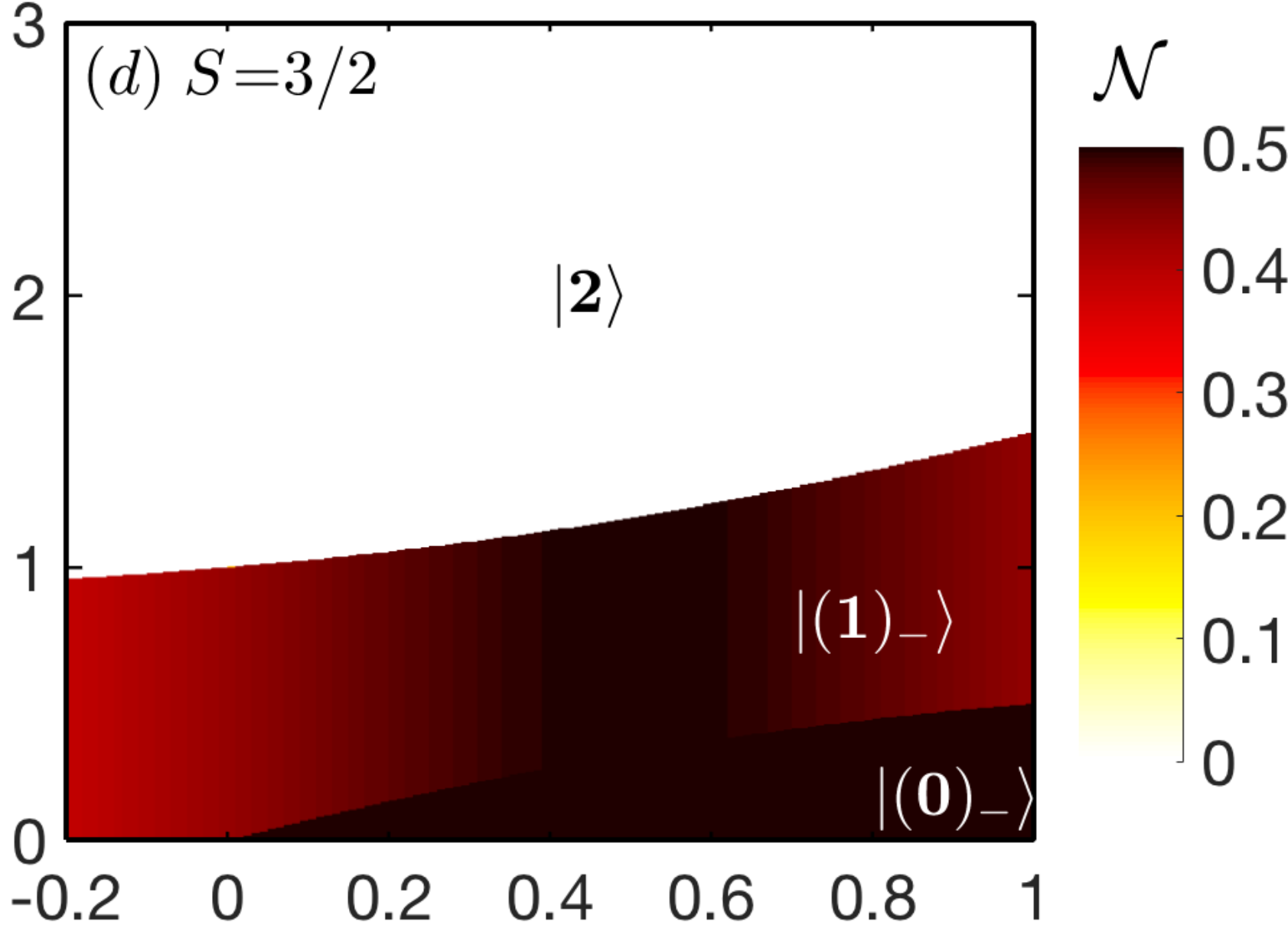}}\\
{\includegraphics[width=.475\columnwidth,trim=0.8cm 9.2cm 4cm 8cm, clip]{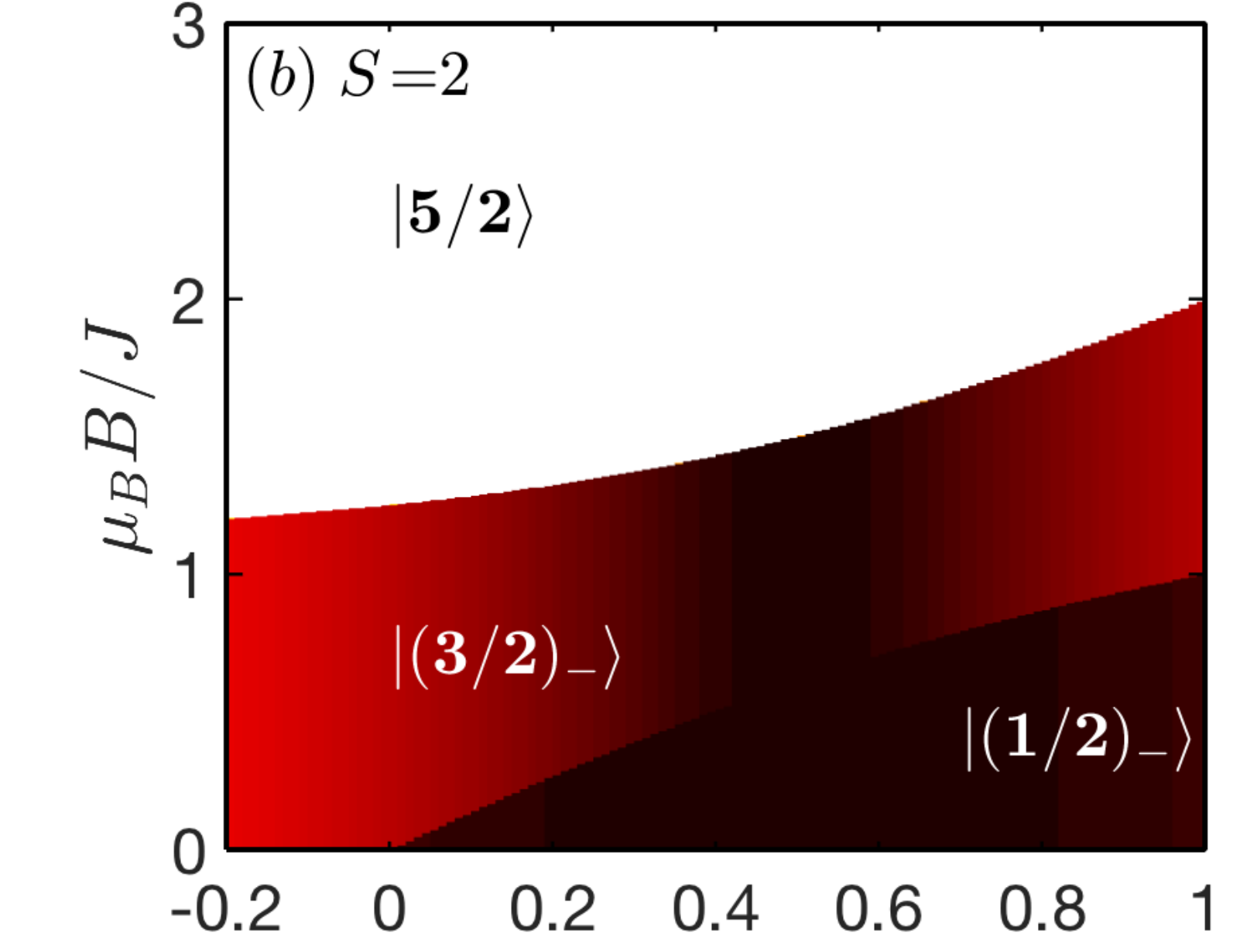}}
{\includegraphics[width=.505\columnwidth,trim=2.9cm 9.2cm 1cm 8cm, clip]{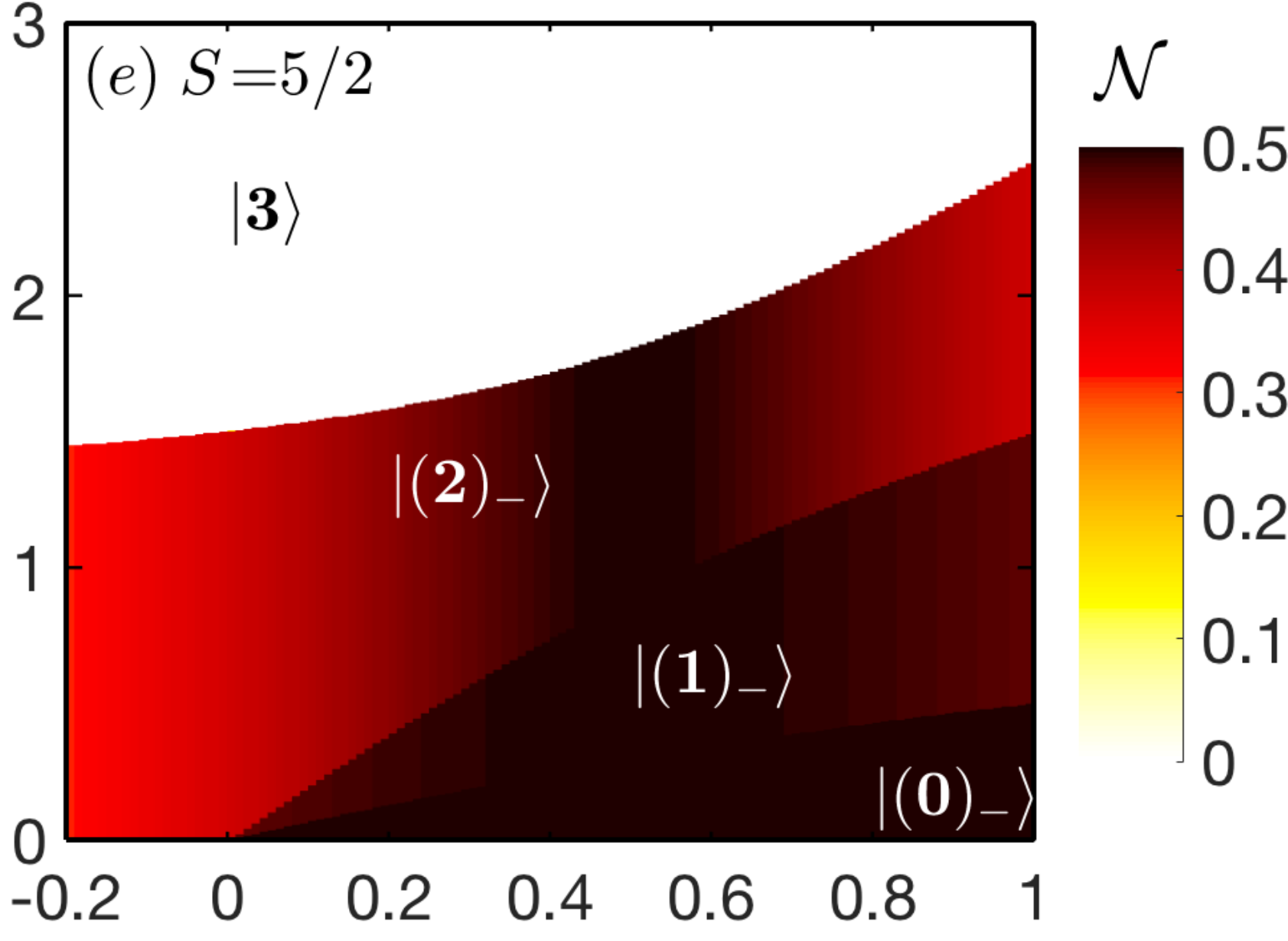}}\\
{\includegraphics[width=.475\columnwidth,trim=0.8cm 7.7cm 4cm 8cm, clip]{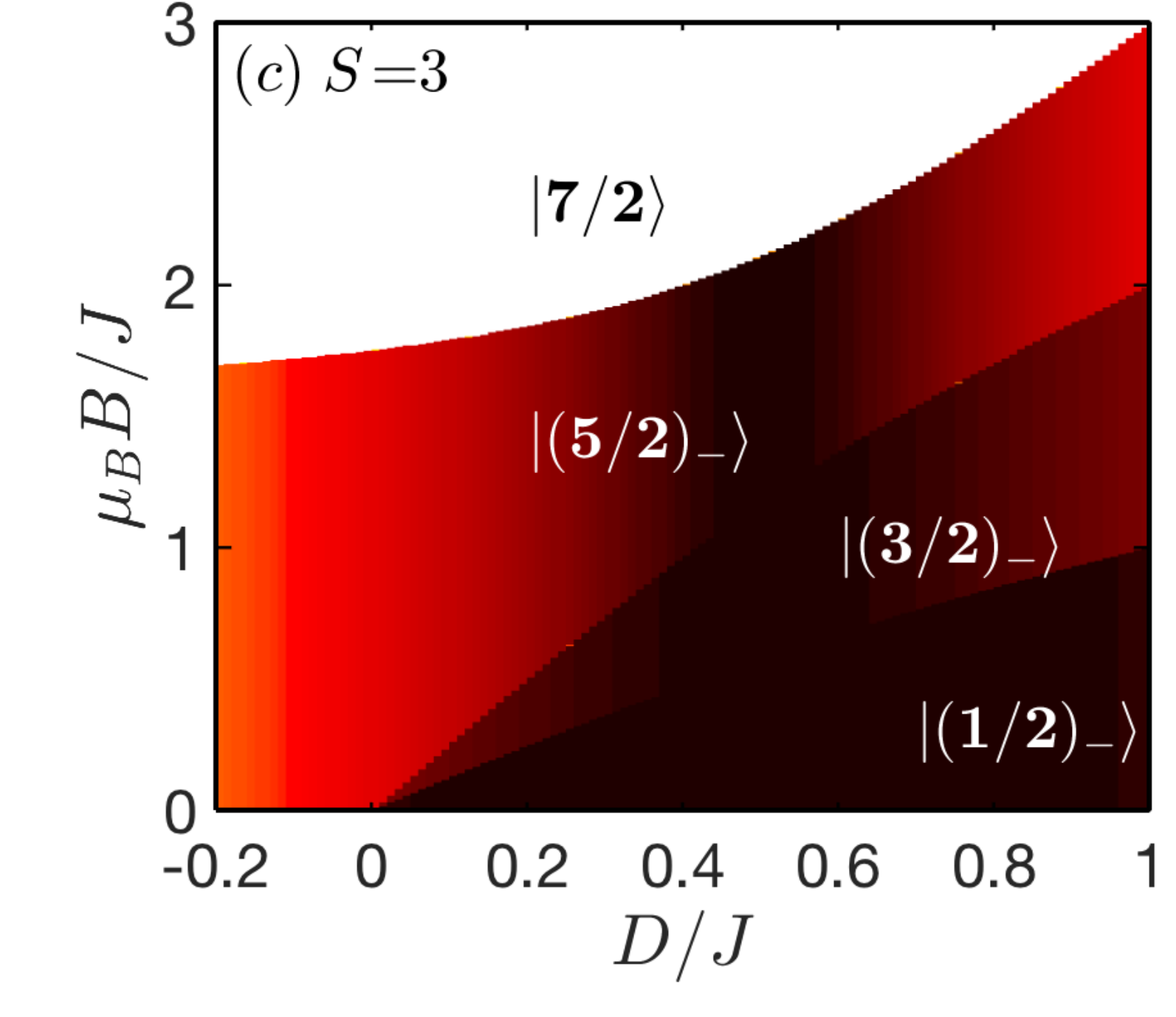}}
{\includegraphics[width=.505\columnwidth,trim=2.9cm 7.7cm 1cm 8cm, clip]{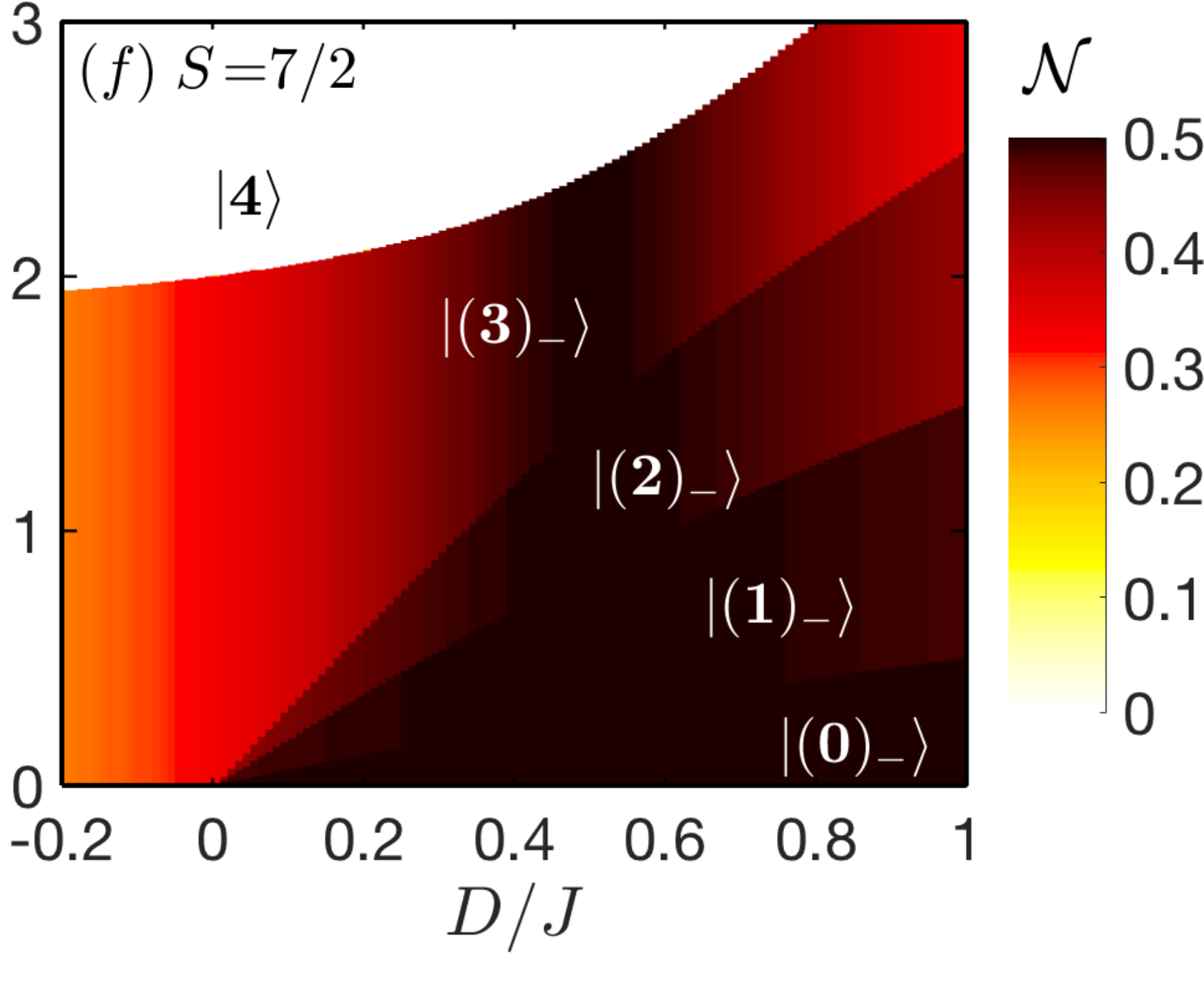}}
\caption{Density plots of a quantum negativity in the $D/J-\mu_BB/J$ plane for the  antiferromagnetic spin-(1/2,$S$)  Heisenberg dimers with the exchange coupling $J\!>\!0$ for an integer (left panels) and half-odd-integer (right panels) spin magnitude and isotropic interaction ($\Delta\!=\!1$). }
\label{fig3a}
\end{figure}

 The behaviour of the quantum negativity of the mixed spin-(1/2,$S$) Heisenberg dimers  with $S\!=\!1,3/2,2,5/2,3,7/2$   is illustrated in Fig.~\ref{fig3a} in the $D/J-\mu_BB/J$ plane by considering the antiferromagnetic exchange coupling $J\!>\!0$. The density plots of quantum negativity simultaneously illustrate  stability regions of all relevant ground states $\vert(S^z_t)_-\rangle$ and $\vert S\!+\!1/2\rangle$. 

It is worthwhile to remark that the negativity at   zero magnetic field was comprehensively  analysed in our preceding paper~\cite{Varga21} and thus, the case $\mu_BB/J\!=\!0$ will be just marginally explored in our subsequent discussions. It has been found that the negativity at $\mu_BB/J\!=\!0$ exhibits qualitatively  different behaviour for  integer  and half-odd-integer spin-$S$ constituents, if the uniaxial single-ion anisotropy $D/J\!>\!0$ of easy-plane type is taken into account.  It was surprisingly detected that the enhancement of a degree of entanglement for the mixed-spin Heisenberg dimers involving an integer spin $S$ emerges a consequence of interplay between the increasing spin magnitude $S$ and uniaxial single-ion anisotropy $D/J\!>\!1/2$. Nevertheless, the highest negativity ${\cal N}\!=\!(\sqrt{5}\!-\!1)/4$ reached for the mixed-spin Heisenberg dimers with integer spin $S$  at the specific value  $D/J\!=\!1/2$  is significantly smaller than the maximal negativity ${\cal N}\!=\!1/2$ detected for the mixed-spin Heisenberg dimers with an arbitrary half-odd-integer spin $S$.

The external magnetic field reduces the  degeneracy of the mixed spin-(1/2,$S$) Heisenberg dimers due to the Zeeman's splitting of energy levels and the quantum negativity of an arbitrary  mixed spin-(1/2,$S$) Heisenberg dimers  in a state with the total spin $S^z_{t}$ can be expressed through the   general formula 
\begin{align}
&{\cal N}\!=
c_{S,S^z_{t}}^+c_{S,S^z_{t}}^- \label{eq:r3}\\
&=
\frac{1}{2}\sqrt{\frac{4S(S\!+\!1)\!-\!(2S^z_{t}\!-\!1)(2S^z_{t}\!+\!1)}{(2S^z_{t})^2(1\!-\!2\tfrac{D}{J})^2\!+\!4S(S\!+\!1)\!-\!(2S^z_{t}\!-\!1)(2S^z_{t}\!+\!1)}}.
\nonumber
\end{align}
It is evident from Eq.~\eqref{eq:r3} and Fig.~\ref{fig3a} that  the quantum negativity of  each $\vert (S^z_t)_-\rangle$ ground state is fully independent of  the external magnetic field  and its magnitude  decreases towards to the  completely separable (${\cal N}\!=\!0$) ferromagnetic state $\vert 1/2\!+\!S\rangle$ with $S^z_t\!=\!S\!+\!1/2$.   However, the increasing magnetic field  is responsible for existence of  discontinuous changes of the quantum negativity at all field-driven   magnetic phase transitions.  It follows from  Fig.~\ref{fig3a} that maximal bipartite entanglement is reached for the  specific value of the uniaxial single-ion anisotropy $D/J\!=\!1/2$ (similar as in the $\mu_BB/J\!=\!0$ case), at which an arbitrary mixed spin-(1/2,$S$) Heisenberg dimer shows, in agreement with Eq.~\eqref{eq:r3}, the highest quantum negativity ${\cal N}\!=\!1/2$  until the sufficiently high magnetic field reorients both spins into its direction. 

The behaviour of the quantum negativity in the regime of easy-axis uniaxial single-ion anisotropy $D/J\!\leq\!0$  confirms previously reported findings~\cite{Li, Huang2008,Hao} that the increasing spin magnitude $S$ enlarges the stability of entangled state with respect to the magnetic field. Nevertheless,  the degree of respective bipartite entanglement is gradually reduced. Contrary to this, the increasing  spin magnitudes $S$ induces the  enhancement of  a quantum negativity for an arbitrary  $\vert ({S^z_t})_-\rangle$ ($S^z_t\!\leq\!S\!-\!1/2$) ground state of the  mixed spin-(1/2,$S$) Heisenberg dimers  assuming  easy-plane single-ion anisotropy $D/J\!>\!0$.  In contradiction to the zero-field case, the enhancement of a negativity is observed even for $0\!<\!D/J\!<\!1/2$ as a consequence of reduction of ground-state degeneracy in respective parametric space. This is a very important observation  from the application perspective, because  variation of a magnetic ion  in the mixed spin-(1/2,$S$) Heisenberg dimers, offers a relative simple alternative how to enhance the bipartite entanglement. 
  The origin of  qualitatively different behaviour of the negativity below and above $D/J\!=\!0$ can be explained   through the respective variation of the  total spin value $S^z_{t}$. In the $\vert (S\!-\!1/2)_-\rangle$ ground state  emergent in easy-axis regime $D/J\!<\!0$, the total spin $S^z_{t}\!=\!S\!-\!1/2$ is gradually enhanced with  increasing spin size $S$, but the difference  $4S(S\!+\!1)\!-\!(2S_{t}^z\!-\!1)(2S_{t}^z\!+\!1)$ entering into  the Eq.~\eqref{eq:r3} remains constant, $8S$. Consequently,  the quantum negativity decreases  as the spin $S$ magnitude enlarges according to the formula
\begin{align}
{\cal N}\!=\!\left[\frac{(1\!-\!2\frac{D}{J})^2(2S\!-\!1)^2}{2S}\!+\!4\right]^{-1/2}.
\label{qs6}
\end{align}
Considering the  fixed value of the total spin $S^z_{t}$  and the easy-plane regime $D/J\!>\!0$ the difference $4S(S\!+\!1)\!-\!(2S_{t}^z\!-\!1)(2S_{t}^z\!+\!1)$ in Eq.~\eqref{eq:r3} is enlarged with an increasing spin size $S$,  which means that the denominator in rewritten form of Eq.~\eqref{eq:r3} 
\begin{align} 
 {\cal N}\!=\!\left\{\frac{1}{\frac{4(2S^z_{t})^2(1\!-\!2\frac{D}{J})^2}{[4S(S\!+\!1)\!-\!(2S_{t}^z\!-\!1)(2S_{t}^z\!+\!1)]}\!+\!4}\right\}^{-1/2}.
 \label{gs7}
 \end{align}
decreases and thus the respective quantum negativity is naturally  enhanced.
In the special case of $D/J\!=\!0$ the negativity is an inverse function of spin magnitude $S$ and thus the  quantum entanglement reduces upon strengthening of the spin size $S$. 
\begin{figure}[t!]
{\includegraphics[width=.45\columnwidth,trim=0.7cm 7.7cm 4cm 8.cm, clip]{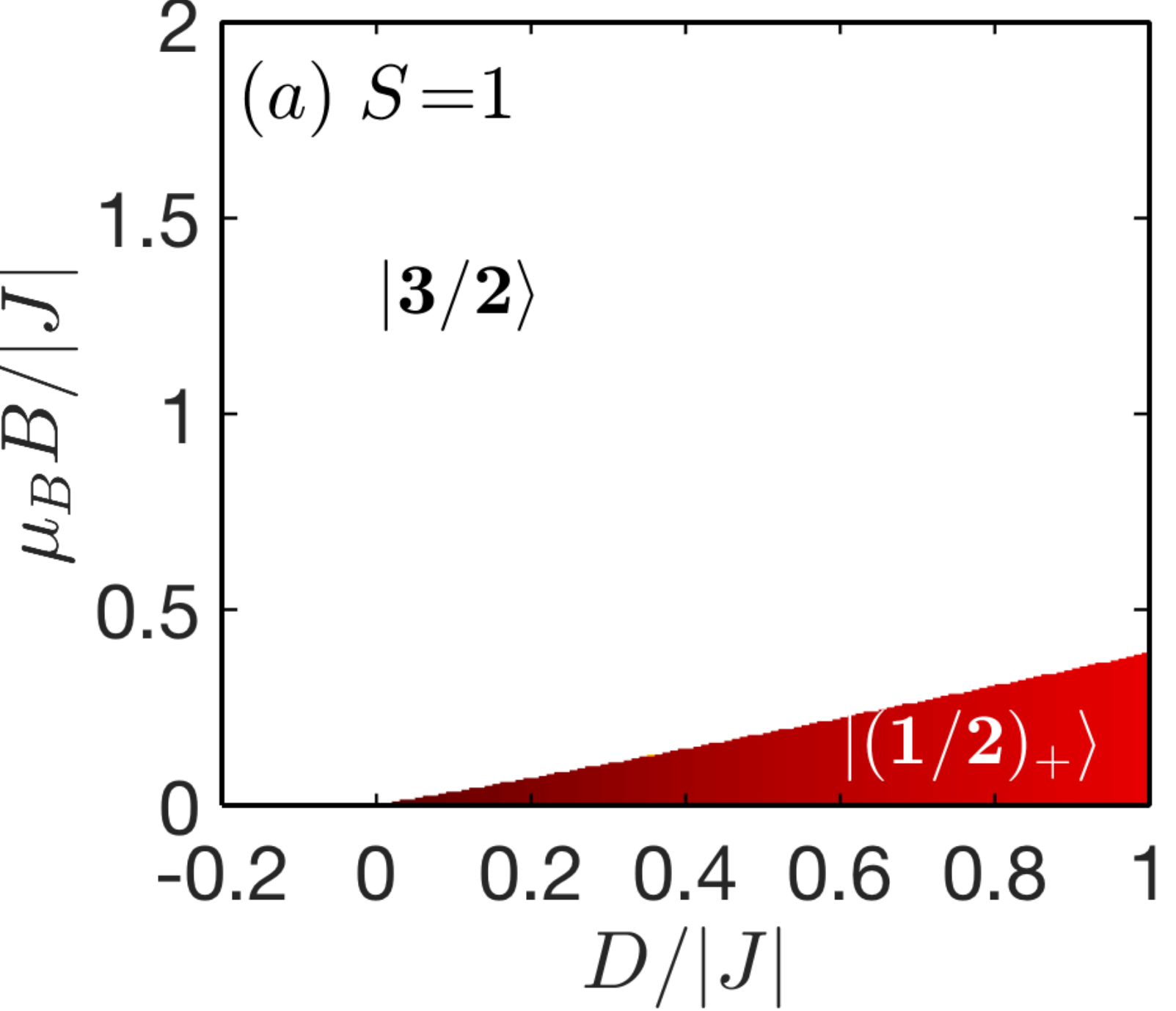}}
{\includegraphics[width=.53\columnwidth,trim=0.7cm 7.7cm 0.8cm 8cm, clip]{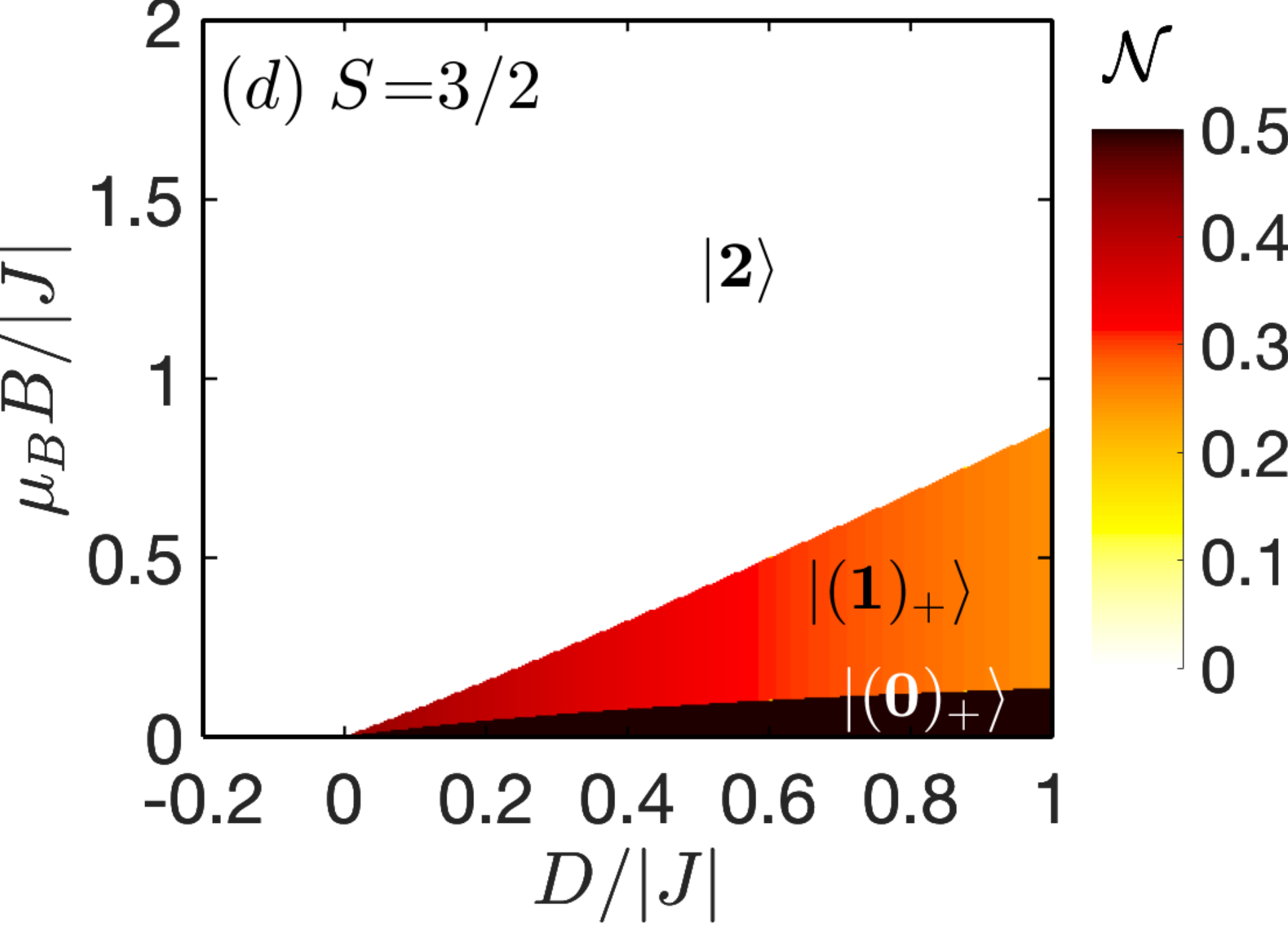}}\\
{\includegraphics[width=.45\columnwidth,trim=0.7cm 7.7cm 4cm 8.cm, clip]{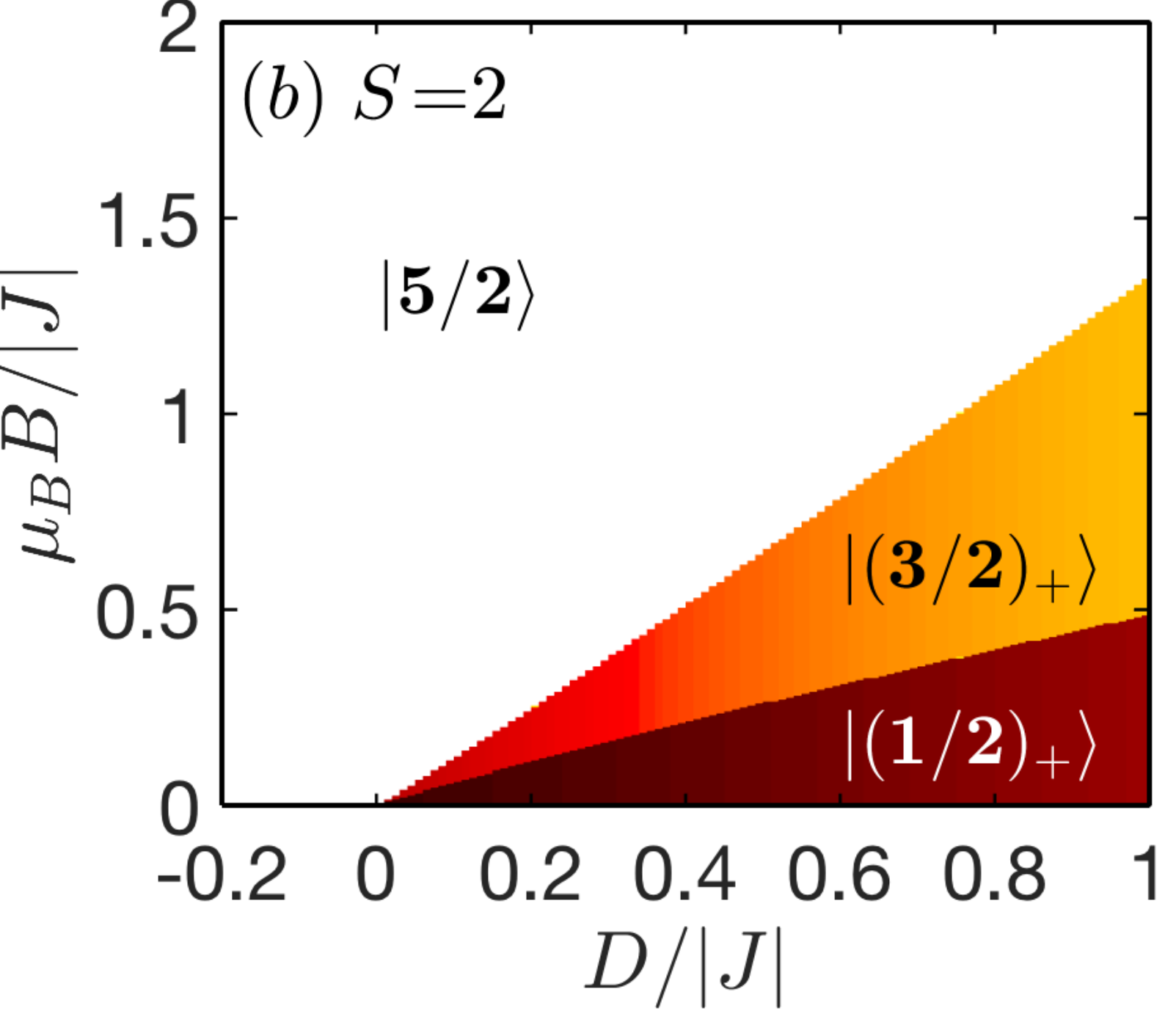}}
{\includegraphics[width=.53\columnwidth,trim=0.7cm 7.7cm 0.8cm 8cm, clip]{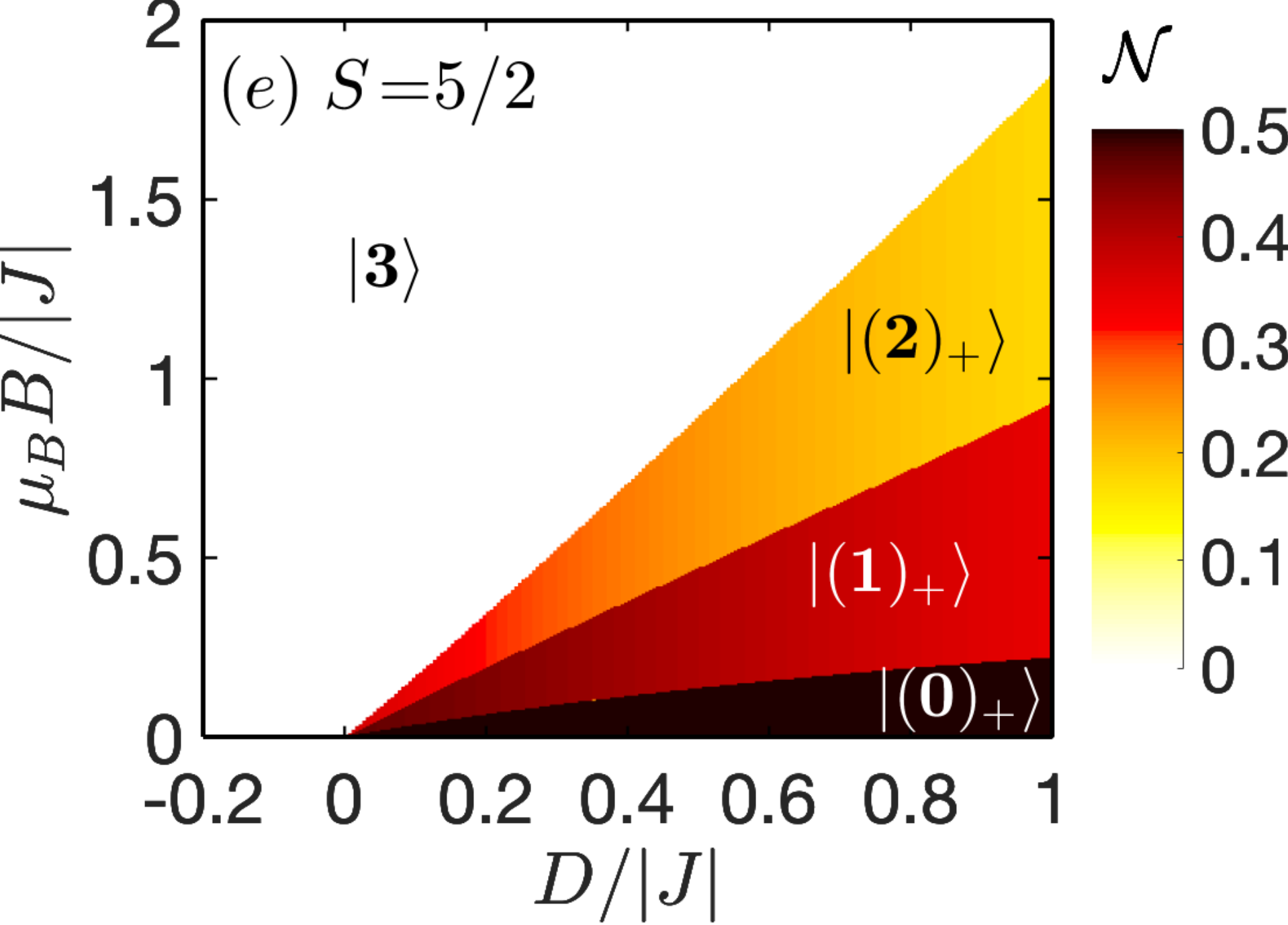}}\\
{\includegraphics[width=.45\columnwidth,trim=0.7cm 7.7cm 4cm 8.cm, clip]{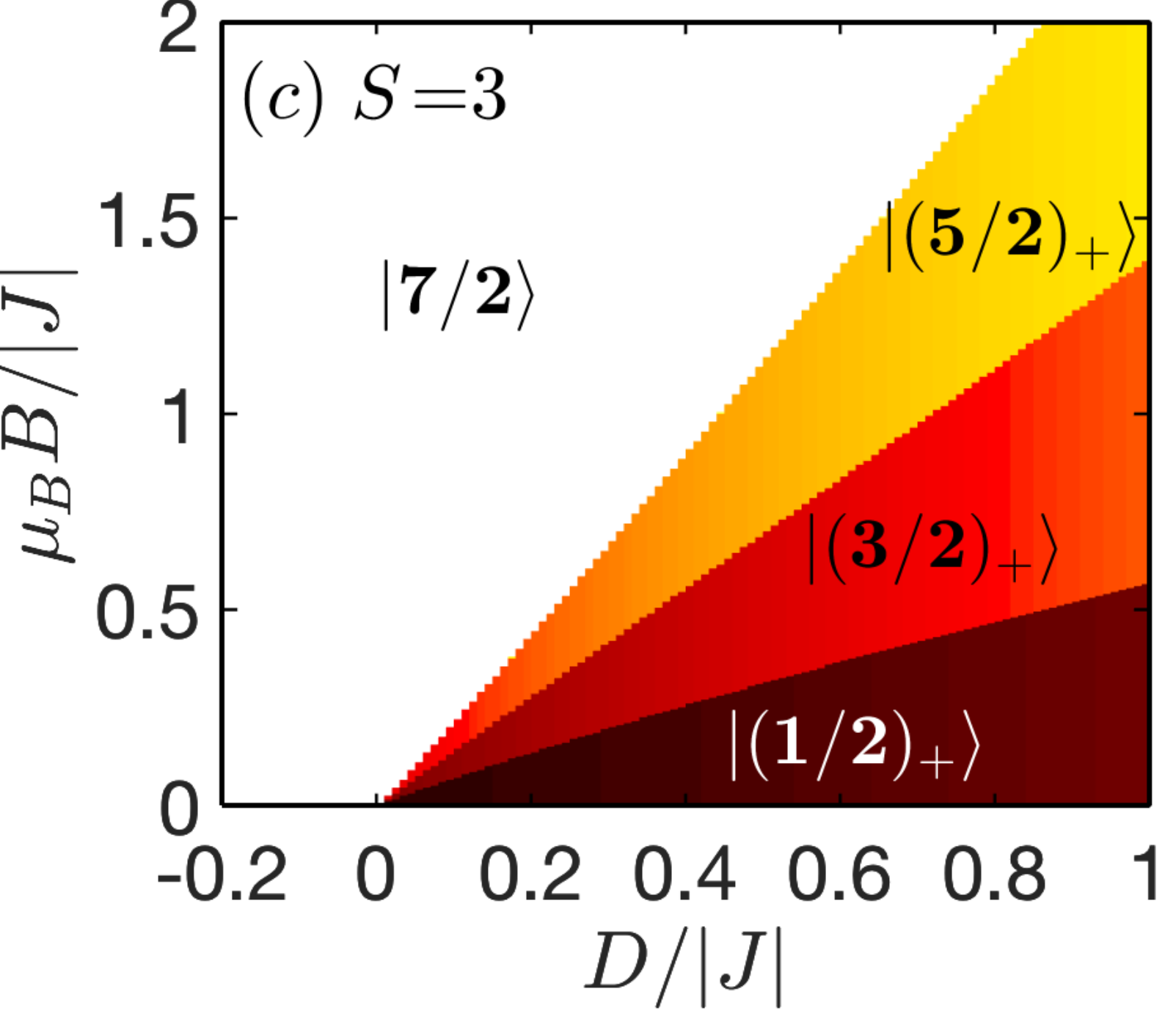}}
{\includegraphics[width=.53\columnwidth,trim=0.7cm 7.7cm 0.8cm 8cm, clip]{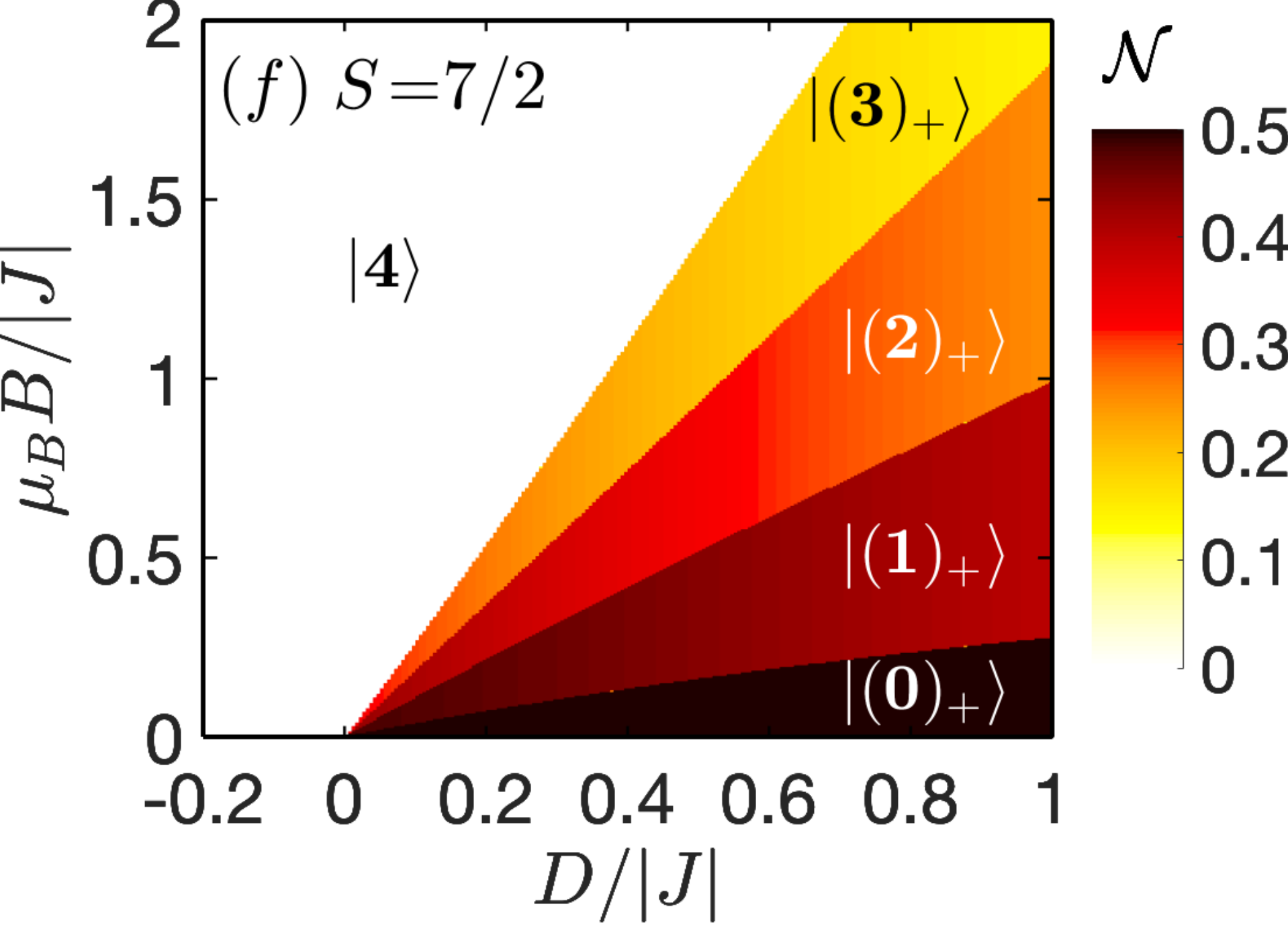}}
\caption{Density plots of a quantum negativity in the $D/|J|-\mu_BB/|J|$ plane for the  ferromagnetic spin-(1/2,$S$)  Heisenberg dimers with the coupling constant $J\!<\!0$ for an integer (left panels) and half-odd-integer (right panels) spin magnitude and isotropic interaction ($\Delta\!=\!1$). }
\label{fig4a}
\end{figure}

 For the ferromagnetic exchange coupling $J\!<\!0$ (Fig.~\ref{fig4a}), the quantum entanglement can be achieved only for the mixed-spin Heisenberg dimers with an easy-plane single-ion anisotropy $D/|J|\!>\!0$ due to possible preference of various ferrimagnetic or antiferromagnetic ground states $\vert (S^z_t)_+\rangle$ ($S^z_t\!\leq\!S\!-\!1/2$). Since the previously derived relation  remains in force, the quantum negativity of each $\vert (S^z_t)_+\rangle$ ground state  with the total spin $S^z_t\!\leq\!S\!-\!1/2$ always increases as the spin  magnitudes  increase. It should be pointed out that the degree of bipartite entanglement  for an arbitrary $\vert (S^z_t)_+\rangle$ ground state is significantly smaller  in comparison to its antiferromagnetic counterpart, which makes the ferromagnetic quantum mixed-spin Heisenberg dimers less attractive for  practical utilizations.
For a completeness, it should be emphasized, that the invariant point at $D/|J|\!=\!1/2$ is completely absent in the ferromagnetic case $J\!<\!0$  and the highest negativity (excluding the ground state $\vert (0)_+\rangle$) can be found in a  proximity of  the isotropic point $D/|J|\!=\!0$. Strictly at $D/|J|\!=\!0$  and $\mu_BB/J\!=\!0$ all $\vert(S^z_t)_+\rangle$ ($|S^z_t|\!\leq\!S\!-\!1/2$) ground states and $\vert\pm(S\!+\!1/2)\rangle$ ones are degenerate and the respective negativity follows the simple relation 
\begin{align}
{\cal N}\!=\!\frac{(S\!-\!1)}{(S\!+\!1)(2S\!+\!1)}.
\label{gs8}
\end{align}

\subsection{The thermal negativity}

\begin{figure}[t]
{\includegraphics[width=.9\columnwidth,trim=0.5cm 7.6cm 2cm 8.cm, clip]{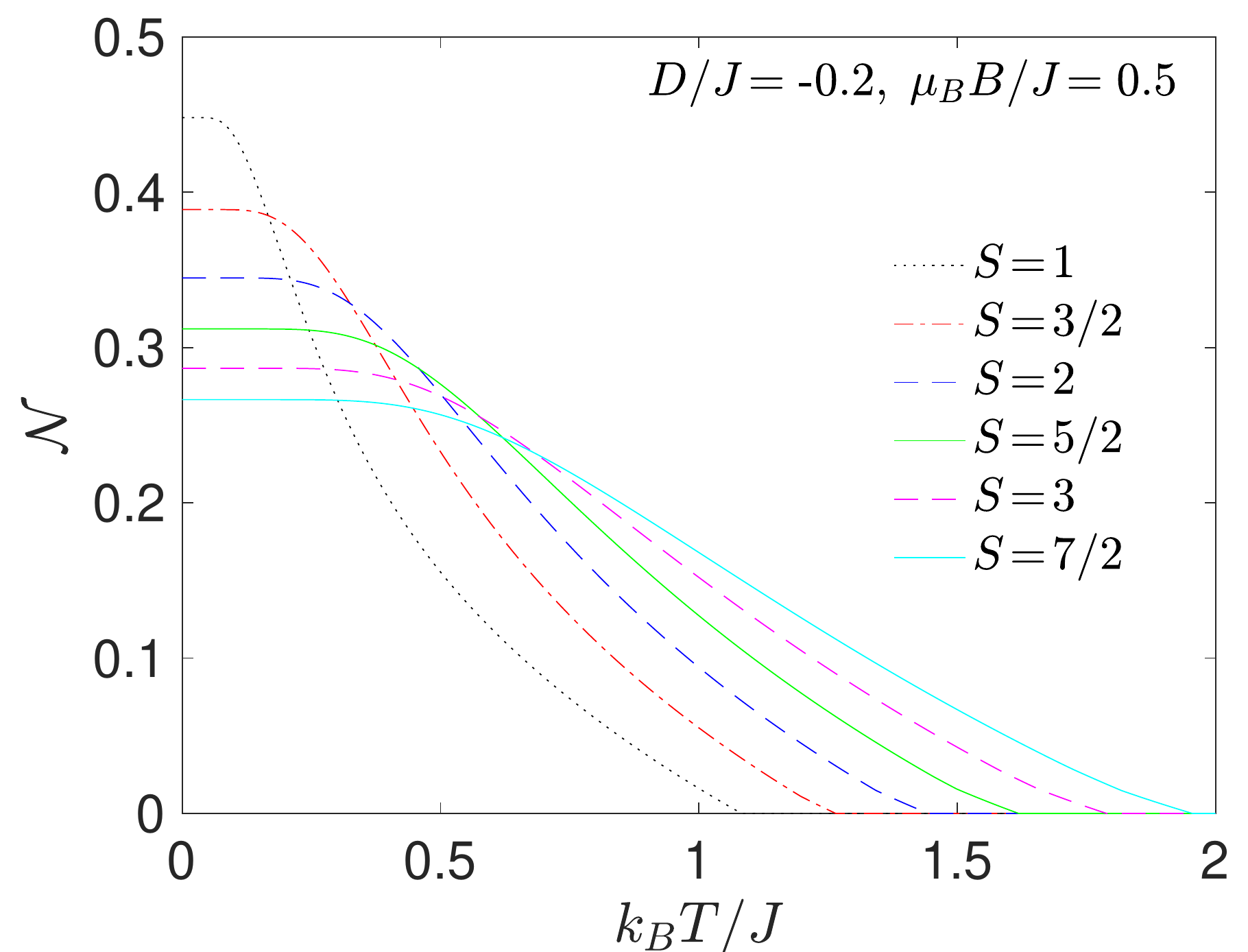}}
\caption{The thermal dependence of the negativity for a few different values of spin  magnitudes $S$ calculated at  $D/J\!=\!-0.2$ and $\mu_BB/J\!=\!0.5$. The choice of model parameters corresponds to the  region where the ground state $\vert (S\!-\!1/2)_-\rangle$ is favoured.}
\label{fig5a}
\end{figure}

\begin{figure}[h]
{\includegraphics[width=.9\columnwidth,trim=0.5cm 7.6cm 2cm 8.cm, clip]{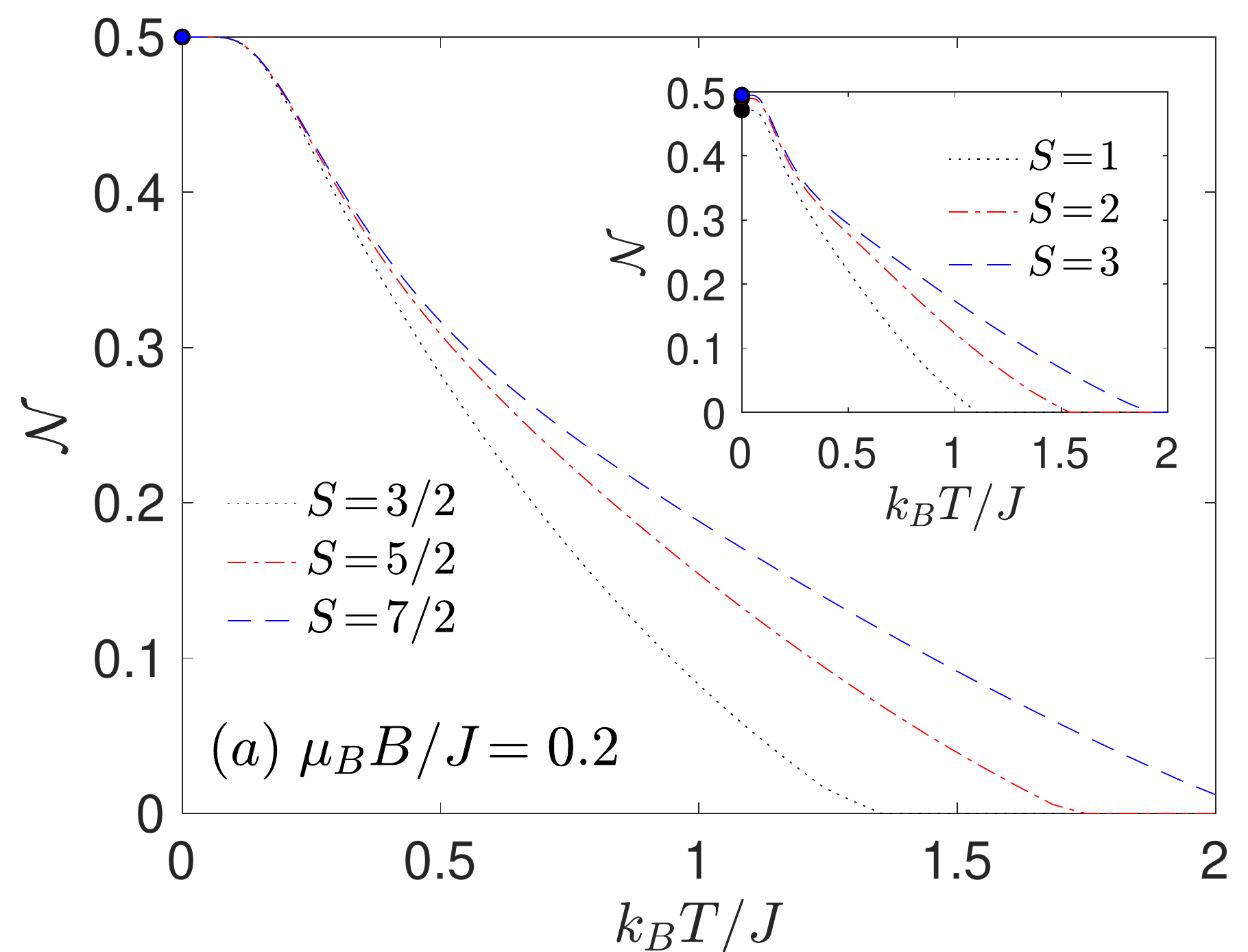}}\\
{\includegraphics[width=.9\columnwidth,trim=0.5cm 7.6cm 2cm 8.cm, clip]{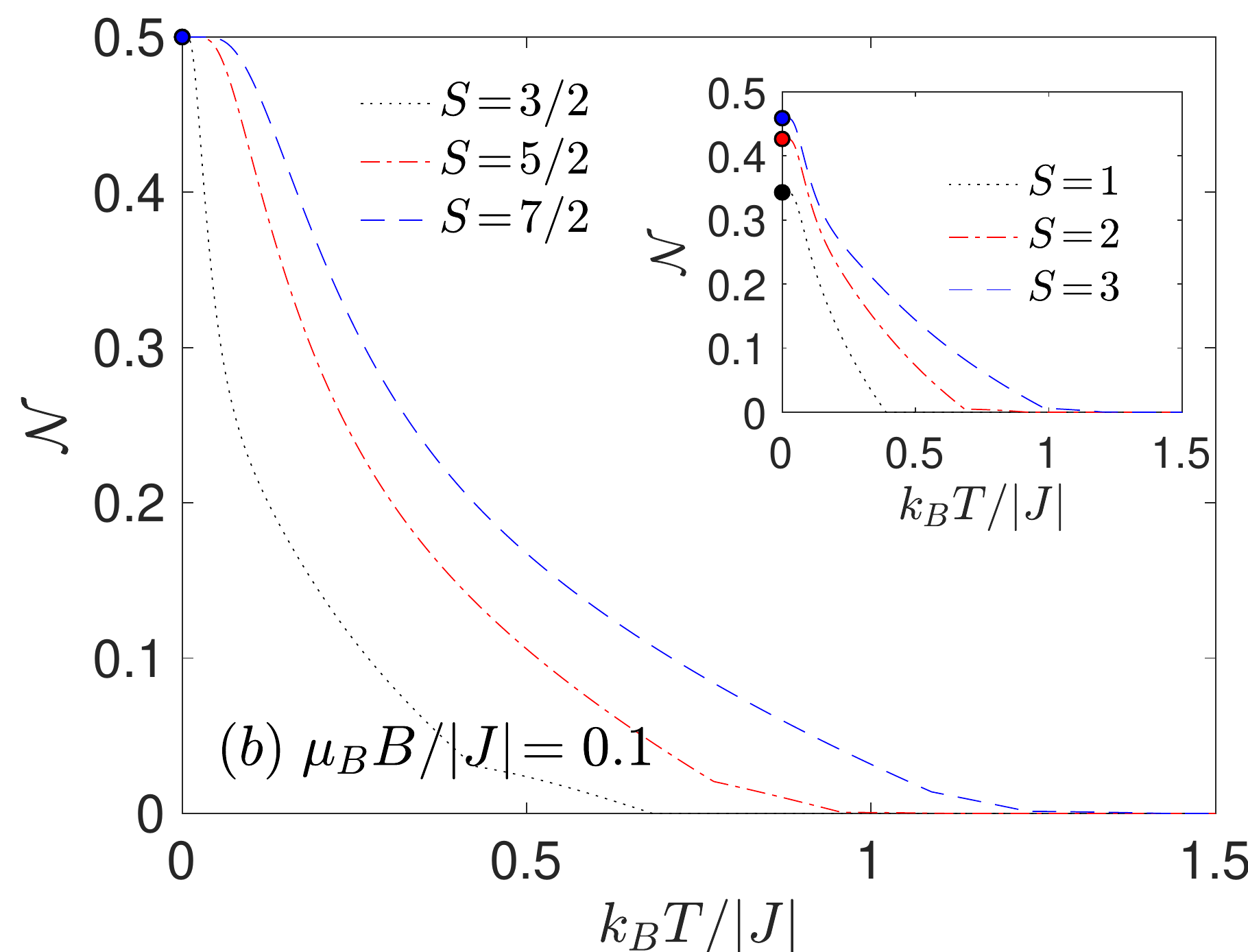}}
\caption{The thermal dependence of the negativity for a few different values of spin  magnitudes $S$ calculated at  $D/J\!=\!1$ and $(a)$ $\mu_BB/J\!=\!0.2$, $J\!>\!0$  or $(b)$ $\mu_BB/|J|\!=\!0.1$, $J\!<\!0$. The choice of model parameters corresponds to the  region where the ground state $\vert (0)_\pm\rangle$ (half-odd-integer spin $S$) or  $\vert (1/2)_\pm\rangle$ (integer spin $S$)  are favoured.
}
\label{fig6a}
\end{figure}

\begin{figure}[t]
{\includegraphics[width=.9\columnwidth,trim=0.5cm 7.6cm 2cm 8.cm, clip]{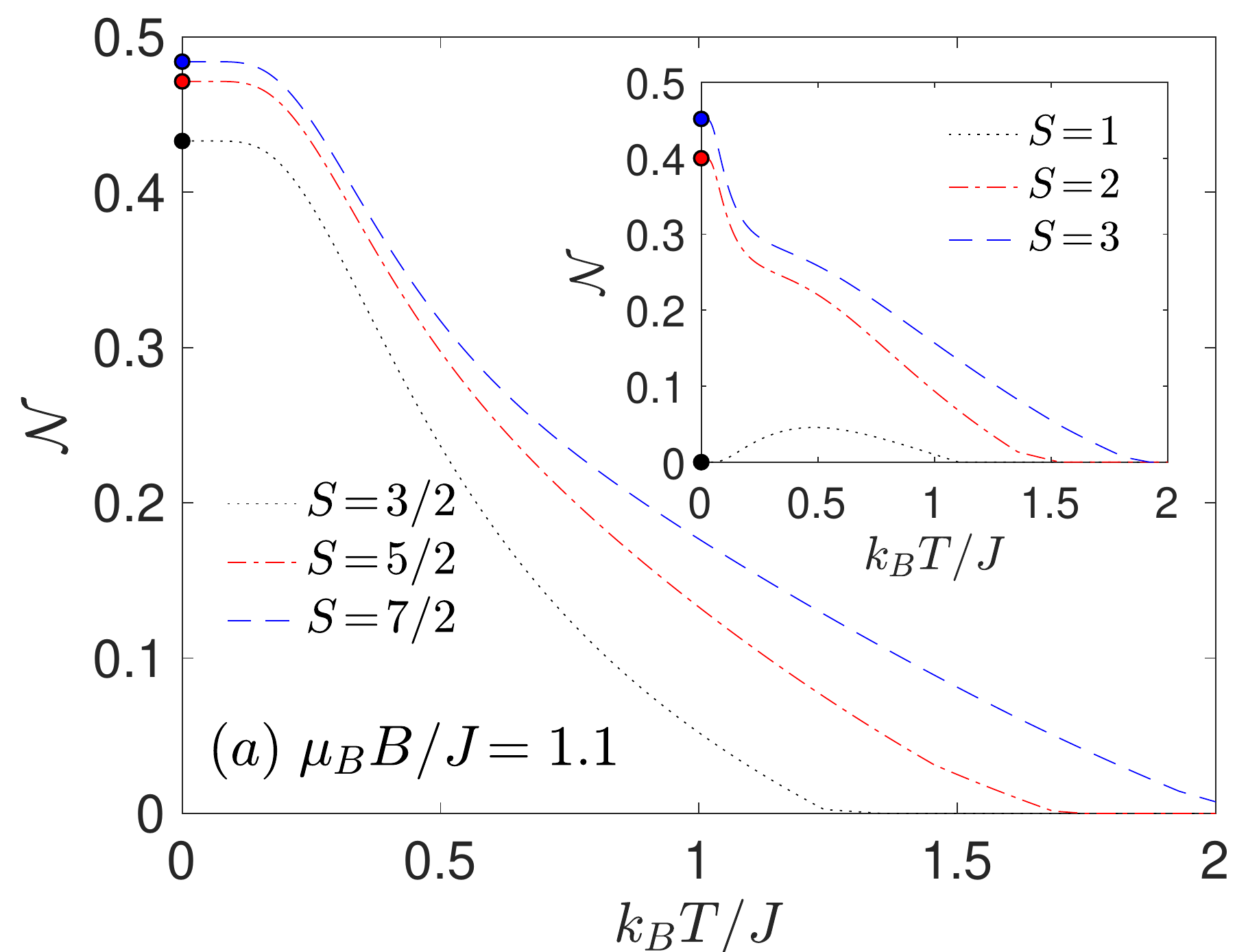}}\\
{\includegraphics[width=.9\columnwidth,trim=0.5cm 7.6cm 2cm 8.cm, clip]{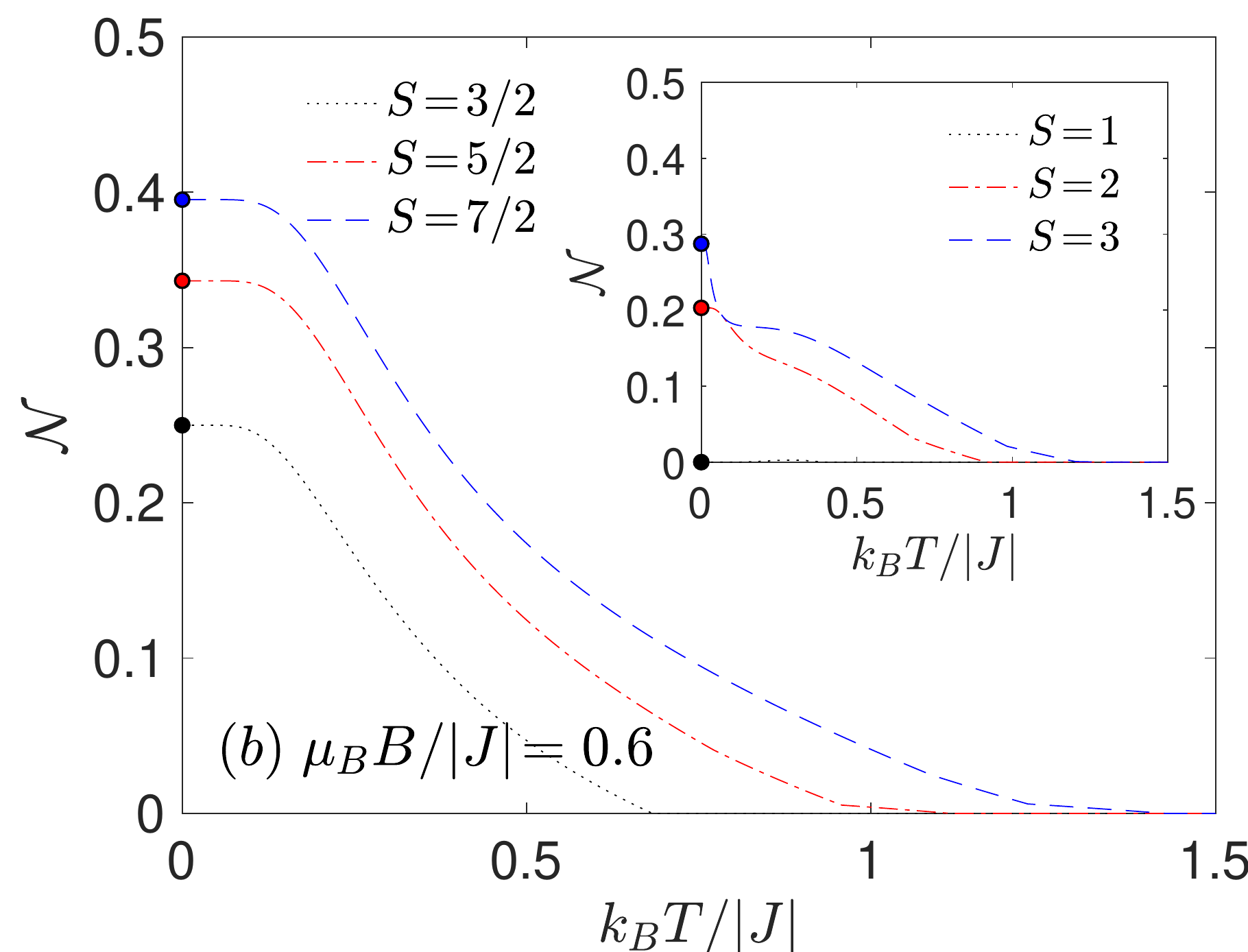}}
\caption{The thermal dependence of the negativity for a few different values of spin  magnitudes $S$ calculated at  $D/J\!=\!1$ and $(a)$ $\mu_BB/J\!=\!1.1$, $J\!>\!0$ or $(b)$ $\mu_BB/|J|\!=\!0.6$, $J\!<\!0$. The choice of model parameters corresponds to the  region where the ground state $\vert (1)_\pm\rangle$ (half-odd-integer spin $S$) or  $\vert (3/2)_\pm\rangle$ (integer spin $S$)  are favoured. In case of $S\!=\!1$ the $\vert S\!+\!1/2\rangle$ ground state is realized in both panels.}
\label{fig7a}
\end{figure}

In order to analyse the thermal behaviour of the negativity we have chosen the specific sets of model parameters under the influence of magnetic field being consistent with $\vert (S\!-\!1)_-\rangle$ ground state (Fig.~\ref{fig5a}),  $\vert (0)_\pm\rangle$ or the $\vert (1/2)_\pm\rangle$ ground state (Fig.~\ref{fig6a}) and finally with the $\vert (1)_\pm\rangle$ or $\vert (3/2)_\pm\rangle$ ones (Fig.~\ref{fig7a}). 

 Focusing on  Fig.~\ref{fig5a} we can generalize our previous zero-temperature conclusion~\cite{Varga21}, which states that the increasing spin magnitude $S$ reduces the quantum as well as low-temperature thermal negativity. On the other hand, the increasing spin magnitude $S$   enlarges the threshold temperature, which subsequently allows us to detect a subtle enhancement of the thermal entanglement at  larger temperatures if the spin size $S$ increases. The obtained results are in a perfect quantitative agreement with  previous observations for the Heisenberg  dimers without the uniaxial single-ion anisotropy $D/J$~\cite{Wang2006,Sun06,Huang2008,Li}. 
The most significant finding follows from  Fig.~\ref{fig6a} and~\ref{fig7a}.
 It is  evident from these figures that the increasing spin  magnitude $S$ can   enhance not only the threshold temperature, but it can also  enhance the degree of thermal entanglement in contrast to previous  knowledge. It should be emphasized that the above statement  holds provided that  the ensemble of integer or half-odd-integer spins $S$ is taken into account separately.  At  low magnetic fields,  the   thermal negativity of Heisenberg  dimers consisting of both half-odd-integer-spins  always saturate in the maximal value ${\cal N}\!=\!1/2$ in the asymptotic limit of absolute zero temperature, whereas   the maximum of  negativity of Heisenberg  dimers composed  of integer and one-half spin converges to the value  ${\cal N}\!=\!\frac{1}{2}\sqrt{\frac{4S(S\!+\!1)}{(1\!-\!2D/J)^2\!+\!4S(S\!+\!1)}}$. Hence, one can immediately conclude that the negativity of the Heisenberg dimers with  integer spins $S$   can reach the maximum value ${\cal N}\!=\!1/2$ just for the special case    $D/J\!=\!1/2$. 

\begin{figure}[t!]
{\includegraphics[width=.9\columnwidth,trim=0.5cm 7.5cm 0.5cm 7cm, clip]{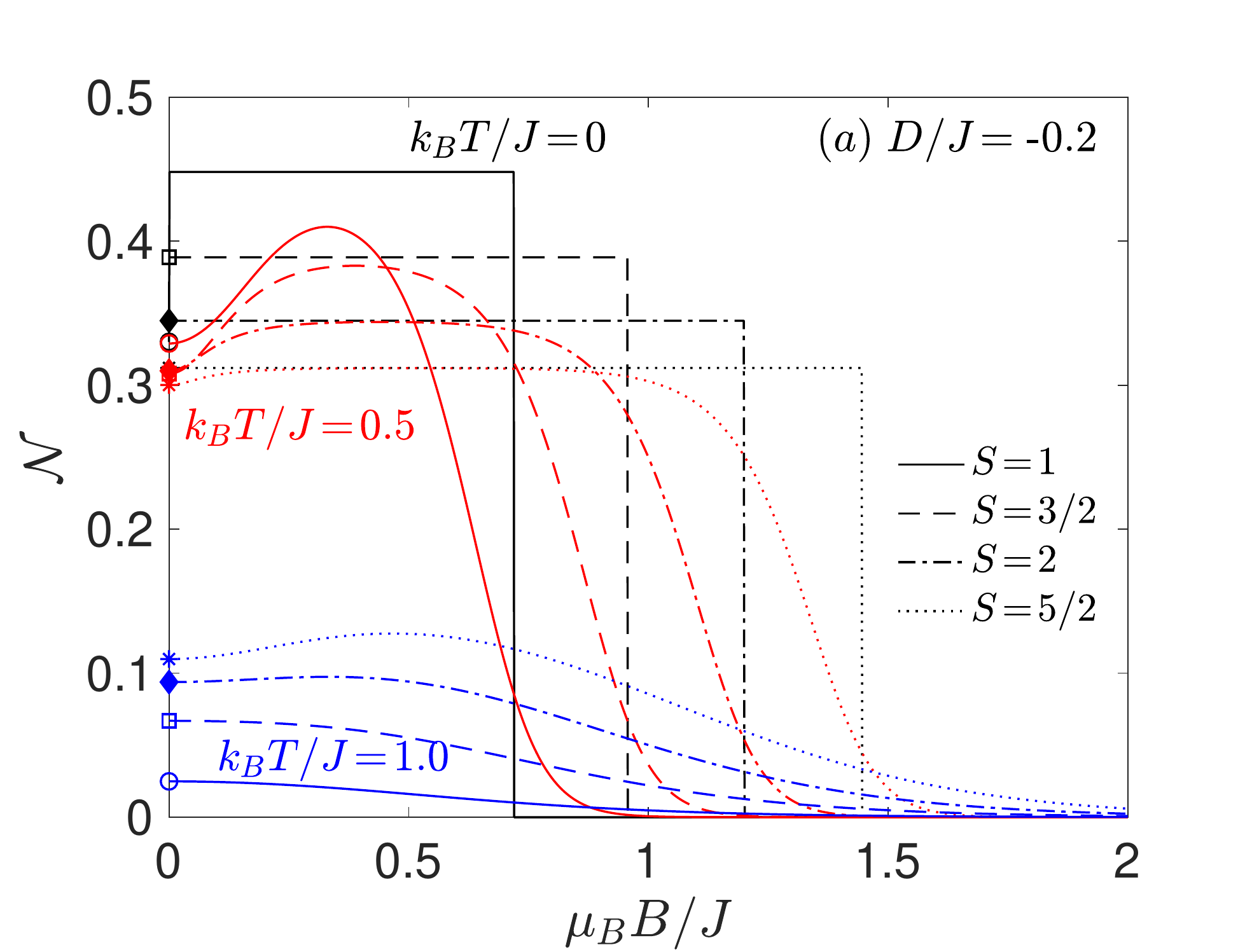}}\\
{\includegraphics[width=.9\columnwidth,trim=0.5cm 7.5cm 0.5cm 7cm, clip]{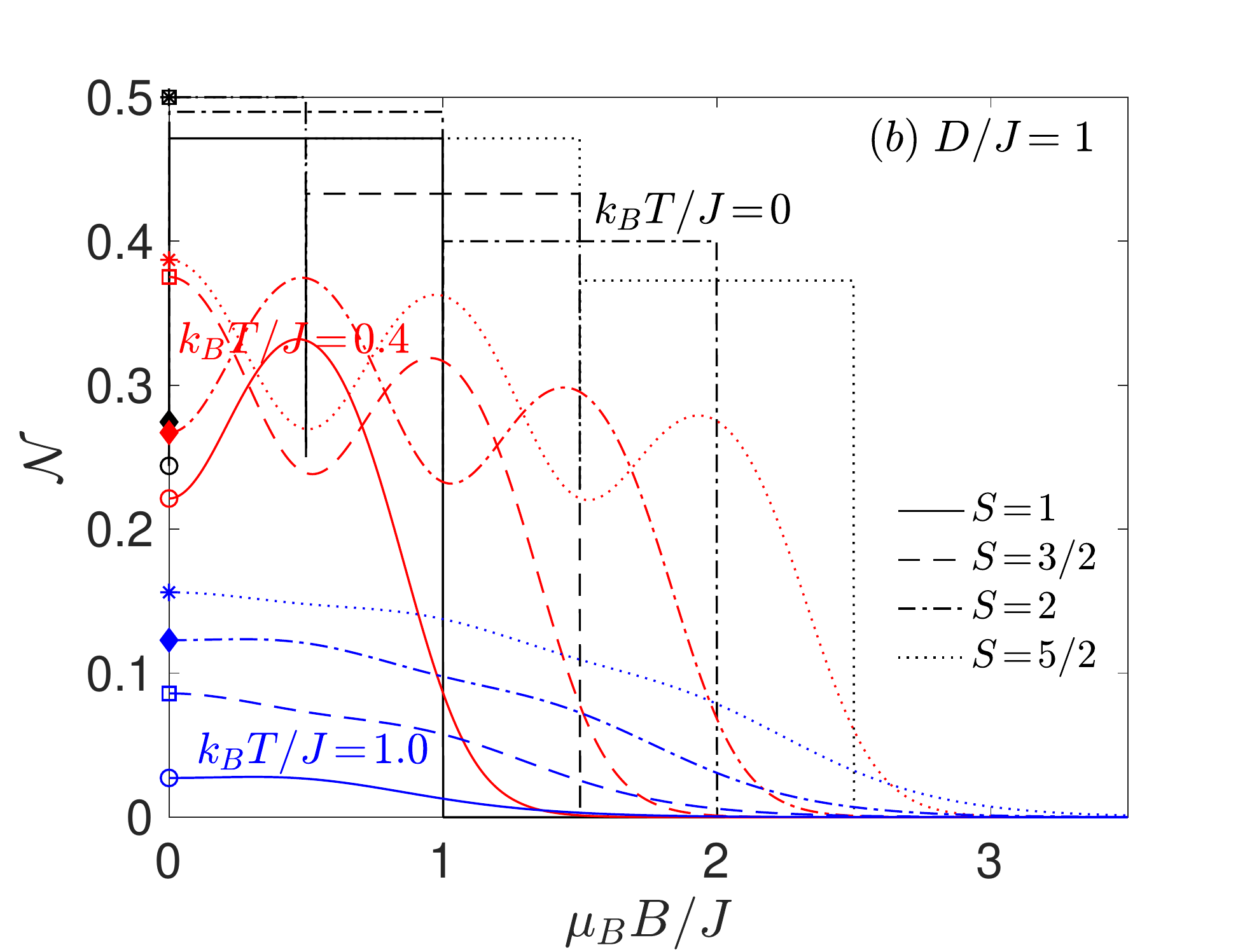}}
\caption{The    negativity as a function of the external magnetic field for four different spin magnitudes, three different values of temperature and two selected values of the uniaxial single-ion anisotropy: $(a)$  $D/J\!=\!-0.2$;  $(b)$ $D/J\!=\!1$.  }
\label{fig8a}
\end{figure}

In Fig.~\ref{fig8a} we present, furthermore, the behaviour of the thermal negativity under the changes of magnetic field.  The same model parameters have been used as in the above analysis.  It should be emphasized that  absence of Zeeman's therm  leads to the ground-state  degeneracy in the zero-temperature limit and thus, the zero-field negativity is always smaller than that   in  an arbitrary small but non-zero magnetic field. This fact is visualized through the symbols on the $y$-axis determining the respective asymptotic values of the negativity in zero-field limit. 
In agreement with general expectation the increasing temperature reduces the bipartite entanglement with a significant drop of the negativity emergent in a proximity of all magnetic-field-induced phase  transitions associated with crossing of energy levels. Around each level-crossing field the interplay between  thermal and quantum fluctuations is the most pronounced and the negativity shows a marked  local maximum located between two neighbouring level-crossing fields. In a consequence of that, the mixed spin-(1/2,$S$) Heisenberg dimers exhibit very specific oscillating changes of the negativity at low and moderate  temperatures (Fig.~\ref{fig8a}$(b)$). It is  worthwhile to remark 
that such oscillating behaviour is possible only for the spin-(1/2,$S$) Heisenberg dimers with higher spin magnitude $S\!>\!1$, because existence of at least  two level-crossing fields has to be  guaranteed. 
In addition it turns out that  the distance  between two local maxima can be tuned through the uniaxial single-ion anisotropy $D/J$, which unambiguously  determines the stability region (field range) of a given magnetic ground state (see Figs.~\ref{fig3a} and~\ref{fig4a}).

\begin{figure}[t!]
{\includegraphics[width=.9\columnwidth,trim=0.5cm 7.5cm 0.5cm 7cm, clip]{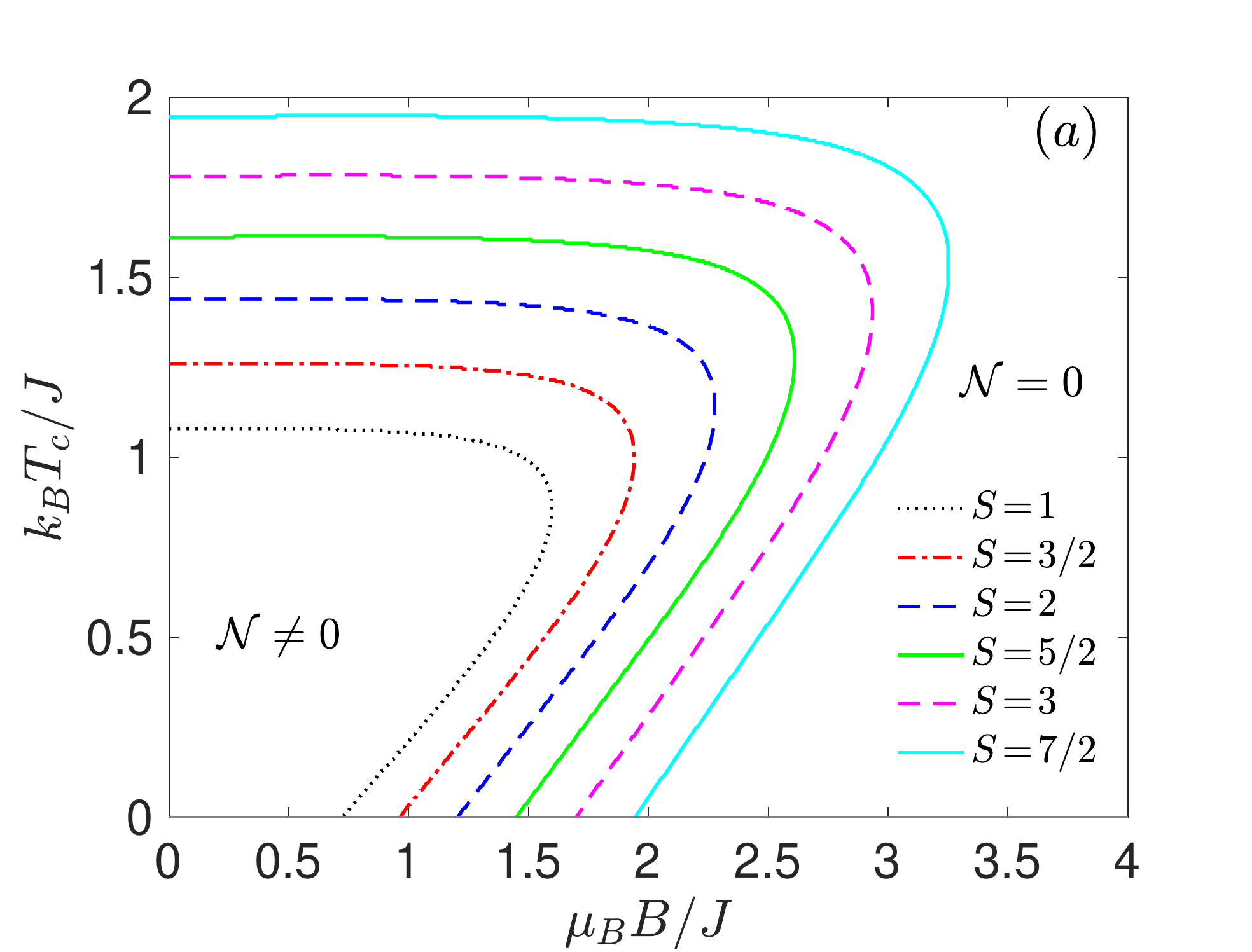}}\\
{\includegraphics[width=.9\columnwidth,trim=0.5cm 7.5cm 0.5cm 7cm, clip]{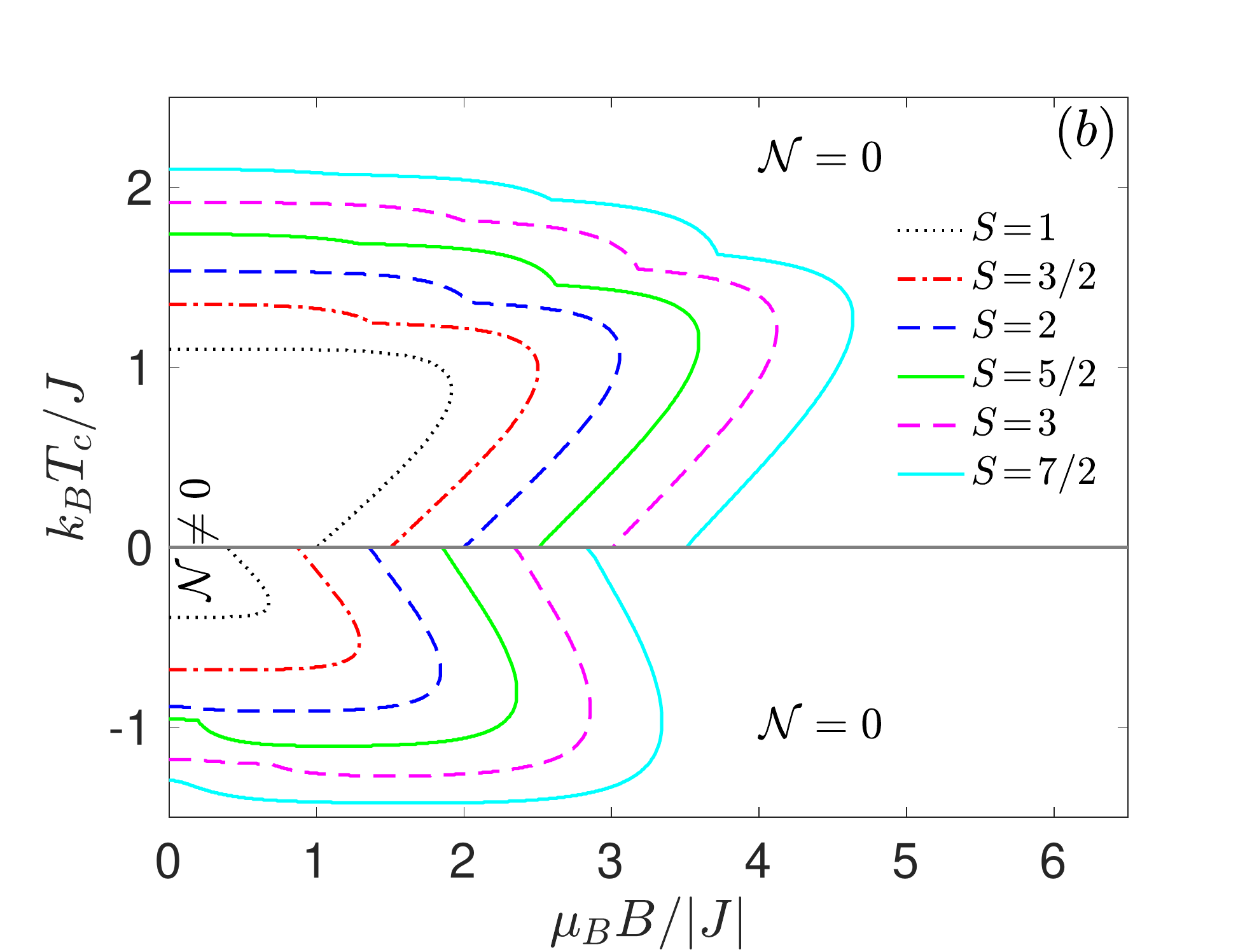}}
\caption{The behaviour of  the threshold temperature as a function of external magnetic field and various spin-$S$ magnitude. Results are calculated for two different values of  uniaxial single-ion anisotropy: $(a)$  $D/J\!=\!-0.2$  and  $(b)$ $D/J\!=\!1$, respectively. }
\label{fig9a}
\end{figure}

Finally, let us turn our attention to  the dependence of the threshold temperature on the spin magnitude $S$  as well as the  external magnetic field. The results presented in Fig.~\ref{fig9a} in the form of the  threshold  temperature  versus magnetic field plot for a few selected spin sizes $S$ confirm previous  conclusions~\cite{Wang2006,Sun06,Huang2008,Li} that the threshold temperature gradually enlarges with an enhancement of the spin magnitude $S$. It is also quite evident from  Fig.~\ref{fig9a}$(b)$ that the antiferromagnetic mixed spin-(1/2,$S$) Heisenberg dimers are more persistent against rising temperature and magnetic field than their ferromagnetic counterparts (the negative values of $k_BT_c/J$ in Fig.~\ref{fig9a}$(b)$ correspond to the mixed-spin Heisenberg dimers with  the ferromagnetic exchange coupling $J\!<\!0$). Another remarkable finding is that 
 all investigated mixed  spin-$(1/2,S)$ Heisenberg dimers exhibit a striking reentrant behaviour of the threshold temperature regardless  of the character and sized of the exchange coupling and the uniaxial single-ion anisotropy.  The origin of the unconventional reentrant phenomenon could be explained by incapability of  the magnetic field to suppress the thermally induced population of the entangled excited states, which is reflected in a thermally stimulated rise of the negativity.  
 It appears worthwhile to remark that existence of  magnetic-field-driven phase transitions  gives rise to a stepwise  dependence  of the respective threshold temperature. The  threshold temperature  slightly decreases at each level-crossing field for the antiferromagnetic Heisenberg dimers ($J\!>\!0$), while   an opposite effect is observed for the ferromagnetic Heisenberg dimers  ($J\!<\!0$), see Fig.~\ref{fig9a}$(b)$.

\section{Conclusions}
\label{sec:conclusion}
In the present paper we have exactly examined the  effect of the  spin magnitude $S$, magnetic field and uniaxial single-ion anisotropy on the quantum and thermal entanglement of the mixed spin-(1/2,$S$) Heisenberg dimers. In particular, it has been verified that the concurrent  interplay  of the  uniaxial single-ion anisotropy and the external magnetic field basically influences bipartite entanglement of the mixed-spin Heisenberg dimers. To  quantify the degree of bipartite entanglement  we have derived the exact analytical expression for the negativity in terms of Peres-Horodecki criterion~\cite{Peres,Horodecki} followed by the mathematical formulation due to  Vidal and Werner~\cite{Vidal}. In the  present study we have  provided first an exhaustive analysis of  all possible ground states of the mixed spin-(1/2,$S$) Heisenberg dimers as a necessarily prerequisite for further  entanglement analysis.  Two different scenarios were observed for both the  antiferromagnetic ($J\!>\!0$) and ferromagnetic ($J\!<\!0$)  coupling constants depending on the character of an uniaxial single-ion anisotropy $D/J$. For the easy-axis single-ion anisotropy $D/J\!<\!0$ the mixed-spin Heisenberg dimers  exhibit either one or none  magnetic-field-driven phase transition, whereas   the increasing magnetic field generates $S\!+\!(2S\mod 2)/2$ consecutive field-driven phase transitions between the ground states with the total spin $|S^z_t|\!\leq\!S\!-\!1/2$ for the easy-plane single-ion anisotropy $D/J\!>\!0$.

As a direct consequence of different effect of easy-axis and easy-plane uniaxial single-ion anisotropy  one detects very different  influence of increasing spin magnitude $S$ on the bipartite  entanglement. In the case of easy-axis uniaxial single-ion anisotropy  an increasing spin $S$   always suppresses the degree of quantum entanglement as dictated by the formula Eq.~\eqref{qs6}. In contrast to this, the increasing spin magnitude $S$  for the easy-plane single-ion anisotropy   $D/J\!>\!0$ leads to the coincidence of regions with a fixed   number of the total spin $S^z_t\!\leq\!S\!-\!1/2$, and in accordance with the formula~\eqref{gs7}, the enhancement of a quantum entanglement can be observed.  Interestingly, two specific  conditions with maximal entanglement invariant on external  stimuli have been identified  for: (i)  antiferromagnetic  mixed spin-(1/2,$S$) Heisenberg dimers with an arbitrary spin-$S$ magnitude for the particular value of the uniaxial single-ion anisotropy $D/J\!=\!1/2$, or (ii) the antiferromagnetic ground state $\vert (0)_-\rangle$  exclusively existing only in the mixed spin-(1/2,$S$) Heisenberg dimers with both half-odd-integer spin constituents.

The comprehensive analysis of a thermal entanglement  above a unique  ground state $\vert (S\!-\!1/2)_-\rangle$ confirms previously reported findings  that the increasing   spin magnitude $S$ enlarges the threshold temperature, but unfortunately reduces the degree of thermal entanglement. However, a completely  novel and unexpected behaviour has been observed for the easy-plane single-ion anisotropy $D/J\!>\!0$, where  the increasing   spin $S$ may  simultaneously  enlarge the threshold temperature as well as    the degree of thermal entanglement if  one compares solely  integer or half-odd-integer  spin-$S$ cases. It has been evidenced that except the particular case with  $D/J\!=\!1/2$  the mixed spin-(1/2,$S$) Heisenberg dimers with both half-odd-integer spins   generally achieve higher degree of entanglement, which makes them more attractive for a practical utilization. Last but not least,  a fascinating oscillating changes of the negativity were observed upon the variation of the external magnetic field at low and moderate temperatures. The unconventional oscillating behaviour of the negativity originates from  existence of consecutive field-driven phase transitions emergent at level-crossing fields, at which the negativity rapidly falls down before it produces  significant local maxima localized in between two level-crossing fields. It is noteworthy that the local maxima of the negativity of the mixed spin-(1/2,$S$) Heisenberg dimers  with integer and half-odd-integer spin $S$  are not equal due to existence   of different ground states  in the parameter region of the easy-plane single-ion anisotropy $D/J\!>\!0$. Of course, the envelope of such oscillations is gradually suppressed upon strengthening of the magnetic field until the fully polarized state is reached.

\begin{acknowledgements}
This work was financially supported by the grant of the Slovak Research and Development Agency provided under the contract No. APVV-20-0150 and by the grant of The Ministry of Education, Science, Research, and Sport of the Slovak Republic provided under the contract No. VEGA 1/0105/20.
\end{acknowledgements}

\begin{widetext}
\appendix
\section{Density matrices}
\label{App A}
\setcounter{equation}{0}
\renewcommand{\theequation}{\thesection.\arabic{equation}}
\subsection{Density matrix of the mixed spin-(1/2,1) Heisenberg dimer}
The density matrix of the  mixed spin-(1/2,1) Heisenberg dimer can be recast into the following block-diagonal form 
\begin{align}
\allowdisplaybreaks
{\hat{\rho}}\!=\!
\begin{blockarray}{r cc cc cc}
 & \vert \tfrac{1}{2},1\rangle & \vert \!-\!\tfrac{1}{2},\!-\!1\rangle &\vert \tfrac{1}{2},0\rangle  &\vert \!-\!\tfrac{1}{2},1\rangle&\vert \tfrac{1}{2},\!-\!1\rangle&\vert \!-\!\tfrac{1}{2},0\rangle\\
\begin{block}{r(cc cc cc)}
\langle \tfrac{1}{2},1\vert \;\;\;&\rho_{1,1} & 0 & 0 & 0 & 0 & 0 \\
\langle \!-\!\tfrac{1}{2},\!-\!1\vert \;\;\;&0 & \rho_{6,6} & 0 & 0 & 0 & 0 \\
\langle \tfrac{1}{2},0\vert\;\;\; & 0 & 0 & \rho_{2,2} &  \rho_{2,4}& 0 & 0 \\
 \langle \!-\!\tfrac{1}{2},1\vert\; \;\;& 0&0 & \rho_{4,2}  & \rho_{4,4} & 0 & 0\\
\langle \tfrac{1}{2},\!-\!1\vert\;\;\;&0 &0 & 0 &0  &  \rho_{3,3} & \rho_{3,5} \\
 \langle \!-\!\tfrac{1}{2},0\vert \;\;\;& 0 & 0 & 0 & 0 & \rho_{5,3} & \rho_{5,5} \\
\end{block}
\end{blockarray}\;\;.
\label{b1}
\end{align}
The individual elements of the density matrix are defined as 
\begin{align}
\rho_{1,1}\!&=\!\langle\tfrac{1}{2},1\rvert {\hat{\rho}}\lvert\tfrac{1}{2},1\rangle\!=\!\frac{1}{\cal Z}{\rm e}^{-\beta \varepsilon_{1,\frac{3}{2}}},
\label{b2}\\
\rho_{2,2}\!&=\!\langle\tfrac{1}{2},0\rvert {\hat{\rho}}\lvert\tfrac{1}{2},0\rangle\!=\!\frac{1}{\cal Z}\left( (c_{1,\frac{1}{2}}^-)^2{\rm e}^{-\beta \varepsilon_{1,\frac{1}{2}}^-} \!+\!(c_{1,\frac{1}{2}}^+)^2{\rm e}^{-\beta \varepsilon_{1,\frac{1}{2}}^+} \right),
\label{b3}\\
\rho_{3,3}\!&=\!\langle\tfrac{1}{2},-1\rvert {\hat{\rho}}\lvert\tfrac{1}{2},-1\rangle\!=\!\frac{1}{\cal Z}\left( (c_{1,{-\frac{1}{2}}}^-)^2{\rm e}^{-\beta \varepsilon_{1,-{\frac{1}{2}}}^-} \!+\!(c_{1,-{\frac{1}{2}}}^+)^2{\rm e}^{-\beta \varepsilon_{1,-{\frac{1}{2}}}^+} \right),
\label{b4}\\
\rho_{4,4}\!&=\!\langle-\tfrac{1}{2},1\rvert {\hat{\rho}}\lvert-\tfrac{1}{2},1\rangle\!=\!\frac{1}{\cal Z}\left( (c_{1,\frac{1}{2}}^+)^2{\rm e}^{-\beta \varepsilon_{1,\frac{1}{2}}^-} \!+\!(c_{1,\frac{1}{2}}^-)^2{\rm e}^{-\beta \varepsilon_{1,\frac{1}{2}}^+} \right),
\label{b5}\\
\rho_{5,5}\!&=\!\langle-\tfrac{1}{2},0\rvert {\hat{\rho}}\lvert-\tfrac{1}{2},0\rangle\!=\!\frac{1}{\cal Z}\left( (c_{1,-{\frac{1}{2}}}^+)^2{\rm e}^{-\beta \varepsilon_{1,-{\frac{1}{2}}}^-} \!+\!(c_{1,-{\frac{1}{2}}}^-)^2{\rm e}^{-\beta \varepsilon_{1,-{\frac{1}{2}}}^+} \right),
\label{b6}\\
\rho_{6,6}\!&=\!\langle-\tfrac{1}{2},-1\rvert {\hat{\rho}}\lvert-\tfrac{1}{2},-1\rangle\!=\!\frac{1}{\cal Z}{\rm e}^{-\beta \varepsilon_{1,-\frac{3}{2}}} ,
\label{b7}\\
\rho_{2,4}\!&=\!\langle\tfrac{1}{2},0\rvert {\hat{\rho}}\lvert-\tfrac{1}{2},1\rangle\!=\!\frac{c_{1,\frac{1}{2}}^+c_{1,\frac{1}{2}}^-}{\cal Z}\left({\rm e}^{-\beta \varepsilon_{1,\frac{1}{2}}^+} \!-\!{\rm e}^{-\beta \varepsilon_{1,\frac{1}{2}}^-} \right),
\hspace*{0.2cm}
\rho_{4,2}\!=\!\langle -\tfrac{1}{2},1\rvert {\hat{\rho}}\lvert\tfrac{1}{2},0\rangle\!=\!\frac{c_{1,\frac{1}{2}}^+c_{1,\frac{1}{2}}^-}{\cal Z}\left({\rm e}^{-\beta \varepsilon_{1,\frac{1}{2}}^+} \!-\!{\rm e}^{-\beta \varepsilon_{1,\frac{1}{2}}^-} \right).
\label{b8}\\
\rho_{3,5}\!&=\!\langle\tfrac{1}{2},-1\rvert {\hat{\rho}}\lvert-\tfrac{1}{2},0\rangle\!=\!\frac{c_{1,-{\frac{1}{2}}}^+c_{1,-{\frac{1}{2}}}^-}{\cal Z}\left({\rm e}^{-\beta \varepsilon_{1,-{\frac{1}{2}}}^+} \!-\!{\rm e}^{-\beta \varepsilon_{1,-{\frac{1}{2}}}^-} \right), \nonumber\\
%\hspace*{0.15cm}
\rho_{5,3}\!&=\!\langle -\tfrac{1}{2},0\rvert {\hat{\rho}}\lvert\tfrac{1}{2},-1\rangle\!=\!\frac{c_{1,-{\frac{1}{2}}}^+c_{1,-{\frac{1}{2}}}^-}{\cal Z}\left({\rm e}^{-\beta \varepsilon_{1,-{\frac{1}{2}}}^+} \!-\!{\rm e}^{-\beta \varepsilon_{1,-{\frac{1}{2}}}^-} \right).
\label{b9}
\end{align}
The partition function ${\cal Z}$  of the mixed spin-(1/2,1) Heisenberg dimer  reads as follows
\begin{align}
{\cal Z}&\!=\!2\left\{{\rm e}^{-\frac{\beta}{2}(J+2D)}\cosh\left(\frac{3\beta h}{2}\right)\!+\!2{\rm e}^{\frac{\beta}{4}(J-2D)}\cosh\left[\frac{\beta}{4}\sqrt{(J\!-\!2D)^2\!+\!8(J\Delta)^2} \right]\cosh\left(\frac{\beta h}{2} \right)\right\}.
\label{b10}
\end{align}
Finally, the probability coefficients $c_{S,S^z_t}^{\pm}$ and the respective energies $\varepsilon^{\pm}_{S,S^z_t}$ ($S^z_t\!=\!-1/2,1/2$) have the explicit forms
\begin{align}
c^{\mp}_{1,-{\frac{1}{2}}}&\!=\!\frac{1}{\sqrt{2}}\sqrt{1\!\pm\!\frac{(J\!-\!2D)}{\sqrt{(J\!-\!2D)^2\!+\!8(J\Delta)^2}}},
\hspace*{1cm}\varepsilon^{\mp}_{1,-{\tfrac{1}{2}}}\!=\!-\frac{1}{4}\left( J\!-\!2D\!-\!2h \right)\!\mp\!\frac{1}{4}\sqrt{(J\!-\!2D)^2\!+\!8(J\Delta)^2},
\label{b11}\\
c^{\mp}_{1,\frac{1}{2}}&\!=\!\frac{1}{\sqrt{2}}\sqrt{1\!\mp\!\frac{J\!-\!2D}{\sqrt{(J\!-\!2D)^2\!+\!8(J\Delta)^2}}},\hspace*{1cm}\varepsilon^{\mp}_{1,{\frac{1}{2}}}\!=\!-\frac{1}{4}\left( J\!-\!2D\!+\!2h \right)\!\mp\!\frac{1}{4}\sqrt{(J\!-\!2D)^2\!+\!8(J\Delta)^2},
\label{b12}\\
c_{1,{\pm}\frac{3}{2}}&\!=\!1, \hspace*{5.5cm}\varepsilon_{1,\pm\frac{3}{2}}\!=\!\frac{1}{2}(J\!+\!2D\!\mp\!3h).
\end{align}
\subsection{Density matrix of the mixed spin-(1/2,3/2) Heisenberg dimer}
The density matrix of the mixed  spin-(1/2,3/2) Heisenberg dimer can be recast into the following block-diagonal form 
\begin{align}
\allowdisplaybreaks
{\hat{\rho}}\!=\!
\begin{blockarray}{r cc cc cc cc}
 & \vert \frac{1}{2},\frac{3}{2}\rangle & \vert \!-\!\frac{1}{2},\!-\!\frac{3}{2}\rangle &\vert \frac{1}{2},\frac{1}{2}\rangle  &\vert \!-\!\frac{1}{2},\frac{3}{2}\rangle&\vert \frac{1}{2},\!-\!\frac{1}{2}\rangle&\vert \!-\!\frac{1}{2},\frac{1}{2}\rangle&\vert \frac{1}{2},\!-\!\frac{3}{2}\rangle&\vert \!-\!\frac{1}{2},\!-\!\frac{1}{2}\rangle\\
\begin{block}{r(cc cc cc cc)}
\langle \frac{1}{2},\frac{3}{2}\vert \;\;\;&\rho_{1,1} & 0 & 0 & 0 & 0 & 0 & 0 & 0\\
\langle \!-\!\frac{1}{2},\!-\!\frac{3}{2}\vert \;\;\;&0 & \rho_{8,8} & 0 & 0 & 0 & 0 & 0 & 0\\
\langle \frac{1}{2},\frac{1}{2}\vert\;\;\; & 0 & 0 & \rho_{2,2} &  \rho_{2,5}& 0 & 0 & 0 & 0\\
 \langle \!-\!\frac{1}{2},\frac{3}{2}\vert\; \;\;& 0&0 & \rho_{5,2}  & \rho_{5,5} & 0 & 0 & 0 & 0\\
\langle \frac{1}{2},\!-\!\frac{1}{2}\vert\;\;\;&0 &0 & 0 &0  &  \rho_{3,3} & \rho_{3,6} & 0 & 0 \\
 \langle \!-\!\frac{1}{2},\frac{1}{2}\vert \;\;\;& 0 & 0 & 0 & 0 & \rho_{6,3} & \rho_{6,6} & 0 & 0 \\
  \langle \frac{1}{2},\!-\!\frac{3}{2}\vert \;\;\;& 0 & 0 & 0 & 0 & 0 & 0 & \rho_{4,4} & \rho_{4,7}\\
   \langle \!-\!\frac{1}{2},\!-\!\frac{1}{2}\vert \;\;\;& 0 & 0 & 0 & 0 & 0 & 0 & \rho_{7,4} & \rho_{7,7}\\
\end{block}
\end{blockarray}\;\;.
\label{bb1}
\end{align}
The individual elements of the density matrix are defined as 
\begin{align}
\rho_{1,1}\!&=\!\langle\tfrac{1}{2},\tfrac{3}{2}\rvert {\hat{\rho}}\lvert\tfrac{1}{2},\tfrac{3}{2}\rangle\!=\!\frac{1}{\cal Z}{\rm e}^{-\beta \varepsilon_{\frac{3}{2},2}},
\label{bb2}\\
\rho_{2,2}\!&=\!\langle\tfrac{1}{2},\tfrac{1}{2}\rvert {\hat{\rho}}\lvert\tfrac{1}{2},\tfrac{1}{2}\rangle\!=\!\frac{1}{\cal Z}\left( (c_{\frac{3}{2},1}^-)^2{\rm e}^{-\beta \varepsilon_{\frac{3}{2},1}^-} \!+\!(c_{\frac{3}{2},1}^+)^2{\rm e}^{-\beta \varepsilon_{\frac{3}{2},1}^+} \right),
\label{bb3}\\
\rho_{3,3}\!&=\!\langle\tfrac{1}{2},-\tfrac{1}{2}\rvert {\hat{\rho}}\lvert\tfrac{1}{2},-\tfrac{1}{2}\rangle\!=\!\frac{1}{\cal Z}\left( (c_{\frac{3}{2},0}^-)^2{\rm e}^{-\beta \varepsilon_{\frac{3}{2},0}^-} \!+\!(c_{\frac{3}{2},0}^+)^2{\rm e}^{-\beta \varepsilon_{\frac{3}{2},0}^+} \right),
\label{bb4}\\
\rho_{4,4}\!&=\!\langle\tfrac{1}{2},-\tfrac{3}{2}\rvert {\hat{\rho}}\lvert\tfrac{1}{2},-\tfrac{3}{2}\rangle\!=\!\frac{1}{\cal Z}\left( (c_{\frac{3}{2},-1}^-)^2{\rm e}^{-\beta \varepsilon_{\frac{3}{2},-1}^-} \!+\!(c_{\frac{3}{2},-1}^+)^2{\rm e}^{-\beta \varepsilon_{\frac{3}{2},-1}^+} \right),
\label{bb5}\\
\rho_{5,5}\!&=\!\langle-\tfrac{1}{2},\tfrac{3}{2}\rvert {\hat{\rho}}\lvert-\tfrac{1}{2},\tfrac{3}{2}\rangle\!=\!\frac{1}{\cal Z}\left( (c_{\frac{3}{2},1}^+)^2{\rm e}^{-\beta \varepsilon_{\frac{3}{2},1}^-} \!+\!(c_{\frac{3}{2},1}^-)^2{\rm e}^{-\beta \varepsilon_{\frac{3}{2},1}^+} \right),
\label{bb6}\\
\rho_{6,6}\!&=\!\langle-\tfrac{1}{2},\tfrac{1}{2}\rvert {\hat{\rho}}\lvert-\tfrac{1}{2},\tfrac{1}{2}\rangle\!=\!\frac{1}{\cal Z}\left( (c_{\frac{3}{2},0}^+)^2{\rm e}^{-\beta \varepsilon_{\frac{3}{2},0}^-} \!+\!(c_{\frac{3}{2},0}^-)^2{\rm e}^{-\beta \varepsilon_{\frac{3}{2},0}^+} \right),
\label{bb7}\\
\rho_{7,7}\!&=\!\langle-\tfrac{1}{2},-\tfrac{1}{2}\rvert {\hat{\rho}}\lvert-\tfrac{1}{2},-\tfrac{1}{2}\rangle\!=\!\frac{1}{\cal Z}\left( (c_{\frac{3}{2},-1}^+)^2{\rm e}^{-\beta \varepsilon_{\frac{3}{2},-1}^-} \!+\!(c_{\frac{3}{2},-1}^-)^2{\rm e}^{-\beta \varepsilon_{\frac{3}{2},-1}^+} \right),
\label{bb8}\\
\rho_{8,8}\!&=\!\langle-\tfrac{1}{2},-\tfrac{3}{2}\rvert {\hat{\rho}}\lvert-\tfrac{1}{2},-\tfrac{3}{2}\rangle\!=\!\frac{1}{\cal Z}{\rm e}^{-\beta \varepsilon_{\frac{3}{2},-2}} ,
\label{bb9}\\
\rho_{2,5}\!&=\!\langle\tfrac{1}{2},\tfrac{1}{2}\rvert {\hat{\rho}}\lvert-\tfrac{1}{2},\tfrac{3}{2}\rangle\!=\!\frac{c_{\frac{3}{2},1}^+c_{\frac{3}{2},1}^-}{\cal Z}\left({\rm e}^{-\beta \varepsilon_{\frac{3}{2},1}^+} \!-\!{\rm e}^{-\beta \varepsilon_{\frac{3}{2},1}^-} \right),
\hspace*{0.2cm}
\rho_{5,2}\!=\!\langle-\tfrac{1}{2},\tfrac{3}{2}\rvert {\hat{\rho}}\lvert\tfrac{1}{2},\tfrac{1}{2}\rangle\!=\!\frac{c_{\frac{3}{2},1}^+c_{\frac{3}{2},1}^-}{\cal Z}\left({\rm e}^{-\beta \varepsilon_{\frac{3}{2},1}^+} \!-\!{\rm e}^{-\beta \varepsilon_{\frac{3}{2},1}^-} \right),
\label{bb10}\\
\rho_{3,6}\!&=\!\langle\tfrac{1}{2},-\tfrac{1}{2}\rvert {\hat{\rho}}\lvert-\tfrac{1}{2},\tfrac{1}{2}\rangle\!=\!\frac{c_{\frac{3}{2},0}^+c_{\frac{3}{2},0}^-}{\cal Z}\left({\rm e}^{-\beta \varepsilon_{\frac{3}{2},0}^+} \!-\!{\rm e}^{-\beta \varepsilon_{\frac{3}{2},0}^-} \right),
%\nonumber\\
\rho_{6,3}\!=\!\langle-\tfrac{1}{2},\tfrac{1}{2}\rvert {\hat{\rho}}\lvert\tfrac{1}{2},-\tfrac{1}{2}\rangle\!=\!\frac{c_{\frac{3}{2},0}^+c_{\frac{3}{2},0}^-}{\cal Z}\left({\rm e}^{-\beta \varepsilon_{\frac{3}{2},0}^+} \!-\!{\rm e}^{-\beta \varepsilon_{\frac{3}{2},0}^-} \right),
\label{bb11}\\
\rho_{4,7}\!&=\!\langle\tfrac{1}{2},-\tfrac{3}{2}\rvert {\hat{\rho}}\lvert-\tfrac{1}{2},-\tfrac{1}{2}\rangle\!=\!\frac{c_{\frac{3}{2},-1}^+c_{\frac{3}{2},-1}^-}{\cal Z}\left({\rm e}^{-\beta \varepsilon_{\frac{3}{2},-1}^+} \!-\!{\rm e}^{-\beta \varepsilon_{\frac{3}{2},-1}^-} \right),
\nonumber\\
\rho_{7,4}\!&=\!\langle-\tfrac{1}{2},-\tfrac{1}{2}\rvert {\hat{\rho}}\lvert\tfrac{1}{2},-\tfrac{3}{2}\rangle\!=\!\frac{c_{\frac{3}{2},-1}^+c_{\frac{3}{2},-1}^-}{\cal Z}\left({\rm e}^{-\beta \varepsilon_{\frac{3}{2},-1}^+} \!-\!{\rm e}^{-\beta \varepsilon_{\frac{3}{2},-1}^-} \right).
\label{bb12}
\end{align}
The partition function ${\cal Z}$  of the mixed spin-(1/2,3/2) Heisenberg dimer  reads as follows
\begin{align}
{\cal Z}&\!=\!2\left\{{\rm e}^{-\frac{3\beta}{4}(J+3D)}\cosh\left(2\beta h\right)
\!+\!2{\rm e}^{\frac{\beta}{4}(J-5D)}\cosh\left(\frac{\beta}{2}\sqrt{(J\!-\!2D)^2\!+\!3(J\Delta)^2} \right)\cosh\left(\beta h \right)
\!+\!{\rm e}^{\frac{\beta}{4}(J-D)}\cosh\left(\beta\sqrt{(J\Delta)^2}\right)\right\}.
\label{bb13}
\end{align}
Finally, the probability coefficients $c_{S,S^z_t}^{\pm}$ and the respective energies $\varepsilon^{\pm}_{S,S^z_t}$ ($S^z_t\!=\!-1,0,1$) have the explicit forms
\begin{align}
c^{\mp}_{\frac{3}{2},-1}&\!=\!\frac{1}{\sqrt{2}}\sqrt{1\!\pm\!\frac{(J\!-\!2D)}{\sqrt{(J\!-\!2D)^2\!+\!3(J\Delta)^2}}},
\hspace*{2cm}
\varepsilon^{\mp}_{\frac{3}{2},-1}\!=\!-\frac{1}{4}\left( J\!-\!5D\!-\!4h \right)\!\mp\!\frac{1}{2}\sqrt{(J\!-\!2D)^2\!+\!3(J\Delta)^2},
\label{bb14}\\
c^{\mp}_{\frac{3}{2},0}&\!=\!\frac{1}{\sqrt{2}},
\hspace*{6.1cm}
\varepsilon^{\mp}_{\frac{3}{2},0}\!=\!-\frac{1}{4}\left( J\!-\!D \right)\!\mp\!\sqrt{(J\Delta)^2},
\label{bb15}\\
c^{\mp}_{\frac{3}{2},1}&\!=\!\frac{1}{\sqrt{2}}\sqrt{1\!\mp\!\frac{(J\!-\!2D)}{\sqrt{(J\!-\!2D)^2\!+\!3(J\Delta)^2}}},
\hspace*{2cm}
\varepsilon^{\mp}_{\frac{3}{2},1}\!=\!-\frac{1}{4}\left( J\!-\!5D\!+\!4h \right)\!\mp\!\frac{1}{2}\sqrt{(J\!-\!2D)^2\!+\!3(J\Delta)^2},
\label{bb16}\\
c_{\frac{3}{2},{\pm}2}&\!=\!1, \hspace*{6.6cm}\varepsilon_{\frac{3}{2},\pm2}\!=\!\frac{3}{4}(J\!+\!3D\!\mp\!2h).
\end{align}
\subsection{Density matrix of the mixed spin-(1/2,2) Heisenberg dimer}
The density matrix of the mixed spin-(1/2,2) Heisenberg dimer can be recast into the following block-diagonal form 
\begin{align}
\allowdisplaybreaks
{\hat{\rho}}\!=\!
\begin{blockarray}{r cc cc cc cc cc}
 & \vert \frac{1}{2},2\rangle & \vert \!-\!\frac{1}{2},\!-\!2\rangle &\vert \frac{1}{2},1\rangle  &\vert \!-\!\frac{1}{2},2\rangle&\vert \frac{1}{2},0\rangle&\vert \!-\!\frac{1}{2},1\rangle&\vert \frac{1}{2},\!-\!1\rangle&\vert \!-\!\frac{1}{2},0\rangle&\vert \frac{1}{2},\!-\!2\rangle&\vert \!-\!\frac{1}{2},\!-\!1\rangle\\
\begin{block}{r(cc cc cc cc cc)}
\langle \frac{1}{2},2\vert \;\;\;&\rho_{1,1} & 0 & 0 & 0 & 0 & 0 & 0 & 0 & 0 & 0\\
\langle \!-\!\frac{1}{2},\!-\!2\vert \;\;\;&0 & \rho_{10,10} & 0 & 0 & 0 & 0 & 0 & 0 & 0 & 0\\
\langle \frac{1}{2},1\vert\;\;\; & 0 & 0 & \rho_{2,2} &  \rho_{2,6}& 0 & 0 & 0 & 0 & 0 & 0\\
 \langle \!-\!\frac{1}{2},2\vert\; \;\;& 0&0 & \rho_{6,2}  & \rho_{6,6} & 0 & 0 & 0 & 0&0 & 0\\
\langle \frac{1}{2},0\vert\;\;\;&0 &0 & 0 &0  &  \rho_{3,3} & \rho_{3,7} &0 & 0& 0 & 0 \\
 \langle \!-\!\frac{1}{2},1\vert \;\;\;& 0 & 0 & 0 & 0 & \rho_{7,3} & \rho_{7,7} & 0 & 0&0&0 \\
  \langle \frac{1}{2},\!-\!1\vert \;\;\;& 0 & 0 & 0 & 0 & 0 & 0 & \rho_{4,4} & \rho_{4,8}&0&0\\
   \langle \!-\!\frac{1}{2},0\vert \;\;\;& 0 & 0 & 0 & 0 & 0 & 0 & \rho_{8,4} & \rho_{8,8}&0&0\\
    \langle \frac{1}{2},\!-\!2\vert \;\;\;& 0&0&0 & 0 & 0 & 0 & 0 & 0 & \rho_{5,5} & \rho_{5,9}\\
        \langle \!-\!\frac{1}{2},\!-\!1\vert \;\;\;& 0&0&0 & 0 & 0 & 0 & 0 & 0 & \rho_{9,5} & \rho_{9,9}\\
\end{block}
\end{blockarray}\;\;.
\label{bbb1}
\end{align}
The individual elements of the density matrix are defined as 
\begin{align}
\rho_{1,1}\!&=\!\langle\tfrac{1}{2},2\rvert {\hat{\rho}}\lvert\tfrac{1}{2},2\rangle\!=\!\frac{1}{\cal Z}{\rm e}^{-\beta \varepsilon_{2,\frac{5}{2}}},
\label{bbb2}\\
\rho_{2,2}\!&=\!\langle\tfrac{1}{2},1\rvert {\hat{\rho}}\lvert\tfrac{1}{2},1\rangle\!=\!\frac{1}{\cal Z}\left( (c_{2,\frac{3}{2}}^-)^2{\rm e}^{-\beta \varepsilon_{2,\frac{3}{2}}^-} \!+\!(c_{2,\frac{3}{2}}^+)^2{\rm e}^{-\beta \varepsilon_{2,\frac{3}{2}}^+} \right),
\label{bbb3}\\
\rho_{3,3}\!&=\!\langle\tfrac{1}{2},0\rvert {\hat{\rho}}\lvert\tfrac{1}{2},0\rangle\!=\!\frac{1}{\cal Z}\left( (c_{2,\frac{1}{2}}^-)^2{\rm e}^{-\beta \varepsilon_{2,\frac{1}{2}}^-} \!+\!(c_{2,\frac{1}{2}}^+)^2{\rm e}^{-\beta \varepsilon_{2,\frac{1}{2}}^+} \right),
\label{bbb4}\\
\rho_{4,4}\!&=\!\langle\tfrac{1}{2},-1\rvert {\hat{\rho}}\lvert\tfrac{1}{2},-1\rangle\!=\!\frac{1}{\cal Z}\left( (c_{2,-\frac{1}{2}}^-)^2{\rm e}^{-\beta \varepsilon_{2,-\frac{1}{2}}^-} \!+\!(c_{2,-\frac{1}{2}}^+)^2{\rm e}^{-\beta \varepsilon_{2,-\frac{1}{2}}^+} \right),
\label{bbb5}\\
\rho_{5,5}\!&=\!\langle\tfrac{1}{2},-2\rvert {\hat{\rho}}\lvert\tfrac{1}{2},-2\rangle\!=\!\frac{1}{\cal Z}\left( (c_{2,-\frac{3}{2}}^-)^2{\rm e}^{-\beta \varepsilon_{2,-\frac{3}{2}}^-} \!+\!(c_{2,-\frac{3}{2}}^+)^2{\rm e}^{-\beta \varepsilon_{2,-\frac{3}{2}}^+} \right),
\label{bbb6}\\
\rho_{6,6}\!&=\!\langle-\tfrac{1}{2},2\rvert {\hat{\rho}}\lvert-\tfrac{1}{2},2\rangle\!=\!\frac{1}{\cal Z}\left( (c_{2,\frac{3}{2}}^+)^2{\rm e}^{-\beta \varepsilon_{2,\frac{3}{2}}^-} \!+\!(c_{2,\frac{3}{2}}^-)^2{\rm e}^{-\beta \varepsilon_{2,\frac{3}{2}}^+} \right),
\label{bbb7}\\
\rho_{7,7}\!&=\!\langle-\tfrac{1}{2},1\rvert {\hat{\rho}}\lvert-\tfrac{1}{2},1\rangle\!=\!\frac{1}{\cal Z}\left( (c_{2,\frac{1}{2}}^+)^2{\rm e}^{-\beta \varepsilon_{2,\frac{1}{2}}^-} \!+\!(c_{2,\frac{1}{2}}^-)^2{\rm e}^{-\beta \varepsilon_{2,\frac{1}{2}}^+} \right),
\label{bb8}\\
\rho_{8,8}\!&=\!\langle-\tfrac{1}{2},0\rvert {\hat{\rho}}\lvert-\tfrac{1}{2},0\rangle\!=\!\frac{1}{\cal Z}\left( (c_{2,-\frac{1}{2}}^+)^2{\rm e}^{-\beta \varepsilon_{2,-\frac{1}{2}}^-} \!+\!(c_{2,-\frac{1}{2}}^-)^2{\rm e}^{-\beta \varepsilon_{2,-\frac{1}{2}}^+} \right),
\label{bbb9}\\
\rho_{9,9}\!&=\!\langle-\tfrac{1}{2},-1\rvert {\hat{\rho}}\lvert-\tfrac{1}{2},-1\rangle\!=\!\frac{1}{\cal Z}\left( (c_{2,-\frac{3}{2}}^+)^2{\rm e}^{-\beta \varepsilon_{2,-\frac{3}{2}}^-} \!+\!(c_{2,-\frac{3}{2}}^-)^2{\rm e}^{-\beta \varepsilon_{2,-\frac{3}{2}}^+} \right),
\label{bbb10}\\
\rho_{10,10}\!&=\!\langle-\tfrac{1}{2},-2\rvert {\hat{\rho}}\lvert-\tfrac{1}{2},-2\rangle\!=\!\frac{1}{\cal Z}{\rm e}^{-\beta \varepsilon_{2,-\frac{5}{2}}} ,
\label{bbb11}\\
\rho_{2,6}\!&=\!\langle\tfrac{1}{2},1\rvert {\hat{\rho}}\lvert-\tfrac{1}{2},2\rangle\!=\!\frac{c_{2,\frac{3}{2}}^+c_{2,\frac{3}{2}}^-}{\cal Z}\left({\rm e}^{-\beta \varepsilon_{2,\frac{3}{2}}^+} \!-\!{\rm e}^{-\beta \varepsilon_{2,\frac{3}{2}}^-} \right),\;
\rho_{6,2}\!=\!\langle-\tfrac{1}{2},2\rvert {\hat{\rho}}\lvert\tfrac{1}{2},1\rangle\!=\!\frac{c_{2,\frac{3}{2}}^+c_{2,\frac{3}{2}}^-}{\cal Z}\left({\rm e}^{-\beta \varepsilon_{2,\frac{3}{2}}^+} \!-\!{\rm e}^{-\beta \varepsilon_{2,\frac{3}{2}}^-} \right),
\label{bbb12}\\
\rho_{3,7}\!&=\!\langle\tfrac{1}{2},0\rvert {\hat{\rho}}\lvert-\tfrac{1}{2},1\rangle\!=\!\frac{c_{2,\frac{1}{2}}^+c_{2,\frac{1}{2}}^-}{\cal Z}\left({\rm e}^{-\beta \varepsilon_{2,\frac{1}{2}}^+} \!-\!{\rm e}^{-\beta \varepsilon_{2,\frac{1}{2}}^-} \right),\;\;
\rho_{7,3}\!=\!\langle-\tfrac{1}{2},1\rvert {\hat{\rho}}\lvert\tfrac{1}{2},0\rangle\!=\!\frac{c_{2,\frac{1}{2}}^+c_{2,\frac{1}{2}}^-}{\cal Z}\left({\rm e}^{-\beta \varepsilon_{2,\frac{1}{2}}^+} \!-\!{\rm e}^{-\beta \varepsilon_{2,\frac{1}{2}}^-} \right),
\label{bbb13}\\
\rho_{4,8}\!&=\!\langle\tfrac{1}{2},-1\rvert {\hat{\rho}}\lvert-\tfrac{1}{2},0\rangle\!=\!\frac{c_{2,-\frac{1}{2}}^+c_{2,-\frac{1}{2}}^-}{\cal Z}\left({\rm e}^{-\beta \varepsilon_{2,-\frac{1}{2}}^+} \!-\!{\rm e}^{-\beta \varepsilon_{2,-\frac{1}{2}}^-} \right),\nonumber\\
\rho_{8,4}\!&=\!\langle-\tfrac{1}{2},0\rvert {\hat{\rho}}\lvert\tfrac{1}{2},-1\rangle\!=\!\frac{c_{2,-\frac{1}{2}}^+c_{2,-\frac{1}{2}}^-}{\cal Z}\left({\rm e}^{-\beta \varepsilon_{2,-\frac{1}{2}}^+} \!-\!{\rm e}^{-\beta \varepsilon_{2,-\frac{1}{2}}^-} \right),
\label{bbb14}\\
\rho_{5,9}\!&=\!\langle\tfrac{1}{2},-2\rvert {\hat{\rho}}\lvert-\tfrac{1}{2},-1\rangle\!=\!\frac{c_{2,-\frac{3}{2}}^+c_{2,-\frac{3}{2}}^-}{\cal Z}\left({\rm e}^{-\beta \varepsilon_{2,-\frac{3}{2}}^+} \!-\!{\rm e}^{-\beta \varepsilon_{2,-\frac{3}{2}}^-} \right),\nonumber\\
\rho_{9,5}\!&=\!\langle-\tfrac{1}{2},-1\rvert {\hat{\rho}}\lvert\tfrac{1}{2},-2\rangle\!=\!\frac{c_{2,-\frac{3}{2}}^+c_{2,-\frac{3}{2}}^-}{\cal Z}\left({\rm e}^{-\beta \varepsilon_{2,-\frac{3}{2}}^+} \!-\!{\rm e}^{-\beta \varepsilon_{2,-\frac{3}{2}}^-} \right).
\label{bbb15}
\end{align}
The partition function ${\cal Z}$  of the mixed spin-(1/2,2) Heisenberg dimer  reads as follows
\begin{align}
{\cal Z}&\!=\!2\left\{{\rm e}^{-\beta(J+4D)}\cosh\left(\frac{5\beta h}{2}\right)
\!+\!2{\rm e}^{\frac{\beta}{4}(J\!-\!10D)}\cosh\left(\frac{\beta}{4}\sqrt{9(J\!-\!2D)^2\!+\!16(J\Delta)^2} \right)\cosh\left(\frac{3\beta h}{2}\right)\right.
\nonumber\\
&\!+\!\left.
2{\rm e}^{\frac{\beta}{4}(J\!-\!2D)}\cosh\left(\frac{\beta}{4}\sqrt{(J\!-\!2D)^2\!+\!24(J\Delta)^2} \right)\cosh\left(\frac{\beta h}{2}\right)
\right\}.
\label{bbb16}
\end{align}
Finally, the probability coefficients $c_{S,S^z_t}^{\pm}$ and the respective energies $\varepsilon^{\pm}_{S,S^z_t}$ ($S^z_t\!=\!-3/2,-1/2,1/2,3/2$) have the explicit forms
\begin{align}
c^{\mp}_{2,-\frac{3}{2}}&\!=\!\frac{1}{\sqrt{2}}\sqrt{1\!\pm\!\frac{3(J\!-\!2D)}{\sqrt{9(J\!-\!2D)^2\!+\!16(J\Delta)^2}}},
\hspace*{1cm}
\varepsilon^{\mp}_{2,-\frac{3}{2}}\!=\!-\frac{1}{4}\left( J\!-\!10D\!-\!6h \right)\!\mp\!\frac{1}{4}\sqrt{9(J\!-\!2D)^2\!+\!16(J\Delta)^2},
\label{bbb17}\\
c^{\mp}_{2,-\frac{1}{2}}&\!=\!\frac{1}{\sqrt{2}}\sqrt{1\!\pm\!\frac{(J\!-\!2D)}{\sqrt{(J\!-\!2D)^2\!+\!24(J\Delta)^2}}},
\hspace*{1cm}
\varepsilon^{\mp}_{2,-\frac{1}{2}}\!=\!-\frac{1}{4}\left( J\!-\!2D\!-\!2h  \right)\!\mp\!\frac{1}{4}\sqrt{(J\!-\!2D)^2\!+\!24(J\Delta)^2},
\label{bbb18}\\
c^{\mp}_{2,\frac{1}{2}}&\!=\!\frac{1}{\sqrt{2}}\sqrt{1\!\mp\!\frac{(J\!-\!2D)}{\sqrt{(J\!-\!2D)^2\!+\!24(J\Delta)^2}}},
\hspace*{1cm}
\varepsilon^{\mp}_{2,\frac{1}{2}}\!=\!-\frac{1}{4}\left( J\!-\!2D\!+\!2h  \right)\!\mp\!\frac{1}{4}\sqrt{(J\!-\!2D)^2\!+\!24(J\Delta)^2},
\label{bbb19}\\
c^{\mp}_{2,\frac{3}{2}}&\!=\!\frac{1}{\sqrt{2}}\sqrt{1\!\mp\!\frac{3(J\!-\!2D)}{\sqrt{9(J\!-\!2D)^2\!+\!16(J\Delta)^2}}},
\hspace*{1cm}
\varepsilon^{\mp}_{2,\frac{3}{2}}\!=\!-\frac{1}{4}\left( J\!-\!10D\!+\!6h \right)\!\mp\!\frac{1}{4}\sqrt{9(J\!-\!2D)^2\!+\!16(J\Delta)^2},
\label{bbb20}\\
c_{2,\pm\frac{5}{2}}&\!=\!1, \hspace*{6cm}\varepsilon_{2,\pm\frac{5}{2}}\!=\!J\!+\!4D\!\mp\!\frac{5h}{2}.
\end{align}
\subsection{Density matrix of the mixed spin-(1/2,5/2) Heisenberg dimer}
The density matrix of the mixed spin-(1/2,5/2) Heisenberg dimer can be recast into the following block-diagonal form 
\begin{align}
\allowdisplaybreaks
{\hat{\rho}}\!=\!
\begin{blockarray}{r cc cc cc cc cc cc}
 & \vert \frac{1}{2},\frac{5}{2}\rangle & \vert \!-\!\frac{1}{2},\!-\!\frac{5}{2}\rangle &\vert \frac{1}{2},\frac{3}{2}\rangle  &\vert \!-\!\frac{1}{2},\frac{5}{2}\rangle&\vert \frac{1}{2},\frac{1}{2}\rangle&\vert \!-\!\frac{1}{2},\frac{3}{2}\rangle&\vert \frac{1}{2},\!-\!\frac{1}{2}\rangle&\vert \!-\!\frac{1}{2},\!-\!\frac{1}{2}\rangle&\vert \frac{1}{2},\!-\!\frac{3}{2}\rangle&\vert \!-\!\frac{1}{2},\!-\!\frac{1}{2}\rangle&\vert \frac{1}{2},\!-\!\frac{5}{2}\rangle&\vert \!-\!\frac{1}{2},\!-\!\frac{3}{2}\rangle\\
\begin{block}{r(cc cc cc cc cc cc)}
\langle \frac{1}{2},\frac{1}{2}\vert \;\;\;&\rho_{1,1} & 0 & 0 & 0 & 0 & 0 & 0 & 0 & 0 & 0&0&0\\
\langle \!-\!\frac{1}{2},\!-\!\frac{5}{2}\vert \;\;\;&0 & \rho_{12,12} & 0 & 0 & 0 & 0 & 0 & 0 & 0 & 0&0&0\\
\langle \frac{1}{2},\frac{3}{2}\vert\;\;\; & 0 & 0 & \rho_{2,2} &  \rho_{2,7}& 0 & 0 & 0 & 0 & 0 & 0&0&0\\
 \langle \!-\!\frac{1}{2},\frac{5}{2}\vert\; \;\;& 0&0 & \rho_{7,2}  & \rho_{7,7} & 0 & 0 & 0 & 0&0 & 0&0&0\\
\langle \frac{1}{2},\frac{1}{2}\vert\;\;\;&0 &0 & 0 &0  &  \rho_{3,3} & \rho_{3,8} &0 & 0& 0 & 0 &0&0\\
 \langle \!-\!\frac{1}{2},\frac{3}{2}\vert \;\;\;& 0 & 0 & 0 & 0 & \rho_{8,3} & \rho_{8,8} & 0 & 0&0&0 &0&0\\
  \langle \frac{1}{2},\!-\!\frac{1}{2}\vert \;\;\;& 0 & 0 & 0 & 0 & 0 & 0 & \rho_{4,4} & \rho_{4,9}&0&0&0 &0\\
   \langle \!-\!\frac{1}{2},\!-\!\frac{1}{2}\vert \;\;\;& 0 & 0 & 0 & 0 & 0 & 0 & \rho_{9,4} & \rho_{9,9}&0&0&0&0\\
    \langle \frac{1}{2},\!-\!\frac{3}{2}\vert \;\;\;& 0&0&0 & 0 & 0 & 0 & 0 & 0 & \rho_{5,5} & \rho_{5,10}&0&0\\
        \langle \!-\!\frac{1}{2},\!-\!\frac{1}{2}\vert \;\;\;& 0&0&0 & 0 & 0 & 0 & 0 & 0 & \rho_{10,5} & \rho_{10,10}&0&0\\
 \langle \frac{1}{2},\!-\!\frac{5}{2}\vert \;\;\;&0&0& 0&0&0 & 0 & 0 & 0 & 0 & 0 & \rho_{6,6} & \rho_{6,11}\\
 \langle \!-\!\frac{1}{2},\!-\!\frac{3}{2}\vert \;\;\;&0&0& 0&0&0 & 0 & 0 & 0 & 0 & 0 & \rho_{11,6} & \rho_{11,11}\\
\end{block}
\end{blockarray}\;\;.
\label{bbb1}
\end{align}
The individual elements of the density matrix are defined as 
\begin{align}
\rho_{1,1}\!&=\!\langle\tfrac{1}{2},\tfrac{5}{2}\rvert {\hat{\rho}}\lvert\tfrac{1}{2},\tfrac{5}{2}\rangle\!=\!\frac{1}{\cal Z}{\rm e}^{-\beta \varepsilon_{\frac{5}{2},3}},
\label{bbbb2}\\
\rho_{2,2}\!&=\!\langle\tfrac{1}{2},\tfrac{3}{2}\rvert {\hat{\rho}}\lvert\tfrac{1}{2},\tfrac{3}{2}\rangle\!=\!\frac{1}{\cal Z}\left( (c_{\frac{5}{2},2}^-)^2{\rm e}^{-\beta \varepsilon_{\frac{5}{2},2}^-} \!+\!(c_{\frac{5}{2},2}^+)^2{\rm e}^{-\beta \varepsilon_{\frac{5}{2},2}^+} \right),
\label{bbbb3}\\
\rho_{3,3}\!&=\!\langle\tfrac{1}{2},\tfrac{1}{2}\rvert {\hat{\rho}}\lvert\tfrac{1}{2},\tfrac{1}{2}\rangle\!=\!\frac{1}{\cal Z}\left( (c_{\frac{5}{2},1}^-)^2{\rm e}^{-\beta \varepsilon_{\frac{5}{2},1}^-} \!+\!(c_{\frac{5}{2},1}^+)^2{\rm e}^{-\beta \varepsilon_{\frac{5}{2},1}^+} \right),
\label{bbbb4}\\
\rho_{4,4}\!&=\!\langle\tfrac{1}{2},-\tfrac{1}{2}\rvert {\hat{\rho}}\lvert\tfrac{1}{2},-\tfrac{1}{2}\rangle\!=\!\frac{1}{\cal Z}\left( (c_{\frac{5}{2},0}^-)^2{\rm e}^{-\beta \varepsilon_{\frac{5}{2},0}^-} \!+\!(c_{\frac{5}{2},0}^+)^2{\rm e}^{-\beta \varepsilon_{\frac{5}{2},0}^+} \right),
\label{bbbb5}\\
\rho_{5,5}\!&=\!\langle\tfrac{1}{2},-\tfrac{3}{2}\rvert {\hat{\rho}}\lvert\tfrac{1}{2},-\tfrac{3}{2}\rangle\!=\!\frac{1}{\cal Z}\left( (c_{\frac{5}{2},-1}^-)^2{\rm e}^{-\beta \varepsilon_{\frac{5}{2},-1}^-} \!+\!(c_{\frac{5}{2},-1}^+)^2{\rm e}^{-\beta \varepsilon_{\frac{5}{2},-1}^+} \right),
\label{bbbb6}\\
\rho_{6,6}\!&=\!\langle\tfrac{1}{2},-\tfrac{5}{2}\rvert {\hat{\rho}}\lvert\tfrac{1}{2},-\tfrac{5}{2}\rangle\!=\!\frac{1}{\cal Z}\left( (c_{\frac{5}{2},-2}^-)^2{\rm e}^{-\beta \varepsilon_{\frac{5}{2},-2}^-} \!+\!(c_{\frac{5}{2},-2}^+)^2{\rm e}^{-\beta \varepsilon_{\frac{5}{2},-2}^+} \right),
\label{bbbb7}\\
\rho_{7,7}\!&=\!\langle -\tfrac{1}{2},\tfrac{5}{2}\rvert {\hat{\rho}}\lvert-\tfrac{1}{2},\tfrac{5}{2}\rangle\!=\!\frac{1}{\cal Z}\left( (c_{\frac{5}{2},2}^+)^2{\rm e}^{-\beta \varepsilon_{\frac{5}{2},2}^-} \!+\!(c_{\frac{5}{2},2}^-)^2{\rm e}^{-\beta \varepsilon_{\frac{5}{2},2}^+} \right),
\label{bbbb8}\\
\rho_{8,8}\!&=\!\langle -\tfrac{1}{2},\tfrac{3}{2}\rvert {\hat{\rho}}\lvert-\tfrac{1}{2},\tfrac{3}{2}\rangle\!=\!\frac{1}{\cal Z}\left( (c_{\frac{5}{2},1}^+)^2{\rm e}^{-\beta \varepsilon_{\frac{5}{2},1}^-} \!+\!(c_{\frac{5}{2},1}^-)^2{\rm e}^{-\beta \varepsilon_{\frac{5}{2},1}^+} \right),
\label{bbbb9}\\
\rho_{9,9}\!&=\!\langle -\tfrac{1}{2},\tfrac{1}{2}\rvert {\hat{\rho}}\lvert-\tfrac{1}{2},\tfrac{1}{2}\rangle\!=\!\frac{1}{\cal Z}\left( (c_{\frac{5}{2},0}^+)^2{\rm e}^{-\beta \varepsilon_{\frac{5}{2},0}^-} \!+\!(c_{\frac{5}{2},0}^-)^2{\rm e}^{-\beta \varepsilon_{\frac{5}{2},0}^+} \right),
\label{bbbb10}\\
\rho_{10,10}\!&=\!\langle -\tfrac{1}{2},-\tfrac{1}{2}\rvert {\hat{\rho}}\lvert-\tfrac{1}{2},-\tfrac{1}{2}\rangle\!=\!\frac{1}{\cal Z}\left( (c_{\frac{5}{2},-1}^+)^2{\rm e}^{-\beta \varepsilon_{\frac{5}{2},-1}^-} \!+\!(c_{\frac{5}{2},-1}^-)^2{\rm e}^{-\beta \varepsilon_{\frac{5}{2},-1}^+} \right),
\label{bbbb11}\\
\rho_{11,11}\!&=\!\langle -\tfrac{1}{2},-\tfrac{3}{2}\rvert {\hat{\rho}}\lvert-\tfrac{1}{2},-\tfrac{3}{2}\rangle\!=\!\frac{1}{\cal Z}\left( (c_{\frac{5}{2},-2}^+)^2{\rm e}^{-\beta \varepsilon_{\frac{5}{2},-2}^-} \!+\!(c_{\frac{5}{2},-2}^-)^2{\rm e}^{-\beta \varepsilon_{\frac{5}{2},-2}^+} \right),
\label{bbbb12}\\
\rho_{12,12}\!&=\!\langle-\tfrac{1}{2},-\tfrac{5}{2}\rvert {\hat{\rho}}\lvert-\tfrac{1}{2},-\tfrac{5}{2}\rangle\!=\!\frac{1}{\cal Z}{\rm e}^{-\beta \varepsilon_{\frac{5}{2},-3}} ,
\label{bbbb13}\\
\rho_{2,7}\!&=\!\langle\tfrac{1}{2},\tfrac{3}{2}\rvert {\hat{\rho}}\lvert-\tfrac{1}{2},\tfrac{5}{2}\rangle\!=\!\frac{c_{\frac{5}{2},2}^+c_{\frac{5}{2},2}^-}{\cal Z}\left({\rm e}^{-\beta \varepsilon_{\frac{5}{2},2}^+} \!-\!{\rm e}^{-\beta \varepsilon_{\frac{5}{2},2}^-} \right),
\rho_{7,2}\!=\!\langle-\tfrac{1}{2},\tfrac{5}{2}\rvert {\hat{\rho}}\lvert\tfrac{1}{2},\tfrac{3}{2}\rangle\!=\!\frac{c_{\frac{5}{2},2}^+c_{\frac{5}{2},2}^-}{\cal Z}\left({\rm e}^{-\beta \varepsilon_{\frac{5}{2},2}^+} \!-\!{\rm e}^{-\beta \varepsilon_{\frac{5}{2},2}^-} \right),
\label{bbbb14}\\
\rho_{3,8}\!&=\!\langle\tfrac{1}{2},\tfrac{1}{2}\rvert {\hat{\rho}}\lvert-\tfrac{1}{2},\tfrac{3}{2}\rangle\!=\!\frac{c_{\frac{5}{2},1}^+c_{\frac{5}{2},1}^-}{\cal Z}\left({\rm e}^{-\beta \varepsilon_{\frac{5}{2},1}^+} \!-\!{\rm e}^{-\beta \varepsilon_{\frac{5}{2},1}^-} \right),
\rho_{8,3}\!=\!\langle-\tfrac{1}{2},\tfrac{3}{2}\rvert {\hat{\rho}}\lvert\tfrac{1}{2},\tfrac{1}{2}\rangle\!=\!\frac{c_{\frac{5}{2},1}^+c_{\frac{5}{2},1}^-}{\cal Z}\left({\rm e}^{-\beta \varepsilon_{\frac{5}{2},1}^+} \!-\!{\rm e}^{-\beta \varepsilon_{\frac{5}{2},1}^-} \right),
\label{bbbb15}\\
\rho_{4,9}\!&=\!\langle\tfrac{1}{2},-\tfrac{1}{2}\rvert {\hat{\rho}}\lvert-\tfrac{1}{2},\tfrac{1}{2}\rangle\!=\!\frac{c_{\frac{5}{2},0}^+c_{\frac{5}{2},0}^-}{\cal Z}\left({\rm e}^{-\beta \varepsilon_{\frac{5}{2},0}^+} \!-\!{\rm e}^{-\beta \varepsilon_{\frac{5}{2},0}^-} \right),
%\nonumber\\
\rho_{9,4}\!=\!\langle-\tfrac{1}{2},\tfrac{1}{2}\rvert {\hat{\rho}}\lvert\tfrac{1}{2},-\tfrac{1}{2}\rangle\!=\!\frac{c_{\frac{5}{2},0}^+c_{\frac{5}{2},0}^-}{\cal Z}\left({\rm e}^{-\beta \varepsilon_{\frac{5}{2},0}^+} \!-\!{\rm e}^{-\beta \varepsilon_{\frac{5}{2},0}^-} \right),
\label{bbbb16}\\
\rho_{5,10}\!&=\!\langle\tfrac{1}{2},-\tfrac{3}{2}\rvert {\hat{\rho}}\lvert-\tfrac{1}{2},-\tfrac{1}{2}\rangle\!=\!\frac{c_{\frac{5}{2},-1}^+c_{\frac{5}{2},-1}^-}{\cal Z}\left({\rm e}^{-\beta \varepsilon_{\frac{5}{2},-1}^+} \!-\!{\rm e}^{-\beta \varepsilon_{\frac{5}{2},-1}^-} \right),\nonumber\\
\rho_{10,5}\!&=\!\langle-\tfrac{1}{2},-\tfrac{1}{2}\rvert {\hat{\rho}}\lvert\tfrac{1}{2},-\tfrac{3}{2}\rangle\!=\!\frac{c_{\frac{5}{2},-1}^+c_{\frac{5}{2},-1}^-}{\cal Z}\left({\rm e}^{-\beta \varepsilon_{\frac{5}{2},-1}^+} \!-\!{\rm e}^{-\beta \varepsilon_{\frac{5}{2},-1}^-} \right),
\label{bbbb17}\\
\rho_{6,11}\!&=\!\langle\tfrac{1}{2},-\tfrac{5}{2}\rvert {\hat{\rho}}\lvert-\tfrac{1}{2},-\tfrac{3}{2}\rangle\!=\!\frac{c_{\frac{5}{2},-2}^+c_{\frac{5}{2},-2}^-}{\cal Z}\left({\rm e}^{-\beta \varepsilon_{\frac{5}{2},-2}^+} \!-\!{\rm e}^{-\beta \varepsilon_{\frac{5}{2},-2}^-} \right),\nonumber\\
\rho_{11,6}\!&=\!\langle-\tfrac{1}{2},-\tfrac{3}{2}\rvert {\hat{\rho}}\lvert\tfrac{1}{2},-\tfrac{5}{2}\rangle\!=\!\frac{c_{\frac{5}{2},-2}^+c_{\frac{5}{2},-2}^-}{\cal Z}\left({\rm e}^{-\beta \varepsilon_{\frac{5}{2},-2}^+} \!-\!{\rm e}^{-\beta \varepsilon_{\frac{5}{2},-2}^-} \right).
\label{bbbb18}
\end{align}
The partition function ${\cal Z}$  of the mixed spin-(1/2,5/2) Heisenberg dimer  reads as follows
\begin{align}
{\cal Z}&\!=\!2\left\{{\rm e}^{-\frac{5\beta}{4}(J+5D)}\cosh\left(3\beta h\right)
\!+\!2{\rm e}^{\frac{\beta}{4}(J-17D)}\cosh\left(\frac{\beta}{2}\sqrt{4(J\!-\!2D)^2\!+\!5(J\Delta)^2} \right)\cosh\left(2\beta h\right)\right.
\nonumber\\
&\!+\!
2{\rm e}^{\frac{\beta}{4}(J-5D)}\cosh\left(\frac{\beta}{2}\sqrt{(J\!-\!2D)^2\!+\!8(J\Delta)^2} \right)\cosh\left(\frac{\beta h}{2}\right)
\!+\!\left.
{\rm e}^{\frac{\beta}{4}(J-D)}\cosh\left(\frac{3\beta}{2}\sqrt{(J\Delta)^2} \right)
\right\}.
\label{bbbb19}
\end{align}
Finally, the probability coefficients $c_{S,S^z_t}^{\pm}$ and the respective energies $\varepsilon^{\pm}_{S,S^z_t}$ ($S^z_t\!=\!-2,-1,0,1,2$) have the explicit forms
\begin{align}
c^{\mp}_{\frac{5}{2},-2}&\!=\!\frac{1}{\sqrt{2}}\sqrt{1\!\pm\!\frac{2(J\!-\!2D)}{\sqrt{4(J\!-\!2D)^2\!+\!5(J\Delta)^2}}},
\hspace*{1cm}
\varepsilon^{\mp}_{\frac{5}{2},-2}\!=\!-\frac{1}{4}\left( J\!-\!17D\!-\!8h \right)\!\mp\!\frac{1}{2}\sqrt{4(J\!-\!2D)^2\!+\!5(J\Delta)^2},
\label{bbbb20}\\
c^{\mp}_{\frac{5}{2},-1}&\!=\!\frac{1}{\sqrt{2}}\sqrt{1\!\pm\!\frac{(J\!-\!2D)}{\sqrt{(J\!-\!2D)^2\!+\!8(J\Delta)^2}}},\hspace*{1.2cm}
\varepsilon^{\mp}_{\frac{5}{2},-1}\!=\!-\frac{1}{4}\left( J\!-\!5D\!-\!4h  \right)\!\mp\!\frac{1}{2}\sqrt{(J\!-\!2D)^2\!+\!8(J\Delta)^2},
\label{bbbb21}\\
c^{\mp}_{\frac{5}{2},0}&\!=\!\frac{1}{\sqrt{2}},\hspace*{5.4cm}
\varepsilon^{\mp}_{\frac{5}{2},0}\!=\!-\frac{1}{4}\left( J\!-\!D \right)\!\mp\!\frac{3}{2}\sqrt{(J\Delta)^2},
\label{bbbb22}\\
c^{\mp}_{\frac{5}{2},1}&\!=\!\frac{1}{\sqrt{2}}\sqrt{1\!\mp\!\frac{(J\!-\!2D)}{\sqrt{(J\!-\!2D)^2\!+\!8(J\Delta)^2}}},\hspace*{1.2cm}
\varepsilon^{\mp}_{\frac{5}{2},1}\!=\!-\frac{1}{4}\left( J\!-\!5D\!+\!4h  \right)\!\mp\!\frac{1}{2}\sqrt{(J\!-\!2D)^2\!+\!8(J\Delta)^2},
\label{bbbb23}\\
c^{\mp}_{\frac{5}{2},2}&\!=\!\frac{1}{\sqrt{2}}\sqrt{1\!\mp\!\frac{2(J\!-\!2D)}{\sqrt{4(J\!-\!2D)^2\!+\!5(J\Delta)^2}}},\hspace*{1cm}
\varepsilon^{\mp}_{\frac{5}{2},2}\!=\!-\frac{1}{4}\left( J\!-\!17D\!+\!8h \right)\!\mp\!\frac{1}{2}\sqrt{4(J\!-\!2D)^2\!+\!5(J\Delta)^2},
\label{bbbb24}\\
c_{\frac{5}{2},{\pm}3}&\!=\!1, \hspace*{5.7cm}\varepsilon_{\frac{5}{2},\pm3}\!=\!\frac{5}{4}(J\!+\!5D)\!\mp\!3h.
\end{align}
\section{The explicit form of the eigenvalues $\lambda_{S^z_{tm}}$}
\label{App B}
\setcounter{equation}{0}
\renewcommand{\theequation}{\thesection.\arabic{equation}}

The explicit form of eigenvalues $\lambda_{S^z_{tm}}$ ($S^z_{tm}\!=\!-S\!+\!1/2,\dots,S\!-\!1/2$) of the partially transposed density matrix $\rho^{T_{1/2}}$ \eqref{eq24}  is as follows\\\\
\underline{if $S^z_{tm}\!=\!-S\!+\!1/2$:}
\begin{align}
\lambda_{S^z_{tm}}^{\mp}&\!=\!\frac{1}{2\cal Z}\left\{(c_{S,S^z_{tm}+1}^-)^2{\rm e}^{-\beta \varepsilon_{S,S^z_{tm}+1}^-}\!+\!(c_{S,S^z_{tm}+1}^+)^2{\rm e}^{-\beta \varepsilon_{S,S^z_{tm}+1}^+}\!+\!{\rm e}^{-\beta \varepsilon_{S,-(S+\frac{1}{2})}}\right\}\nonumber\\
&\hskip -0.5cm\!\mp\!\frac{1}{2\cal Z}\sqrt{\left[(c_{S,S^z_{tm}+1}^-)^2{\rm e}^{-\beta \varepsilon_{S,S^z_{tm}+1}^-}\!+\!(c_{S,S^z_{tm}+1}^+)^2{\rm e}^{-\beta \varepsilon_{S,S^z_{tm}+1}^+}\!-\!{\rm e}^{-\beta \varepsilon_{S,-(S+\frac{1}{2})}}\right]^2\!+\!4
\left[
c_{S,S^z_{tm}}^-c_{S,S^z_{tm}}^+
\left({\rm e}^{-\beta \varepsilon_{S,S^z_{tm}}^-}\!-\!{\rm e}^{-\beta \varepsilon_{S,S^z_{tm}}^+}\right)
\right]^2
}.
\end{align}
\\\\
\underline{if $S^z_{tm}\!=\!S\!-\!1/2$:}
\begin{align}
\lambda_{S^z_{tm}}^{\mp}&\!=\!\frac{1}{2\cal Z}\left\{{\rm e}^{-\beta \varepsilon_{S,(S+\frac{1}{2})}}\!+\!(c_{S,S^z_{tm}-1}^+)^2{\rm e}^{-\beta \varepsilon_{S,S^z_{tm}-1}^-}\!+\!(c_{S,S^z_{tm}-1}^-)^2{\rm e}^{-\beta \varepsilon_{S,S^z_{tm}-1}^+}\right\}\nonumber\\
&\hskip -0.5cm\!\mp\!\frac{1}{2\cal Z}\sqrt{\left[{\rm e}^{-\beta \varepsilon_{S,(S+\frac{1}{2})}}\!-\!(c_{S,S^z_{tm}-1}^+)^2{\rm e}^{-\beta \varepsilon_{S,S^z_{tm}-1}^-}\!-\!(c_{S,S^z_{tm}-1}^-)^2{\rm e}^{-\beta \varepsilon_{S,S^z_{tm}-1}^+}\right]^2\!+\!4
\left[
c_{S,S^z_{tm}}^-c_{S,S^z_{tm}}^+
\left({\rm e}^{-\beta \varepsilon_{S,S^z_{tm}}^-}\!-\!{\rm e}^{-\beta \varepsilon_{S,S^z_{tm}}^+}\right)
\right]^2
}.
\end{align}
\\\\
\underline{if $S^z_{tm}\!=\!-S\!+\!3/2,\dots,S\!-\!3/2$:}
\begin{align}
\lambda_{S^z_{tm}}^{\mp}&\!=\!\frac{1}{2\cal Z}\left\{
(c_{S,S^z_{tm}+1}^-)^2{\rm e}^{-\beta \varepsilon_{S,S^z_{tm}+1}^-}\!+\!(c_{S,S^z_{tm}+1}^+)^2{\rm e}^{-\beta \varepsilon_{S,S^z_{tm}+1}^+}
\!+\!(c_{S,S^z_{tm}-1}^+)^2{\rm e}^{-\beta \varepsilon_{S,S^z_{tm}-1}^-}\!+\!(c_{S,S^z_{tm}-1}^-)^2{\rm e}^{-\beta \varepsilon_{S,S^z_{tm}-1}^+}\right\}\nonumber\\
&\!\mp\!\frac{1}{2\cal Z}
\left\{\left[(c_{S,S^z_{tm}+1}^-)^2{\rm e}^{-\beta \varepsilon_{S,S^z_{tm}+1}^-}\!+\!(c_{S,S^z_{tm}+1}^+)^2{\rm e}^{-\beta \varepsilon_{S,S^z_{tm}+1}^+}
\!-\!(c_{S,S^z_{tm}-1}^+)^2{\rm e}^{-\beta \varepsilon_{S,S^z_{tm}-1}^-}\!-\!(c_{S,S^z_{tm}-1}^-)^2{\rm e}^{-\beta \varepsilon_{S,S^z_{tm}-1}^+}\right]^2\right.
\nonumber\\
&\hskip1cm\!+\!\left.4
\left[
c_{S,S^z_{tm}}^-c_{S,S^z_{tm}}^+
\left({\rm e}^{-\beta \varepsilon_{S,S^z_{tm}}^-}\!-\!{\rm e}^{-\beta \varepsilon_{S,S^z_{tm}}^+}\right)
\right]^2
\right\}^{1/2}.
\end{align}

\section{ Density matrices partially transposed with respect the spin-1/2 subsystem}
\label{App C}
\setcounter{equation}{0}
\renewcommand{\theequation}{\thesection.\arabic{equation}}
\subsection{Partially transposed density matrix  of the mixed spin-(1/2,1) Heisenberg dimer}
The block diagonal structure of the partially transposed density matrix with respect to the spin-1/2 subsystem of the spin-(1/2,1) Heisenberg dimer reads
\begin{align}
\allowdisplaybreaks
{\hat{\rho}}^{T_{1/2}}\!=\!
%\left(
\begin{blockarray}{r cc cc cc}
& \vert \tfrac{1}{2},\!-\!1\rangle & \vert \!-\!\tfrac{1}{2},1\rangle &\vert \tfrac{1}{2},1\rangle  &\vert \!-\!\tfrac{1}{2},0\rangle&\vert \tfrac{1}{2},0\rangle&\vert \!-\!\tfrac{1}{2},\!-\!1\rangle\\
\begin{block}{r(cc cc cc)}
\langle \tfrac{1}{2},\!-\!1\vert \;\;\;&\rho_{3,3} & 0 & 0 & 0 & 0 & 0 \\
\langle \!-\!\tfrac{1}{2},1\vert \;\;\;&0 & \rho_{4,4} & 0 & 0 & 0 & 0 \\
\langle \tfrac{1}{2},1\vert \;\;\;&0 & 0 & \rho_{1,1} &  \rho_{4,2}& 0 & 0 \\
\langle \!-\!\tfrac{1}{2},0\vert \;\;\;&0&0 & \rho_{2,4}  & \rho_{5,5} & 0 & 0\\
\langle \tfrac{1}{2},0\vert \;\;\;&0 &0 & 0 &0  &  \rho_{2,2} & \rho_{5,3} \\
\langle \!-\!\tfrac{1}{2},\!-\!1\vert \;\;\;&0 & 0 & 0 & 0 & \rho_{3,5} & \rho_{6,6} \\
\end{block}
\end{blockarray}%\right)
\;\;.
\label{c1}
\end{align}
The non-zero elements of $\rho_{i,j}^{T_{1/2}}$ are explicitly defined in Eqs.~\eqref{b2}-\eqref{b9}.
\subsection{Partially transposed density matrix of the mixed spin-(1/2,3/2) Heisenberg dimer}
The block diagonal structure of the partially transposed density matrix with respect to the spin-1/2 subsystem of the spin-(1/2,3/2) Heisenberg dimer reads
\begin{align}
\allowdisplaybreaks
{\hat{\rho}}^{T_{1/2}}\!=\!
%\left(
\begin{blockarray}{r cc cc cc cc}
& \vert \tfrac{1}{2},\!-\!\tfrac{3}{2}\rangle & \vert \!-\!\tfrac{1}{2},\tfrac{3}{2}\rangle &\vert \tfrac{1}{2},\tfrac{3}{2}\rangle  &\vert \!-\!\tfrac{1}{2},\tfrac{1}{2}\rangle&\vert \tfrac{1}{2},\tfrac{1}{2}\rangle&\vert \!-\!\tfrac{1}{2},\!-\!\tfrac{1}{2}\rangle&\vert \tfrac{1}{2},\!-\!\tfrac{1}{2}\rangle&\vert \!-\!\tfrac{1}{2},\!-\!\tfrac{3}{2}\rangle\\
\begin{block}{r(cc cc cc cc)}
\langle \tfrac{1}{2},\!-\!\frac{3}{2}\vert \;\;\;&\rho_{4,4} &0&0& 0 & 0 & 0 & 0 & 0 \\
\langle \!-\!\tfrac{1}{2},\frac{3}{2}\vert \;\;\;&0 & \rho_{5,5} &0&0& 0 & 0 & 0 & 0 \\
\langle \tfrac{1}{2},\frac{3}{2}\vert \;\;\;&0 & 0 & \rho_{1,1} &  \rho_{5,2}& 0 & 0 & 0 & 0 \\
\langle \!-\!\tfrac{1}{2},\frac{1}{2}\vert \;\;\;&0&0 & \rho_{2,5}  & \rho_{6,6} & 0 & 0& 0 & 0 \\
\langle \tfrac{1}{2},\frac{1}{2}\vert \;\;\;&0 &0 & 0 &0  &  \rho_{2,2} & \rho_{6,3} & 0 & 0 \\
\langle \!-\!\tfrac{1}{2},\!-\!\frac{1}{2}\vert \;\;\;&0 & 0 & 0 & 0 & \rho_{3,6} & \rho_{7,7} & 0 & 0 \\
\langle \tfrac{1}{2},\!-\!\frac{1}{2}\vert \;\;\;&0 & 0 & 0 &0 & 0 &0  &  \rho_{3,3} & \rho_{7,4} \\
\langle \!-\!\tfrac{1}{2},\!-\!\frac{3}{2}\vert \;\;\;&0 & 0 & 0 & 0 & 0 & 0 & \rho_{4,7} & \rho_{8,8} \\
\end{block}
\end{blockarray}%\right)
\;\;.
\label{c2}
\end{align}
The non-zero elements of $\rho_{i,j}^{T_{1/2}}$ are explicitly defined in Eqs.~\eqref{bb2}-\eqref{bb12}.
\subsection{Partially transposed density matrix of the mixed spin-(1/2,2) Heisenberg dimer}
The block diagonal structure of the partially transposed density matrix with respect to the spin-1/2 subsystem of the spin-(1/2,2) Heisenberg dimer reads
\begin{align}
\allowdisplaybreaks
{\hat{\rho}}^{T_{1/2}}\!=\!
%\left(
\begin{blockarray}{r cc cc cc cc cc}
 &\vert \tfrac{1}{2},\!-\!2\rangle&  \vert \!-\!\tfrac{1}{2},2\rangle& \vert \tfrac{1}{2},2\rangle & \vert \!-\!\tfrac{1}{2},1\rangle &\vert \tfrac{1}{2},1\rangle  &\vert \!-\!\tfrac{1}{2},0\rangle&\vert \tfrac{1}{2},0\rangle&\vert \!-\!\tfrac{1}{2},\!-\!1\rangle& \vert \tfrac{1}{2},\!-\!1\rangle &  \vert \!-\!\tfrac{1}{2},\!-\!2\rangle\\
\begin{block}{r(cc cc cc cc cc)}
\langle \tfrac{1}{2},\!-\!2\vert \;\;\;&\rho_{5,5} &0&0& 0 & 0 & 0 & 0 & 0 & 0 & 0 \\
\langle \!-\!\tfrac{1}{2},2\vert \;\;\;&0 & \rho_{6,6} &0&0& 0 & 0 & 0 & 0 & 0 & 0 \\
\langle \tfrac{1}{2},2\vert \;\;\;&0 & 0 & \rho_{1,1} &  \rho_{6,2}& 0 & 0 & 0 & 0& 0 & 0  \\
\langle \!-\!\tfrac{1}{2},1\vert \;\;\;&0&0 & \rho_{2,6}  & \rho_{7,7} & 0 & 0& 0 & 0 & 0 & 0 \\
\langle \tfrac{1}{2},1\vert \;\;\;&0 &0 & 0 &0  &  \rho_{2,2} & \rho_{7,3} & 0 & 0& 0 & 0  \\
\langle \!-\!\tfrac{1}{2},0\vert \;\;\;&0 & 0 & 0 & 0 & \rho_{3,7} & \rho_{8,8} & 0 & 0 & 0 & 0 \\
\langle \tfrac{1}{2},0\vert \;\;\;&0 & 0 & 0 &0 & 0 &0  &  \rho_{3,3} & \rho_{8,4}& 0 & 0  \\
\langle \!-\!\tfrac{1}{2},\!-\!1\vert \;\;\;&0 & 0 & 0 & 0 & 0 & 0 & \rho_{4,8} & \rho_{9,9} & 0 & 0 \\
\langle \tfrac{1}{2},\!-\!1\vert \;\;\;&0 & 0 & 0 & 0 & 0 &0 & 0 &0  &  \rho_{4,4} & \rho_{9,5} \\
\langle \!-\!\tfrac{1}{2},\!-\!2\vert \;\;\;&0 & 0 & 0 & 0 & 0 & 0 & 0 & 0 & \rho_{5,9} & \rho_{10,10} \\
\end{block}
\end{blockarray}%\right)
\;\;.
\label{c3}
\end{align}
The non-zero elements of $\rho_{i,j}^{T_{1/2}}$ are explicitly defined in Eqs.~\eqref{bbb2}-\eqref{bbb15}.
\subsection{Partially transposed density matrix of the mixed spin-(1/2,5/2) Heisenberg dimer}
The block diagonal structure of the partially transposed density matrix with respect to the spin-1/2 subsystem of the spin-(1/2,5/2) Heisenberg dimer reads
\begin{align}
\allowdisplaybreaks
{\hat{\rho}}^{T_{1/2}}\!=\!
%\left(
\begin{blockarray}{r cc cc cc cc cc cc}
& \vert \tfrac{1}{2},\!-\!\tfrac{5}{2}\rangle& \vert \!-\!\tfrac{1}{2},\tfrac{5}{2}\rangle& \vert \tfrac{1}{2},\tfrac{5}{2}\rangle & \vert \!-\!\tfrac{1}{2},\tfrac{3}{2}\rangle &\vert \tfrac{1}{2},\tfrac{3}{2}\rangle  &\vert \!-\!\tfrac{1}{2},\tfrac{1}{2}\rangle&\vert \tfrac{1}{2},\tfrac{1}{2}\rangle&\vert \!-\!\tfrac{1}{2},\!-\!\tfrac{1}{2}\rangle&\vert \tfrac{1}{2},\!-\!\tfrac{1}{2}\rangle&\vert \!-\!\tfrac{1}{2},\!-\!\tfrac{3}{2}\rangle&\vert \tfrac{1}{2},\!-\!\tfrac{3}{2}\rangle&\vert \!-\!\tfrac{1}{2},\!-\!\tfrac{5}{2}\rangle\\
\begin{block}{r(cc cc cc cc cc cc)}
\langle \tfrac{1}{2},\!-\!\frac{5}{2}\vert \;\;\;&\rho_{6,6} &0&0& 0 & 0 & 0 & 0 & 0 & 0 & 0& 0 & 0 \\
\langle \!-\!\tfrac{1}{2},\frac{5}{2}\vert \;\;\;&0 & \rho_{7,7} &0&0& 0 & 0 & 0 & 0 & 0 & 0 & 0 & 0\\
\langle \tfrac{1}{2},\frac{5}{2}\vert \;\;\;&0 & 0 & \rho_{1,1} &  \rho_{7,2}& 0 & 0 & 0 & 0& 0 & 0  & 0 & 0\\
\langle \!-\!\tfrac{1}{2},\frac{3}{2}\vert \;\;\;&0&0 & \rho_{2,7}  & \rho_{8,8} & 0 & 0& 0 & 0 & 0 & 0 & 0 & 0\\
\langle \tfrac{1}{2},\frac{3}{2}\vert \;\;\;&0 &0 & 0 &0  &  \rho_{2,2} & \rho_{8,3} & 0 & 0& 0 & 0 & 0 & 0 \\
\langle \!-\!\tfrac{1}{2},\frac{1}{2}\vert \;\;\;&0 & 0 & 0 & 0 & \rho_{3,8} & \rho_{9,9} & 0 & 0 & 0 & 0 & 0 & 0\\
\langle \tfrac{1}{2},\frac{1}{2}\vert \;\;\;&0 & 0 & 0 &0 & 0 &0  &  \rho_{3,3} & \rho_{9,4}& 0 & 0 & 0 & 0 \\
\langle \!-\!\tfrac{1}{2},\!-\!\frac{1}{2}\vert \;\;\;&0 & 0 & 0 & 0 & 0 & 0 & \rho_{4,9} & \rho_{10,10} & 0 & 0 & 0 & 0\\
\langle \tfrac{1}{2},\!-\!\frac{1}{2}\vert \;\;\;&0 & 0 & 0 & 0 & 0 &0 & 0 &0  &  \rho_{4,4} & \rho_{10,5} & 0 & 0\\
\langle \!-\!\tfrac{1}{2},\!-\!\frac{3}{2}\vert \;\;\;&0 & 0 & 0 & 0 & 0 & 0 & 0 & 0 & \rho_{5,10} & \rho_{11,11} & 0 & 0\\
\langle \tfrac{1}{2},\!-\!\frac{3}{2}\vert \;\;\;&0 & 0 & 0& 0 & 0 & 0 & 0 &0 & 0 &0  &  \rho_{5,5} & \rho_{11,6} \\
\langle \!-\!\tfrac{1}{2},\!-\!\frac{5}{2}\vert \;\;\;&0 & 0 & 0& 0 & 0 & 0 & 0 & 0 & 0 & 0 & \rho_{6,11} & \rho_{12,12} \\
\end{block}
\end{blockarray}%\right)
\;\;.
\label{c4}
\end{align}
The non-zero elements of $\rho_{i,j}^{T_{1/2}}$ are explicitly defined in Eqs.~\eqref{bbbb2}-\eqref{bbbb18}.
\end{widetext}


\begin{thebibliography}{99}
\bibitem{Loss} D. Loss and D. P. DiVincenzo, Phys. Rev. A {\bf 57} (1998) 120.
\bibitem{Hayashi} T. Hayashi, T. Fujisawa, H. D. Cheong, Y. H. Jeong, and Y. Hirayama, Phys. Rev. Lett. {\bf 91} (2003) 226804.
\bibitem{Shulman}M. D. Shulman, O. E. Dial, S. P. Harvey, H. Bluhm, V.
Umansky, and A. Yacoby, Science {\bf 336} (2012) 202.
\bibitem{Shi} Z. Shi, C. B. Simmons, D. R. Ward, J. R. Prance, X. Wu, T. S.
Koh, J. K. Gamble, D. E. Savage, M. G. Lagally, M. Friesen,
S. N. Coppersmith, and M. A. Eriksson, Nat. Commun. {\bf 5} (2014) 3020.
\bibitem{Jones} N. C. Jones, R. Van Meter, A. G. Fowler, P. L. McMahon, J.
Kim, T. D. Ladd, and Y. Yamamoto, Phys. Rev. X {\bf 2} (2012) 031007.
\bibitem{Delbecq} M. R. Delbecq, T. Nakajima, P. Stano, T. Otsuka, S. Amaha, J. Yoneda, K. Takeda, G. Allison, A. Ludwig, A. D.Wieck, and S.
Tarucha, Phys. Rev. Lett. {\bf 116} (2016) 046802.
\bibitem{Watson} T. F. Watson, S. G. J. Philips, E. Kawakami, D. R. Ward,
P. Scarlino, M. Veldhorst, D. E. Savage, M. G. Lagally, M.
Friesen, S. N. Coppersmith, M. A. Eriksson, and L. M. K.
Vandersypen, Nature (London) {\bf 555} (2018) 633.
\bibitem{Berger} C. Berger, U. Huttner, M. Mootz, M. Kira, S. W. Koch, J. S.
Tempel, M. Assmann, M. Bayer, A. M. Mintairov, and J. L.
Merz, Phys. Rev. Lett. {\bf 113} (2014) 093902.
\bibitem{Li17} Y. Li, G. W. Holloway, S. C. Benjamin, G. A. D. Briggs,
J. Baugh, and J. A. Mol, Phys. Rev. B {\bf 96} (2017) 075446.
\bibitem{Gunlycke} D. Gunlycke, V.~M. Kendon, V. Vedral, and S. Bose, Phys. Rev. A. {\bf 64} (2001) 423021.
\bibitem{Arnesen}M. C. Arnesen, S. Bose, and V. Vedral, Phys. Rev. Lett. {\bf 87} (2001) 017901. 
\bibitem{Wang} X. Wang, Phys. Rev. A {\bf 64} (2001) 123131.
\bibitem{Wang3} X. Wang, H. Fu, and A.~I. Solomon, J. Phys. A {\bf 34} (2001) 11307.
\bibitem{Wang2} X. Wang, Phys. Rev. A {\bf 66} (2002) 044305.
\bibitem{Zhou}  L. Zhou, H.~S. Song, Y.~Q. Guo, and C. Li, Phys. Rev. A
{\bf 68} (2003) 024301.
%------
\bibitem{Asoudeh} M. Asoudeh and V. Karimipour, Phys. Rev. A {\bf 70} (2004) 052307.
\bibitem{Canosa} N. Canosa and R. Rossignoli, Phys. Rev. A {\bf 73} (2006) 022347.
%\bibitem{Rojas12} O. Rojas, M. Rojas, N.~S. Ananikian, and S.~M. de Souza, Phys. Rev. A {\bf 86} (2012) 042330.
\bibitem{Ananikian}  N.~S. Ananikian, L.~N. Ananikyan, L.~A. Chakhmakhchyan, and O. Rojas, J. Phys.: Condens. Matter {\bf 24} (2012) 256001.
\bibitem{Torrico}  J. Torrico, M. Rojas, S.~M. De Souza, O. Rojas, and N.~S. Ananikian, EPL {\bf 108} (2014) 50007.
%\bibitem{Strecka} J. Strečka, O. Rojas, T. Verkholyak, M.~L. Lyra, Phys. Rev E {\bf 89} (2014) 022143.
%\bibitem{Rojas14} M. Rojas, S.~M. de Souza, and O. Rojas, Phys. Rev. A {\bf 89} (2014) 032336.
\bibitem{Luitz} D.~J. Luitz, N. Laflorencie, and F. Alet, Phys. Rev. B {\bf 91} (2015) 081103.
\bibitem{Rojas} O. Rojas, M. Rojas, S.~M. de Souza, J. Torrico, J. Strečka, and M.~L. Lyra, Physica A {\bf 486} (2017) 367.
\bibitem{Arian}  H. Arian Zad and N. Ananikian, Sol. State Commun. {\bf 276} (2018) 24.
\bibitem{Souza19} F. Souza, M.~L. Lyra, J. Strečka, and M.~S.~S. Pereira, J. Magn. Magn. Mater {\bf 471} (2019) 423.
\bibitem{Verissimo} L.~M. Veríssimo, M.~S.~S. Pereira, J. Strečka, and M.~L. Lyra, Phys. Rev. B {\bf 99} (2019) 134408.
\bibitem{Karlova19}  K. Karl\!$'$ov\'a, J. Stre\v{c}ka, and M.~L. Lyra, Phys. Rev. E {\bf 100} (2019) 042127.
\bibitem{Wang2006} X. Wang and Z.~D. Wang, Phys. Rev. A {\bf 73} (2006) 064302.
\bibitem{Cenci2} H. \v{C}en\v{c}arikov\'a and J. Stre\v{c}ka, Phys. Rev. B {\bf 102} (2020) 184419.
%\bibitem{Cenci2020} H. \v{C}en\v{c}arikov\'a and J. Stre\v{c}ka, {\it Conventional and inverse magnetocaloric and electrocaloric effect ...}, arXiv:xxx.
\bibitem{Varga21} H. Vargov\'a, J. Stre\v{c}ka and N. Toma\v{s}ovi\v{c}ov\'a, arXiv: 2107.14620. 
%------

\bibitem{Sun06} Z. Sun, X. Wang, A. Hu, and Y.-Q. Li, Physica A {\bf 370} (2006) 483.
\bibitem{Hao} X. Hao and S. Zhu, Phys. Lett. A {\bf 366} (2007) 206.
\bibitem{Guo} J.-L. Guo, X.-L. Huang, and H.-S. Song, Phys. Scr. {\bf 76} (2007) 327.
\bibitem{Wang08} F. Wang, L.-P. Fu, and K.-T. Guo, Commun. Theor. Phys. {\bf 50} (2008) 341.
\bibitem{Huang2008} H. Huang, X. Wang, Z. Sun, and G. Yang, Physica A {\bf 387} (2008) 2736.
\bibitem{Yang} G.-H. Yang and L. Zhou, Phys. Scr. {\bf 78} (2008) 025703.
\bibitem{Wang09} F. Wang, H. Jia, H. Zhang, X. Zhang, and S. Chang, Sci. China Ser G. {\bf 52} (2009) 1919.
\bibitem{Zhu} G.-Q. Zhu, Cent. Eur. J. Phys. {\bf 7} (2009) 135.
\bibitem{Sun09} Z. Shun, X. Wang, and H.-N. Xiong, Physica A {\bf 388} (2009) 1337.
\bibitem{Guo10} K.-T. Guo, M.-Ch. Liang, H.-Y. Xu, and Ch.-B. Zhu, J. Phys. A: Math. Theor. {\bf 43} (2010) 505301.
\bibitem{Guo11} K.-T. Guo, M.-Ch. Liang, H.-Y. Xu, and Ch.-B. Zhu, Sci. China Phys. Mech. Astron. {\bf 54} (2011) 491.
\bibitem{Solano1} E. Solano-Carrilo, R. Franco, and J. Silva-Valencia, Physica A {\bf 390} (2011) 2208.
\bibitem{Solano2} E. Solano-Carrilo, R. Franco, and J. Silva-Valencia, Phys. Lett. A {\bf 375} (2011) 1032.
\bibitem{Li} S.-S. Li, T.-Q. Ren, X.-M. Kong,and K. Liu, Physica A {\bf 391} (2012) 35.
\bibitem{Xu} S. Xu, X. Song, and L. Ye, Quantum Inf. Process. {\bf 13} (2014) 1013.
\bibitem{Guo2014} K.-T. Guo, S.-H. Xiang, H.-Y. Xu, and X.-H. Li,  Quantum Inf. Process. {\bf 13} (2014) 1511.
\bibitem{Zhou15} Ch.-B. Zhou, S.-Y. Xiao, C. Zhang, G. Wu, and Y.-Q. Ran, Physica B {\bf 477} (2015) 40.
\bibitem{Zhou16} Ch.-B. Zhou, S.-Y. Xiao, C. Zhang, G. Wu, and Y.-Q. Ran, Int. J. Theor. Phys. {\bf 55} (2016) 875.
\bibitem{Han} S.~D. Han, T. T\"{u}fek\c{c}i, and T.~P. Spiller, Int. J. Theor. Phys. {\bf 56} (2017) 1474.
\bibitem{Adamyan} Z. A. Adamyan, S. A. Muradyana, and V. R. Ohanyan,  Contemp. Phys. {\bf 55} (2020) 292.
\bibitem{Holleitner} A.W. Holleitner, K. Knotz, R.C. Myers, A.C. Gossard, and D. D. Awschalom, Appl. Phys. Lett. {\bf 85} (2004) 5622.
%-------
%\bibitem{Pei} Y. Pei, M. Verdaguer, O. Kahn, J. Sletten, and J.~P. Renard,   Inorg. Chem. {\bf 26} (1987)  138.
\bibitem{Kahn} O. Kahn, Y. Pei, M. Verdaguer, J.~P. Renard, and J. Sletten, J. Am. Chem. Soc. {\bf 110} (1988) 782.
\bibitem{Hagiwara99} M. Hagiwara, K. Minami, Y. Narumi, K. Tatani, and K. Kindo, J. Phys. Soc. Jpn. {\bf 69} (1999) 2214.
\bibitem{Hagiwara98} M. Hagiwara, K. Minami, Y. Narumi, K. Tatani, and K. Kindo, J. Phys. Soc. Jpn. {\bf 67} (1998) 2209.
%%%%%%%%%%

\bibitem{Peres} A. Peres, Phys. Rev. Lett. {\bf 77} (1996) 1413.
\bibitem{Horodecki} M. Horodecki, P. Horodecki, and R. Horodecki, Phys. Lett. A {\bf 223} (1996) 1.
\bibitem{Vidal} G. Vidal and R.~F. Werner, Phys. Rev. A {\bf 65} (2002) 032314.

\end{thebibliography}
\end{document}